# Uncoupled high-latitude wave models in COAMPS

W. Erick Rogers
Timothy J. Campbell
Jie Yu
Richard A. Allard

Naval Research Laboratory, Code 7322, Stennis Space Center, MS, USA

Corresponding author: W. Erick Rogers (w.e.rogers.civ@us.navy.mil)

# CONTENTS

**Executive Summary**

This report describes six demonstration cases of two numerical ocean wave models in high latitude regions where waves interact with sea ice. The two wave models are SWAN (Simulating WAves Nearshore) and WW3 (WAVEWATCH III), run in uncoupled mode within the Navy's coupled regional modeling system, COAMPS® (Coupled Ocean Atmosphere Mesoscale Prediction System). The COAMPS software handles a large majority of the tasks associated with the setup, running, and post-processing of the wave models. All six cases are cycling runs with 12-hour increments, each providing a continuous hindcast of four to 26 days duration. SWAN is applied in a Bering Strait case and two Gulf of Bothnia cases. WW3 is applied in a Sea of Okhotsk case and two Barents Sea cases. Verification is performed by visual inspection of model output fields, and by comparing model runs with alternative settings. In the standard configuration, forcing comes from archived global model output, including information on surface wind vectors, sea ice concentration and thickness, and surface current vectors, and the wave model uses a new empirical formula for dissipation of wave energy by sea ice that is dependent on ice thickness. The Barents Sea cases are compared to spectral wave data from satellite (SWIM instrument on CFOSAT) and from motion sensors deployed on the ice by the Norwegian Meteorological Institute. Experiments are performed with non-standard settings, 1) disabling the dissipation by sea ice, 2) using an older formula for dissipation by sea ice which does not depend on ice thickness, 3) using higher resolution wind and sea ice concentration forcing fields, and 4) omitting surface currents. The impact of these settings on model skill is quantified by comparison to the observations. The skill is also compared to that from a global (thus, lower resolution) ocean wave model.

# 1. Introduction

COAMPS®[1] (Coupled Ocean Atmosphere Mesoscale Prediction System) is the US Navy's propriety regional coupled modeling system, e.g., Campbell et al. (2010); Smith et al. (OM 2013); Doyle et al. (TOS 2014). New routines for treatment of sea ice have been implemented in each of the two wave models used by the system, which are WAVEWATCH III® (WW3, Tolman 1991, WW3GD 2019) and SWAN (Booij et al. 1999). These new routines are available for tightly coupled modeling, coupled to mesoscale models:
- Atmospheric model: "COAMPS-atmosphere" (Hodur 1997)
- Ocean model: NCOM (Navy Coastal Ocean Model, Martin (2000))
- Ice model: CICE (Community ice code, Hunke et al. (2015))

However, this report does not deal with tightly coupled modeling. Instead, in all demonstration cases shown in this report, the atmosphere, ocean, and ice forcing are provided in "data" mode, meaning that the fields are created from data files associated with a model running separately (global models, in this case), rather than by running a regional implementation of an atmosphere, ocean, or ice model within the same COAMPS implementation. This simplicity is a core strategy: we reduce the complexity such that new wave model implementations can be created with small effort, recognizing current manpower constraints at operational centers. The new product will permit FNMOC to easily stand up new wave model implementations which include ice and ocean forcing. This will be a significant advance relative to existing standalone SWAN models which omit ice and ocean forcing and existing standalone WW3 models which omit ocean forcing.

In an operational context, global forcing is available as follows:
- Atmospheric model: "NAVGEM" (Hogan et al. 2014). Fields are currently available at 0.18° (20 km) resolution which is significantly coarser than a COAMPS-atmosphere model run with a tightly coupled air-ocean-ice-wave COAMPS implementation.
- Ocean and Ice model: Fields are currently available Earth System Prediction Capability (EPSC) Deterministic v2 at 1/25° (4.4 km) resolution[2]. This is somewhat coarser than a typical SWAN implementation (1 to 2 km resolution), but is comparable to, or finer than, a typical regional WW3 implementation (4 to 7 km resolution).

# 2. Code changes

## 2.1. SWAN code changes

In the case of SWAN (Booij et al. 1999), NRL introduced the initial and subsequent code which allows reading and use of sea ice fields and added a parameterization for dissipation of wave energy by sea ice, $S_{ice}$, that is dependent on wave frequency. Prior to this, SWAN did not have any sea ice capability. This was first available in the public release version 41.31.

We made our first changes using an SVN repository maintained by the U. Washington. All coding and verification was performed by NRL, while U. Washington performed initial validation. This work was published in the gray literature: Rogers (2019) and Kumar et al. (2020). We made changes to the official SWAN code at TU Delft, via their code manager (Dr. Zijlema). To make those changes available to COAMPS, we needed to port them to the version of SWAN used in COAMPS.

---

[1] COAMPS is a registered trademark of the Naval Research Laboratory.
[2] The Global Ocean Forecasting System (GOFS) 3.1 system was turned off earlier this year (2024).

Follow-up changes by NRL to the official Delft version of SWAN included new empirical parameterizations for dissipation of wave energy by sea ice, $S_{ice}$, which are dependent on both wave frequency and ice thickness, SWAN version 41.31AB. Relevant dissipation schemes are Doble et al. (2015), Meylan et al. (2018), and Rogers et al. (2021b). This was an update on the previous implementation by Rogers in which $S_{ice}$ is dependent on wave frequency alone, SWAN version 41.31. It was necessary to port the most recent updates (41.31AB) from the Delft version to the NRL trunk. The code was evaluated using idealized tests. NRL also merged in later updates to the Delft version, 41.45, into a branch created from the NRL trunk. However, we observed a problem when using the code for idealized tests. The cause of the problem is unknown, but it is apparently associated with MPI (i.e., SWAN parallelization). We determined that problem also exists in the Delft version of the code. Because of this, the NRL trunk was not updated to version 41.45. The changes from 41.31AB to 41.45 are not relevant to wave-ice interaction, but some features may be considered for use as new options in later version of COAMPS.

Rogers et al.'s (2021b) empirical formula for $S_{ice}$ is illustrated in Figure 1.

During FY23, we noticed a problem in the SWAN code which causes the model to run slowly (roughly doubling the run time) when ice-related parameters (e.g. ice concentration, ice thickness, ice-induced dissipation) are requested as output from SWAN. The cause of the problem was identified by running profiler software on the DSRC for a simple test case. The code was adapted to prevent the slow-down for the case of output of ice concentration and thickness, verified using a simple test case, and committed to the NRL-SSC SVN repository and the NRL-MRY github repository. A second code modification was made to prevent slow-down for the case of output of integrated dissipation. This applies to other dissipation types, e.g., integrated dissipation from whitecapping, and so it is useful for other applications also. This second modification was verified using a simple test case and committed to the NRL repository.

Here is the timeline relevant to the official SWAN code:
- Jan.-Mar 2019: We added $S_{ice}$ to a development version of 41.20AB and tested it. This $S_{ice}$ routine did not depend on ice thickness. Though the code allowed ice thickness to be read in by SWAN, the model did not use the parameter.
- May 2019: We worked with Zijlema (TU Delft) on user's manual (documentation) for the new features.
- June 2020: TU Delft release version with $S_{ice}$, 41.31.
- May 2021: We provided SWAN with new $S_{ice}$ and test cases to Zijlema (TU Delft). It was published as "patch B" to 41.31 not long after. This $S_{ice}$ included options for dependency on ice thickness.
- March 2023: We noticed irregularities in output from 41.45, associated with MPI.
- April 2023: We decided not to update NRL version of SWAN to 41.45
- June 2023: Zijlema provided a temporary workaround for problem with 41.45 (specifically, avoid use of first order propagation scheme, "PROP BSBT").

For reference, on the official SWAN website:
1. Description of major version changes, denoted by changes to version number may be found here: https://swanmodel.sourceforge.io/modifications/modifications.htm
2. Description of minor version changes, denoted by addition of letters after the version number, like "a", "A", "AB". These are referred to as "patches" even though they may be new features (not necessarily bug fixes, as is implied by the web address). Documentation on these can be found here: https://swanmodel.sourceforge.io/modifications/modifications.htm#bugfixes

Here are changes that we made to the SWAN code in the NRL repository (this is the version of SWAN used in COAMPS):
- March 2023: Update from 41.31 to 41.45. This was done to make the NRL changes to the $S_{ice}$ routines (specifically, adding $S_{ice}$ which depends on ice thickness), available in the version of SWAN used by COAMPS.
- April 2023: Update to 41.31AB, by rolling back the prior update, taking COAMPS SWAN from 41.45 to 41.31, and then updating from 41.31 to 41.31AB.
   - ➔ In SWAN, command syntax "IC4M2" in 41.31 is replaced with syntax "SICE". The latter has four sub-options, one of which is equivalent to the earlier "IC4M2" option.
- June 2023:
   - ➔ We updated SWAN to prevent subroutine ADDDIS from being called for output parameters that don't require ADDDIS (specifically: NPL, AICE, HICE). This is important, because when ADDDIS is called, total computation time can increase by factor 1.5 to 2.5. For the most part, ADDDIS is only required when computing output parameters that are based on integration of source terms, like wave-to-ocean stress, or total dissipation, or dissipation from bottom friction, etc.
- Jan 2024:
   - ➔ We changed SWAN such that AICE and HICE are not allowed to be negative. By definition these should never be negative, but we found some negative values in CICE fields, and without the limiter, the non-physical values can cause the wave models to fail.

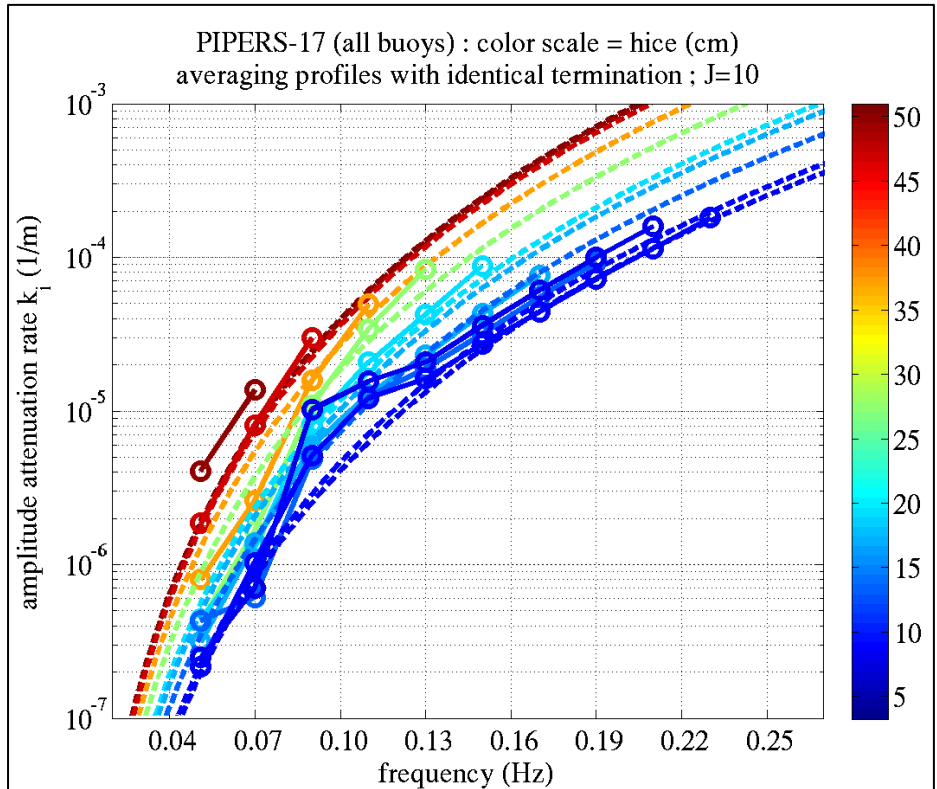

Figure 1. Dissipation rate. Dependency on wave frequency and ice thickness is shown. Circle-lines: inversion results from PIPERS experiment, Rogers et al. (2021a). Dashed lines: parametric model of Rogers et al. (2021b) ('R21B') and Yu et al. (2022), which is based on Reynolds length-scale normalization proposed by Yu et al. (2019), calibrated to this PIPERS dataset. This is the default setting of method known as R21B in SWAN and IC4M9 in WW3.

## 2.2. WW3 code changes

Sea ice had previously been implemented in WW3 (Tolman 1991, WW3DG 2019) by Tolman (2003). However, there was no dissipation of wave energy by ice. Instead, there was a "partial blocking" of wave energy by sea ice, according to ice concentration. This was a feature of the model's propagation, rather than a physics parameterization, and it did not depend on wave frequency, and thus did not produce the well-known low-pass filter effect of sea ice on ocean waves. The dissipation of wave energy by sea ice, using the wave model source/sink term $S_{ice}$ (i.e., via model dynamics) was first implemented by Rogers and Orzech (2013). This included the $S_{ice}$ routines denoted as "IC1" and "IC2". The latter has since been updated by Ifremer (France) (Stopa et al. 2016). NRL and Clarkson University implemented the visco-elastic model of Wang and Shen (2010) and it became known as "IC3" (Rogers and Zieger (2014), Cheng et al. (2017)). The IC3 model depends on ice thickness and rheological parameters. This routine was quite complex, requiring root-finding and root-selection. It has a positive feature of quantifying the impact of sea ice on the real part of the wavenumber, but this had a negative consequence in practice: increases to group velocities could result in violation of CFL. Seeking a more practical approach, NRL later implemented an option for empirical/parametric computations of $S_{ice}$ denoted "IC4". This was originally published in Collins and Rogers (2017) with several "sub-methods" (IC4M1 through IC4M6) and has since expanded with additional "sub-methods", IC4M7 through IC4M9.

The work of Rogers et al. (2016) and Rogers et al. (2018a,b) were adapted for sub-methods IC4M2, IC4M5, and IC4M6. These parameterizations for $S_{ice}$ are <u>not dependent on ice thickness</u>.

In IC4M6, the dissipation by sea ice $k_i(f)$ is described as a step function which is provided to WW3 using a namelist. Where IC4M6 is used in this report, it refers to the following settings:

```
$ IC4METHOD = 6 - Simple ki step function via namelist
 &SIC4 IC4METHOD = 6,
IC4FC=0.045,0.055,0.10,0.15,0.20,0.25,0.30,0.35,0.40,99.0,
IC4KI=1.0e-6,2.0e-6,2.94e-06,4.27e-06,7.95e-06,2.95e-05,
      1.12e-04,2.74e-04,4.95e-04,8.94e-04
       /
```

This step-function is denoted as "IC4M6H" in Figure 90 of Rogers et al. (2018a) and "IC4M6H1" in Table 1 of Rogers et al. (2018b). It is based on $k_i(f)$ from the pancake and frazil ice case of Rogers et al. (2016).

The only reference for the code for the IC4 expansions (IC4M7 to IC4M9) are the WW3 user manual (e.g., WW3DG 2019), and the original source material upon which the methods are based. All three new sub-methods (IC4M7 through IC4M9) are <u>dependent on ice thickness</u>:
- IC4M7: Doble et al. (2015) (here "M7" denotes "sub-method 7")
- IC4M8: Liu et al. (2020).
- IC4M9: Parametric model of Rogers et al. (2021b) ('R21B') and Yu et al. (2022), which is based on Reynolds length-scale normalization proposed by Yu et al. (2019), calibrated to the dissipation estimates by Rogers et al. (2021a) for the PIPERS field experiment, for broken ice floes in the Antarctic, north of the Ross Sea. A more recent calibration, not used here, is given by Yu (2022); this is a refinement to give a better fit to laboratory observations while still providing a good match to field observations.

IC4M9 is illustrated in Figure 1. In Figure 2, we verify that the implementation in WW3 is correct, and compare with IC4M6.

The following is a timeline of changes to the WW3 code which are relevant to the "IC4M9" implementation.
- Dec 2022 (COAMPS WW3)
  - This was in branch '1-R21B-in-IC4'.
  - We started by porting in the latest modifications to IC4M2 through IC4M7 from another WW3 branch.
  - We then ported IC4 with sub options IC4M8 and IC4M9 from SWAN
    - https://github.nrlmry.navy.mil/COAMPS/WW3_v6.07/issues/1
    - commit log: "Introduce R21B variant of Sice, as an option, to IC4 subroutine of WW3. This feature already exists in SWAN v41.31AB."
- Sep 2023 (COAMPS WW3)
  - We integrated $S_{ice}$ in output field post-processing ww3_outf (and therefore COAMPS 'ffout' (flat file output) post-processing).

- Nov 2023 (COAMPS WW3)
    - We changed the default value of ICEHMIN to 0. Without the change, the default ice thickness $h_{ice}$ globally was 20 cm. This did not affect $S_{ice}$ in locations where ice concentration $a_{ice}$ was zero, but it did result in strange plots (e.g., non-zero ice thickness in the tropics).
- Jan. 2024 (Official WW3).
    - The IC4M9 code was ported to the official WW3 code. (https://github.com/NOAA-EMC/WW3/pull/1176)
- Feb 2024 (COAMPS WW3)
    - Ice thickness is changed so that it is not allowed to be negative. This is similar to another commit to the SWAN code.
- May 2024 (COAMPS WW3). Here we committed two sets of bug fixes. Both are related to the program ww3_sprst which is used to generate boundary forcing using WW3 restart files.
    - BUGFIX SET #1. Primary symptom: Program ww3_sprst fails during run time, but only when the program is compiled in "debugging mode". Secondary symptom (not known a priori), ww3_sprst would sometime use the wrong water depth to calculate the Jacobian for converting wave spectra from one type to another (energy spectra vs. wave action spectra). Summary of changes:
        - Source code file 'w3iobcmd.ftn':
            - 1A) Change to use file unit 6 instead of NDSE, NDST during call to W3SET0 since NDSE NDST have not necessarily been set prior to the call. This addresses a complaint raised by the compiler when using the "debugging" options.
            - 1B) ZB (water depth) used in wavenumber calculations was not valid, since W3SETG was missing. This is a significant bug which causes incorrect calculations but doesn't necessarily cause the model to stop with an error.
        - Source code file 'w3iogrmd.ftn':
            - 2A) Change to use file unit 6 instead of NDSE, NDST during call to W3SET0 since NDSE NDST have not necessarily been set prior to the call. This addresses a complaint raised by the compiler when using the "debugging" options.
        - Source code file 'ww3_sprst.ftn':
            - 3A) Fix problem where water depth was set in cases where water depth is not valid (mapsta=0)
            - 3B) Unit numbers are set to NDSO=6, NDSE=6, following settings in other WW3 programs. NDSE is no longer pulled from W3ODATMD, since the value there is not necessarily valid.
    - BUGFIX SET #2.
        - Symptom: The ww3_sprst program fails during run time. Most critically, the program fails even for cases when the "debugging" options are not enabled during compile. The program fails with a cryptic message "free(): invalid pointer".
        - Debug tracing reveals the subroutine containing the error, but does not reveal the actual cause.

- We determined that the ww3_sprst program was trying to pull spectra using incorrect grid information and failing when the arrays were insufficiently sized.
- Our fix: at the end of subroutine W3IOBC, we changed the code to set the pointers WDATAS (via subroutine W3SETW) and GRIDS (via subroutine W3SETG) back to what they were when the program enters this subroutine. In the case of W3SETW, it is to avoid the cryptic error. In the case of W3SETG, it is just for "abundance of caution", since only W3SETW affects "invalid pointer" behavior.

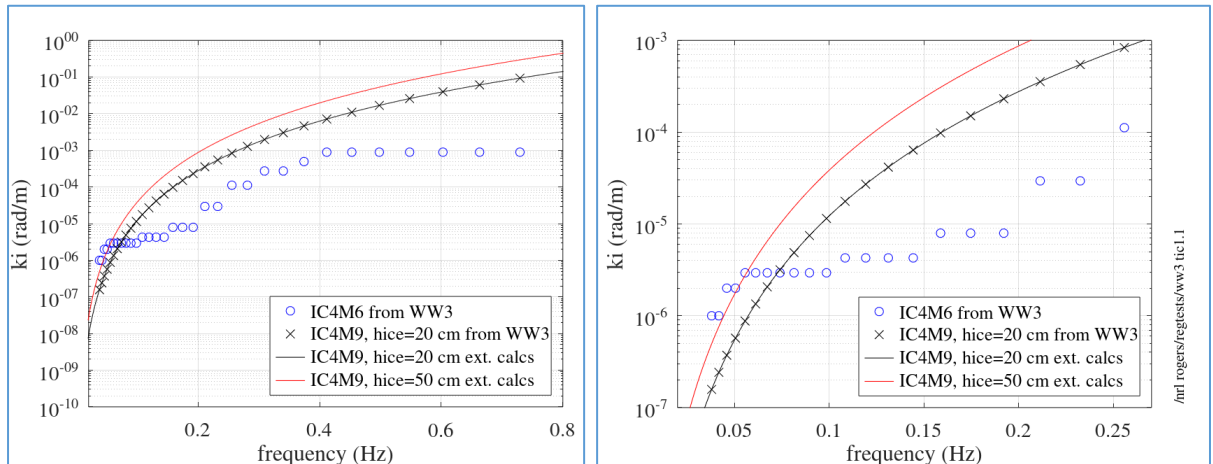

Figure 2. Verification of IC4M9 code in WW3, and comparison to IC4M6. Right panel is identical to left panel, except with different axis limits (zoomed in). IC4M6 is described in the text and is indicated here with blue circles.

## 3. COAMPS instruction methods

When the wave models are run within the COAMPS system, many of the tasks that would normally be performed by the user to set up, perform, and post-process the wave model simulations are automated or semi-automated. The following are some examples of tasks that are performed by the system:
- Pre-processing forcing files.
- Post-processing output to .nc format.
- Setting up the model instruction files.
- Organizing output and log files.
- Cycling the system, including transferring restart information from one run cycle to the next.

The following are examples of tasks that must be performed by the user:
- The user must provide information on where to find the forcing files (e.g., winds, boundary forcing). Utilities are provided to make the process of preparing these files easier.
- The user must provide information about the desired geographic grid. Utilities are available to make this easier, including:

- o A utility to calculate grid-related characteristics, such as the maximum time step for stable propagation (CFL).
- o A utility to create a kml file which can be loaded into google earth to visualize the grid.
- Plotting results (though the ncview utility can generate simple plots directly from .nc files).
- If the user desires non-default settings (e.g., spectral grid, physics parameterization selection and settings, numerics), those of course must be provided.
- If the user desires non-default output, information of course must be provided.
- The user must provide instructions about the time frame and run cycle interval.

The COAMPS system is not a GUI. In principle, a GUI can be created or adapted as a wrapper to the COAMPS system. COAMPS On-Scene (COAMPS-OS) packages a version of COAMPS with a GUI. COAMPS-OS is not used here.

### 3.1. COAMPS Configuration files

COAMPS stores configuration information in ascii files with extension .rc. Below we summarize the .rc files most relevant to our use of COAMPS here, i.e., running wave models in uncoupled mode. These descriptions are not intended as usage guides. The reader is referred to COAMPS (2022a,b) for specific information about these files.

`./input/atm.rc`
This file has instructions for the uncoupled wave model to find atmospheric forcing (U and V components of 10-m wind speed), and the time intervals for the files and fields within the files. In the "vanilla" case of running COAMPS, these are binary files from NAVGEM staged in advance using a script provided by Tim Campbell, 'do_windprep_NAVGEM_wind.s'. The script uses local 0.5° NAVGEM archives to create files in a specific format, with names like 'wnd_ucmp.2024050100.bin', which are then usable by COAMPS. In cases where we use other atmospheric forcing (namely, higher resolution NAVGEM) by processing to WW3 format outside of COAMPS, the atm.rc file is not used. On the DSRC, NAVGEM fields are typically available at resolution better than 0.5°; available resolution depends on date.

`./input/ice.rc and ./input/ocn.rc`
When the wave model is run in uncoupled mode (which is the topic of this report), these files point to local archives of netcdf files containing output from GOFS, 1/12° resolution, "GLBy0.08". On the DSRC, better resolution (1/25°) is available for some dates.

`./input/wav.rc`
This file contains information on wave model settings. Some are required, such as grid information. Other settings are optional, e.g., selection and calibration for physics parameterization: if these settings are not specified, the model will use default settings, specified in the wav.rc file in the /defaults/ directory of the COAMPS installation. The file contains information on the number of processors, the types of forcing to include, and how to receive the forcing (which is from files, in our case). The file also specifies the method of receiving boundary forcing, where to find the relevant files, and information about the files.

```
./inc/suite_proj.rc
```
This file is automatically generated.

```
./suite.rc
```
This file specifies the first and last run cycle of the sequence, it is created using the script 'project_create_suite' which is provided in the COAMPS installation.

```
./job.rc
```
Though it is generally brief, this is the primary .rc file. It contains the project name and (optionally) instructions to skip the processing of forcing from other component models. In the "vanilla" cases, this processing is fully enabled, but with some cases, we disable the processing to perform experiments with alternative forcing.

### 3.2. Cylc

Cylc is a scheduling system developed by NIWA (National Institute of Water and Atmospheric Research of New Zealand) and is used in USN operations for systems such as ESPC, COAMPS, and the standalone global wave model.

The COAMPS cycling system is initiated and monitored using cylc commands. A few examples of cylc commands are given here:
- `cylc register barents2`
- `cylc run barents2`
- `cylc monitor barents2`
- `cylc stop --kill barents2`
- `cylc log --file=e barents2 ucyc_wav_setup.20240404T0000Z`
- `cylc trigger barents2 ucyc_wav_setup.20240404T0000Z`

### 3.3. COAMPS output

The primary output from the wave models consists of wave parameters on the model computational grid. During runtime, COAMPS stores this information in flat binary files with a separate file for each time and parameters in <workdir>/wav/ffout/, and at the end of each cycle, it combines the parameters and stores the output in netcdf files, one per time (tau), in <workdir>/wav/ncout/.

### 4. Demo cases

We created and ran six COAMPS cycling demos for four areas: 1) Bering Strait (SWAN, 2.3 km), 2) Gulf of Bothnia (SWAN, 1 km and 2 km), 3) Okhotsk Sea (WW3 6 km), and 4) the Barents Sea (WW3 6 km). In all cases, the atmosphere, ocean, and ice forcing are provided in "data" mode, as described in the Introduction section.

The Bering Strait (SWAN) case receives boundary forcing from a point-output spectral file created from a global WW3 (IRI-1/4) hindcast. The Okhotsk Sea and Barents Sea cases are both regional WW3 cases, receiving boundary forcing from the Spritzer software in COAMPS which generates boundary forcing from WW3 restart files; the latter files are again created from a global WW3 (IRI-1/4) hindcast. Thus, two modes of providing boundary forcing are demonstrated. The Gulf of Bothnia (SWAN) case is run without boundary forcing.

All three demos utilize sea ice concentration and thickness fields provided by CICE. The ice concentration is used to proportionally reduce the deep-water source functions (e.g., the wind input source function). The ice concentration and thickness are both used to estimate the dissipation of wave energy by sea ice. For that dissipation, we use the method described in Rogers et al. (2021b), calibrated primarily using the study of Rogers et al. (2021a). This is an advance over existing dissipation methods presently in use by all Navy wave models (including "legacy" standalone WW3 (a 10-grid global system previously run at FNMOC-SSC, described in Rogers et al. (2012, 2014), IRI standalone WW3 (Fan et al. 2021), WW3 in ESPC (Metzger et al. 2023), and WW3/SWAN in older versions of COAMPS, in that it uses ice thickness information.

The wave model implementations produced field output of parameters such as re-gridded ice fraction, re-gridded ice thickness, significant wave height, peak period, peak wave direction, and spectral-integrated dissipation of wave energy by sea ice. We verified that these outputs were reasonable. We also evaluated the spectral distribution of dissipation by sea ice at specific points, again to verify that results conformed to expectations.

During the process of creating the demos, default COAMPS settings associated with the wave model were modified. Some settings were associated with the new "waves in sea ice" features, while others were more general. Some technical problems were identified, such as the problems with ww3_sprst described earlier, and the COAMPS and wave model codes were updated to address them.

### 4.1. Compute resources

Unless otherwise noted, computation times given here are for computations on a linux workstation with 48 physical cores, Intel(R) Xeon(R) Gold 5318N CPU @ 2.10GHz. In the case of SWAN, we did not use all 48 physical cores, as noted below.

### 4.2. Settings common to all SWAN demos

Here we list setting common to all SWAN demos.
- number of directions: 36; thus, directional resolution is 10°.
- deepwater physics (wind input and whitecapping source terms): ST6 physics with settings: 'GEN3 ST6 2.8E-6 3.5E-5 4.0 4.0 UP HWANG VECTAU U10PROXY 32.0 AGROW' [We recognize that some of these descriptions are very cryptic, but for purposes of brevity, we refer the reader to SWAN (2019) rather than explaining their meaning here.]
- Swell dissipation with setting: 'SSWELL ARDHUIN 1.2'
- Bottom friction with setting: 'FRIC JON 0.019'
- Four-wave nonlinear interactions with setting: 'QUAD iquad=8'
- Dissipation by sea ice with setting: 'SICE R21B': this gives $k_i = 2.9 h_{ice}^{1.25} f^{4.5}$ using the Rogers et al. (2021a) dataset: see Rogers et al. (2021b) for details. It is equivalent to the default calibration of IC4M9 in WW3.
- Nonstationary computation with time step of 6 minutes.
- Processors used for the model, 'fcst_proc_count': 32
- Processors used for post-processing, 'post_proc_count': 24

- SWAN model version: 41.31AB-NRL4.4b

### 4.3. Settings common to all WW3 demos

Here we list settings common to all WW3 demos. These are the settings used *unless otherwise noted*; some settings are changed in non-vanilla cases for specific demo, e.g., experimenting with a different $S_{ice}$ formula.
- WW3 model version: 6.07 (available with COAMPS at time of writing)
- number of directions: 36; thus, directional resolution=10°
- lowest computed frequency: 0.038 Hz. This setting is used for consistency with ESPC WW3.
- highest computed frequency: 0.729 Hz. Again, this setting is used for consistency with ESPC WW3.
- frequency increment: factor 1.1
- number of frequencies: 32
- nominal resolution: 6 km
- propagation and anti-GSE. [Again, for purposes of brevity, we refer the reader to WW3DG (2019) for explanations of these settings.]
   ⇨ PR3 with GSE averaging.
   ⇨ WDTHCG = 4.8, WDTHTH = 5.8
   ⇨ Averaging area factor $C_g$:  4.80
   ⇨ Averaging area factor $\theta$:  5.80
- ST4 wind input setting:  Z0MAX = 1.002
- Dissipation by sea ice with setting for $S_{ice}$ routine:
   ⇨ IC4METHOD=9, also known as "IC4M9"
   ⇨ IC4CN=  2.90,  4.50, 3*0.00000000,
   ⇨ this gives $k_i = 2.9 h_{ice}^{1.25} f^{4.5}$ using the Rogers et al. (2021a) dataset: see Rogers et al. (2021b) for details.
- time steps:
   ⇨ Maximum global time step (s):  600.00
   ⇨ Maximum CFL time step X-Y (s):  200.00
   ⇨ Maximum CFL time step k-theta (s):  200.00
   ⇨ Minimum source term time step (s):  10.00
- Processors used for the model, wav.rc fcst_proc_count: 48
- Processors used for post-processing, wav.rc post_proc_count: 24

### 4.4. Demo case 1: Bering Strait (SWAN), Dec. 2022

*4.4.1. Settings and run time*

The following settings were used for this demo case. In some places, we refer to local directories which are only available to those on the NRL 7300 linux network, and they are given here for internal documentation; other readers may disregard this information.
- local directory: /net/chosin/export/backup_rogers/Bering_Strait/
- "work directory" of record: /work_Oct6A/
- Duration: 10 run cycles 20221201T0000Z to 20221205T1200Z, 12 hours each

- Surface currents, water levels, ice fraction, and ice thickness from local HYCOM/CICE archives: /u/HYCOM/GLBy0.08/nc
- Surface winds from local NAVGEM archives, 0.5°
- lowest computed frequency: 0.038 Hz. This setting is used for consistency with ESPC WW3.
- highest computed frequency: 0.971 Hz
- frequency increment: factor 1.1
- number of frequencies: 35
- longitudes: 185 to 201, nx=321, spacing = 0.05°, so approximately 750 km at 2.3 km spacing
- latitudes: 62.5 to 67.5, ny=251, spacing = 0.02°, so 555 km with spacing between 2.1 and 2.6 km.
- 45219 sea points
- Boundary forcing from a global WW3 run (IRI-1/4).
- BOUN WW3 '../../BCs.20221115.20230101.ascii' FRE OPEN
- anti-GSE diffusion strength: 1.0 HR
- run time: approximately 50 minutes
- Timings:
  - total time from one cycle to another (12-hour cycle) = 304 sec
  - 72-hour forecast: 32.1 minutes
  - 120-hour (5 day) forecast: 51.5 minutes
- sample input file: ./work_Oct6A/run/2022120512/swan.inp
- sample output file: ./work_Oct6A/run/2022120512/swan.prt-002
- non-vanilla variants of Demo 1: none

*4.4.2. Verification*

We verified that the model is producing reasonable results for this demo by inspecting plots of various parameters. In Figure 3 to Figure 6, we show examples of these plots.

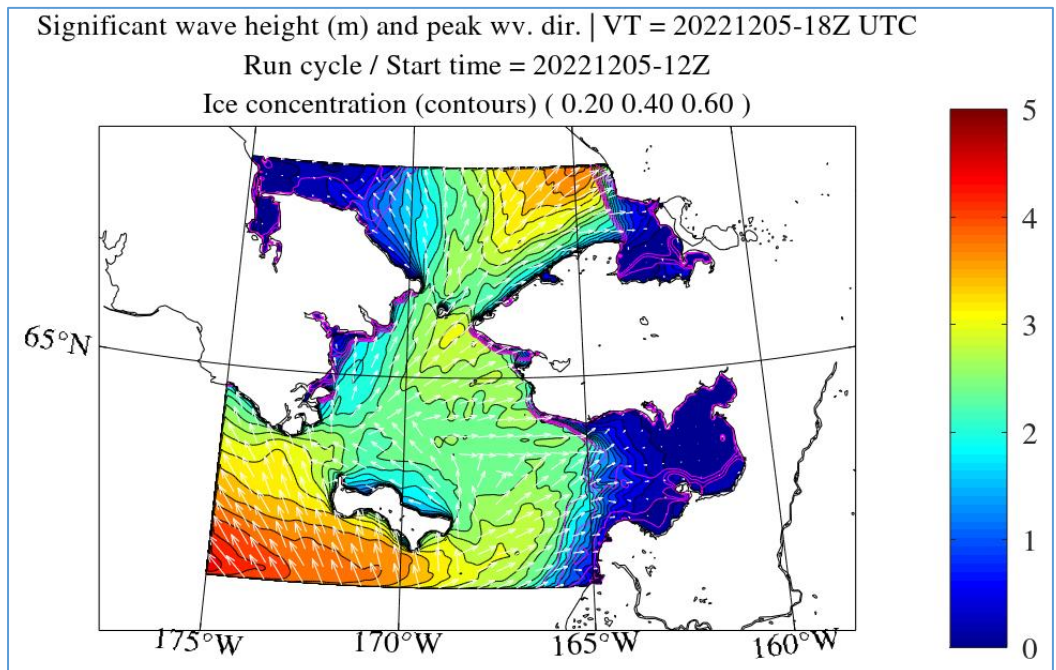

Figure 3. Example of a wave forecast from the Bering Strait SWAN demo on December 5, 2022. Ice concentration contours are shown as magenta lines. Color scale is significant wave height $H_s$ in meters. Peak wave direction is indicated with white arrows. The valid time (VT) of the plot is 1800 UTC 5 Dec. 2022, being the 6-hour forecast from run cycle 1200 UTC 5 Dec. 2022.

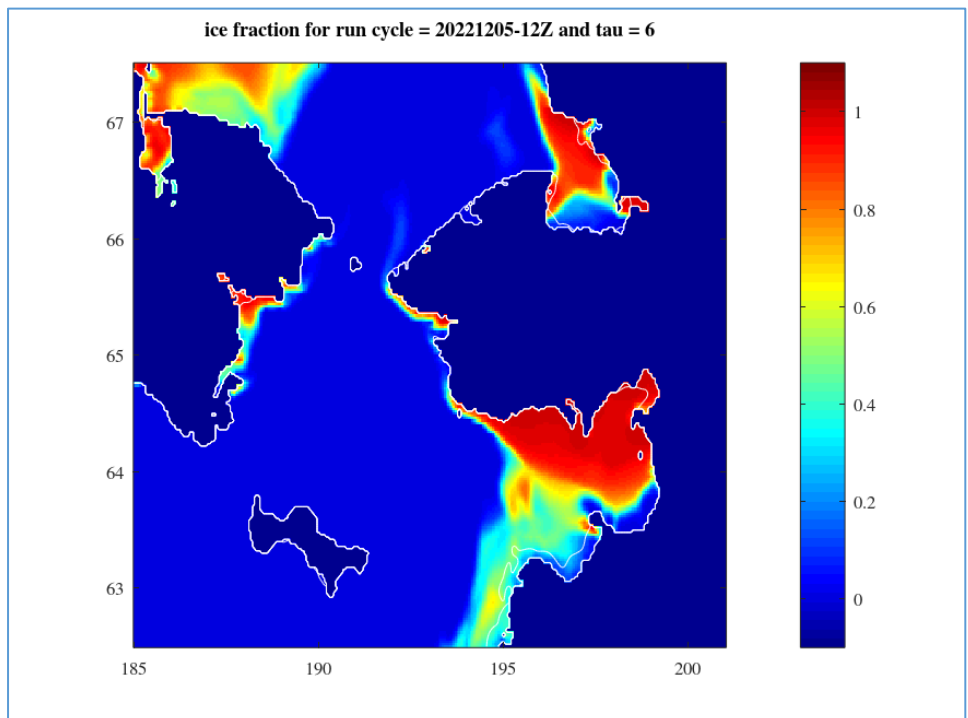

Figure 4. Like Figure 3, but showing the ice fraction, $a_{ice}$. The forcing is taken from the ice model, but the $a_{ice}$ field shown is on the SWAN model's computational grid, i.e., after re-gridding.

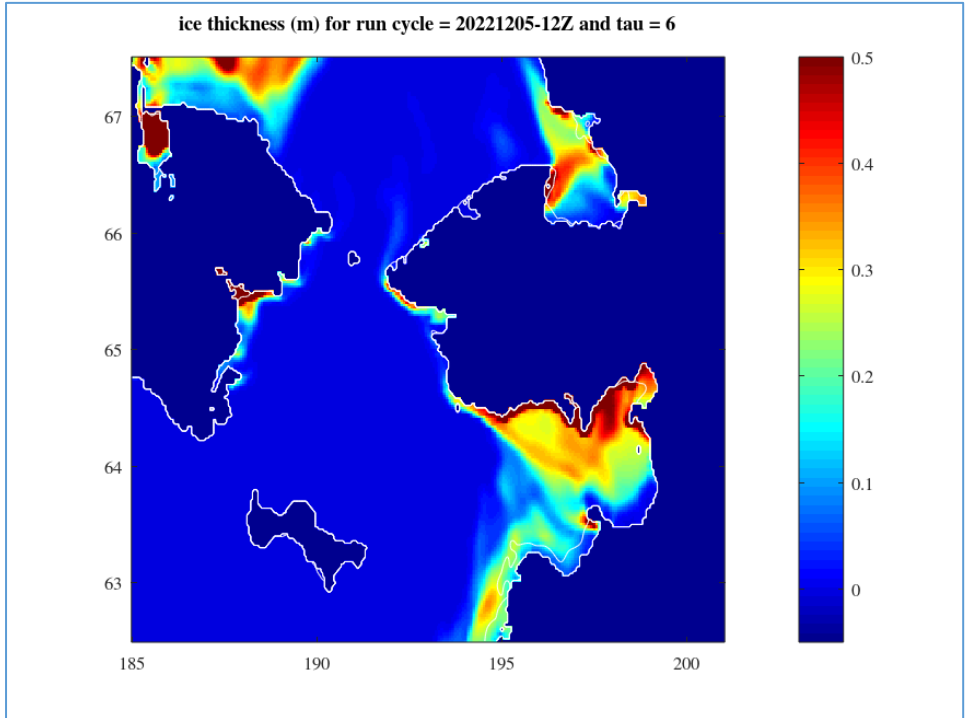

Figure 5. Like Figure 3, but showing ice thickness $h_{ice}$.

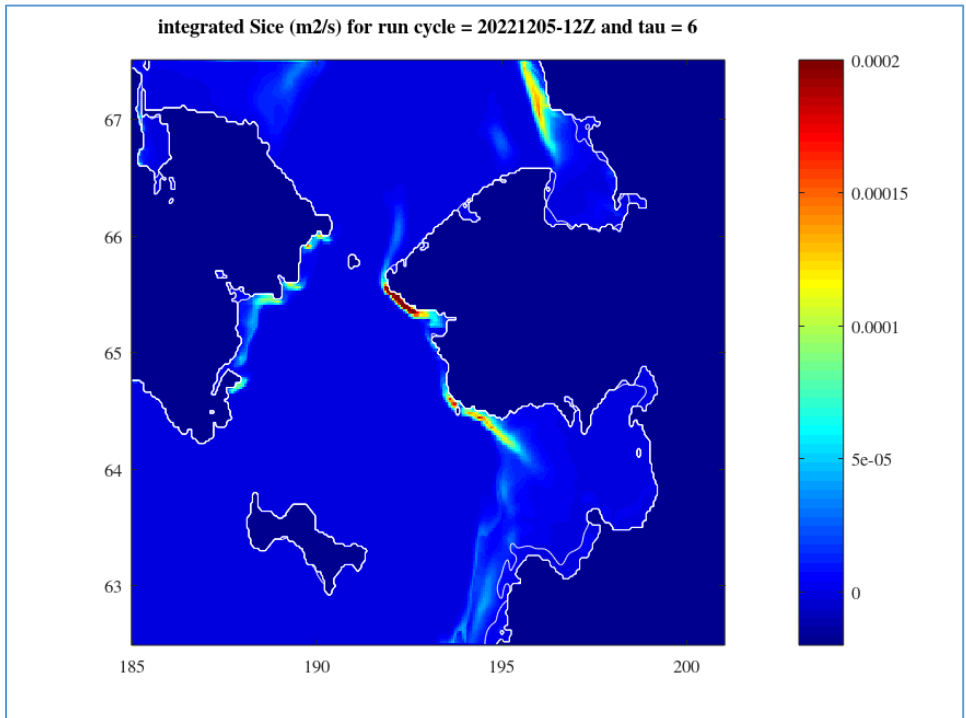

Figure 6. Like Figure 3, but showing the spectrum-integrated value of the source term for dissipation of wave energy by sea ice.

**4.5. Demo case 2A: Gulf of Bothnia (1 km SWAN), March 2023**

   *4.5.1. Settings and run times*

As mentioned in Section 4.4.1, directory names are for internal documentation and may be disregarded by readers outside NRL 7320.
- Local directory /net/chosin/export/data/rogers/Bothnia_1km_COAMPS/
- "work directory" of record: /work/
- Duration: 12 run cycles: 20230309T0000Z to 20230314T1200Z, 12 hours each
- Surface currents, water levels, ice fraction, and ice thickness from local HYCOM/CICE archives: /u/HYCOM/GLBy0.08/nc
- Surface winds from local NAVGEM archives, 0.5°
- lowest computed frequency: 0.0418 Hz
- highest computed frequency: 0.97 Hz
- frequency increment: factor 1.1
- number of frequencies: 34
- regular grid:
    - longitudes: 17° to 25.5°, nx=426, spacing=0.02° (0.90 to 1.10 km)
    - latitudes: 60.25° to 66°, ny=576, spacing=0.01° (1.11 km)
- 88127 sea points
- Boundary forcing: none
- anti-GSE diffusion strength: 0.2 hours
- run time: approximately 159 minutes
- total time from one cycle to another:
    - 746 sec (for 12-hour cycle), of which 729 sec is the main compute
    - 72-hour forecast: 74.6 minutes (a 72-hour forecast is chosen as an example, since a typical maximum forecast range for a regional wave model is from 72 to 84 hours)
    - 120-hour (5 day) forecast: 124.3 minutes
- sample input file:  ./work/run/2023031400/swan.inp
- sample output file:  ./work/run/2023031400/swan.prt-012
- non-vanilla variants of Demo 2A: none

*4.5.2. Verification*

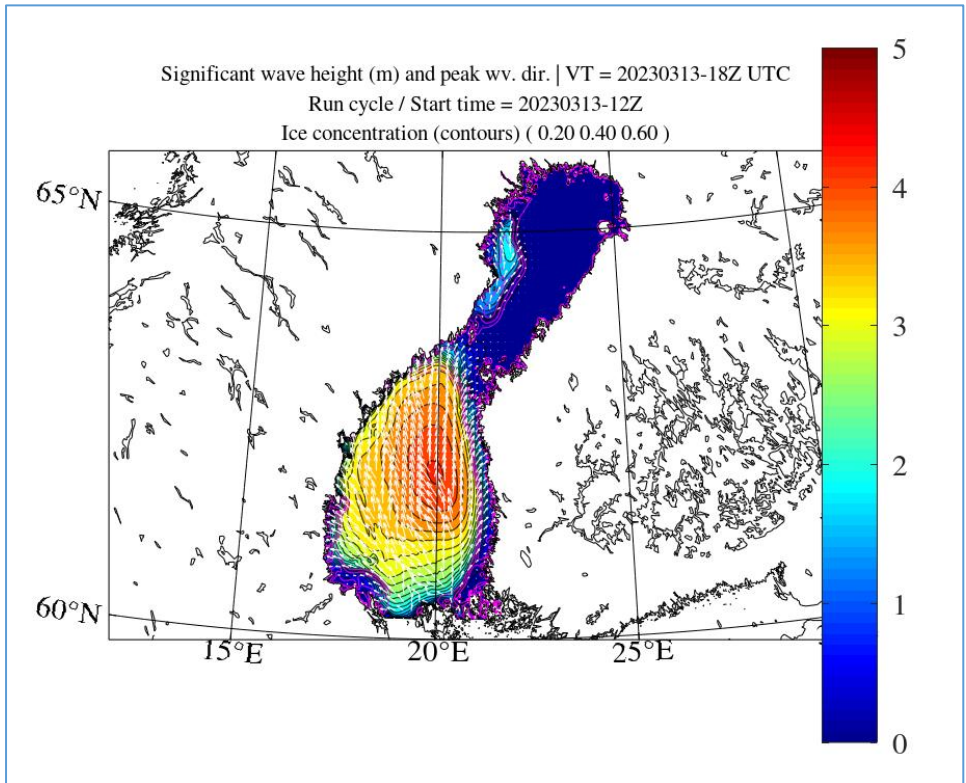

Figure 7. Example of a wave forecast from the 1 km Gulf of Bothnia SWAN demo. Ice concentration contours are shown as magenta lines. Color scale is significant wave height $H_s$ in meters. Peak wave direction is indicated with white arrows. The valid time (VT) of the plot is 1800 UTC 13 March 2023, being the 6-hour forecast from run cycle 1200 UTC 13 March 2023.

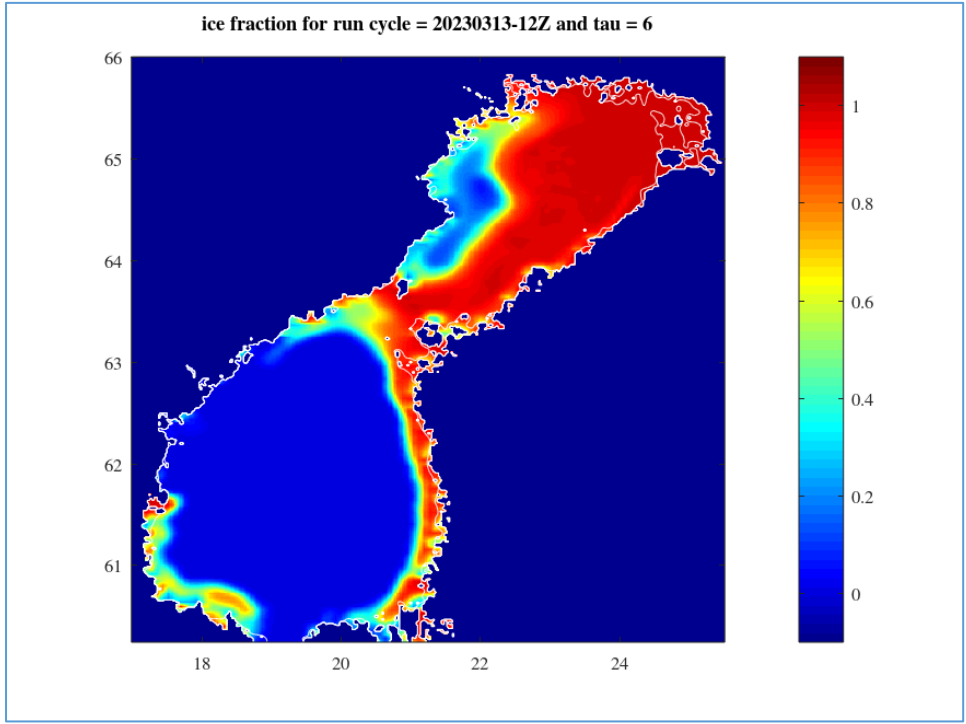

Figure 8. Like Figure 7, but showing the ice fraction, $a_{ice}$. The forcing is taken from the ice model, but the $a_{ice}$ field shown is on the SWAN model's computational grid, i.e., after re-gridding.

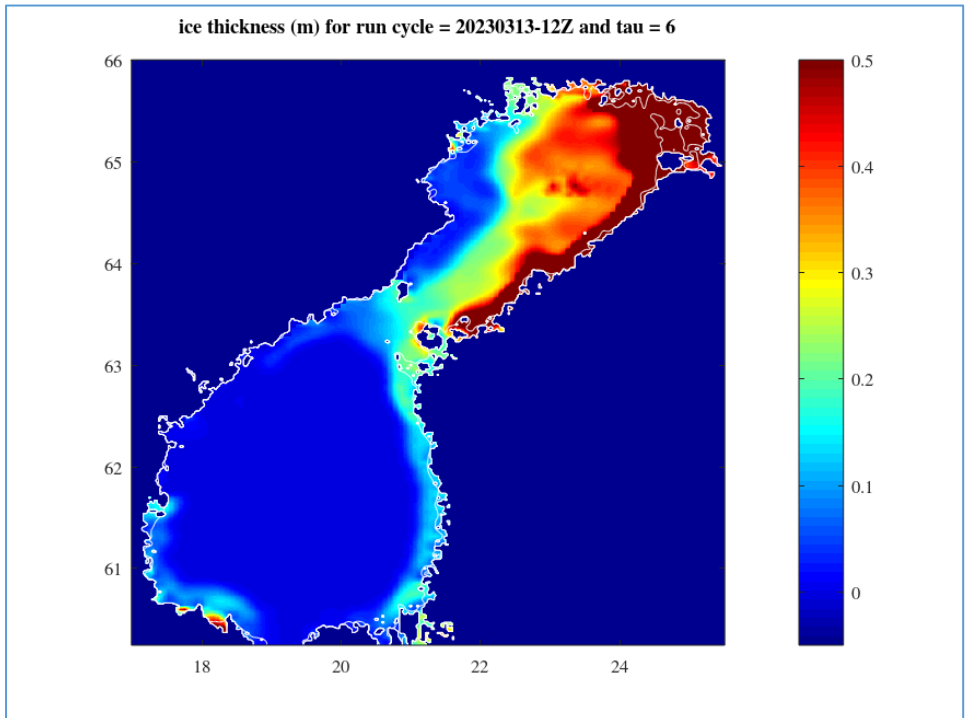

Figure 9. Like Figure 8, but showing ice thickness $h_{ice}$.

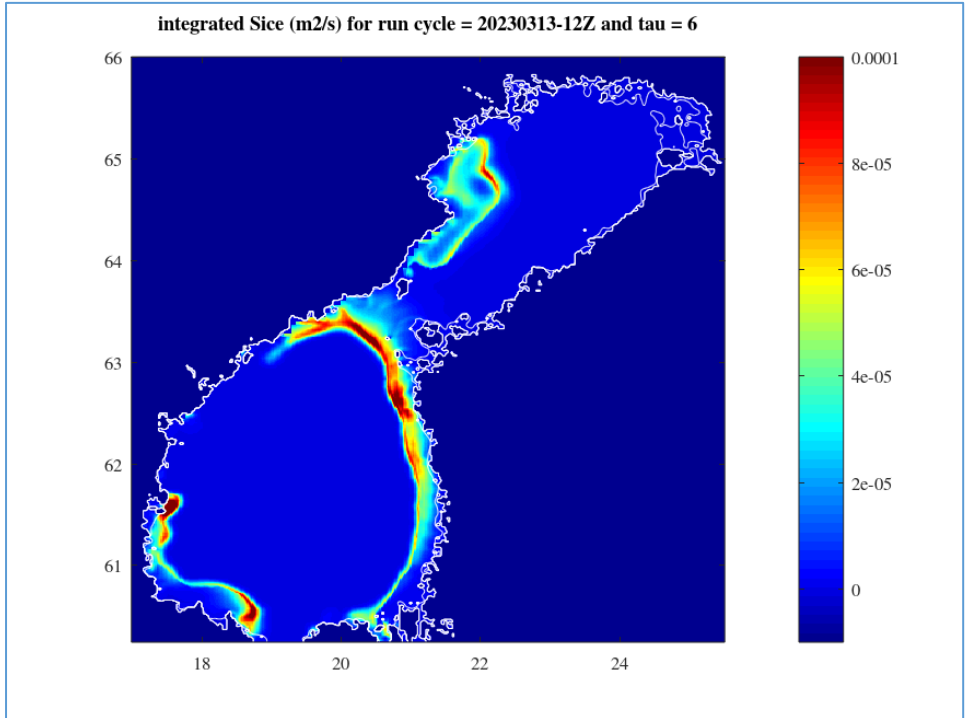

Figure 10. Like Figure 7, but showing the spectrum-integrated value of the source term for dissipation of wave energy by sea ice.

### 4.6. Demo case 2B: Gulf of Bothnia (2 km SWAN), March 2023

   *4.6.1. Settings and run times*

As mentioned in Section 3.4.1, directory names are for internal documentation and may be disregarded by readers outside NRL 7320.
- Local directory /net/chosin/export/backup_rogers/Bothnia_2km_COAMPS/
- "work directory" of record: /work/
- Duration: 12 run cycles: 20230309T0000Z to 20230314T1200Z, 12 hours each (identical to 2A)
- Surface currents, water levels, ice fraction, and ice thickness from local HYCOM/CICE archives: /u/HYCOM/GLBy0.08/nc
- Surface winds from local NAVGEM archives, 0.5° resolution.
- lowest computed frequency: 0.038 Hz. This setting is used for consistency with ESPC WW3.
- highest computed frequency: 0.97 Hz
- frequency increment: factor 1.1
- number of frequencies: 33
- regular grid
    - longitudes: 17 to 25.48, nx=213, spacing = 0.04°, or 1.81 to 2.21 km
    - latitudes: 60.25 to 65.99, ny=288, spacing = 0.02°, or 2.22 km
- 21758 sea points
- Boundary forcing: none
- anti-GSE diffusion strength: 0.7 HR
- Timings
    - run time: approximately 45 minutes
    - total time from one cycle to another: 216 sec (for 12-hour cycle), of which 202 sec is the main compute.
    - 72-hour forecast: 21.7 minutes.
    - 120-hour (5 day) forecast: 37.1 minutes.
- sample input file: ./work/run/2023030900/swan.inp
- sample output file: ./work/run/2023031412/swan.prt-002
- non-vanilla variants of Demo 2B: none

### 4.6.2. *Verification*

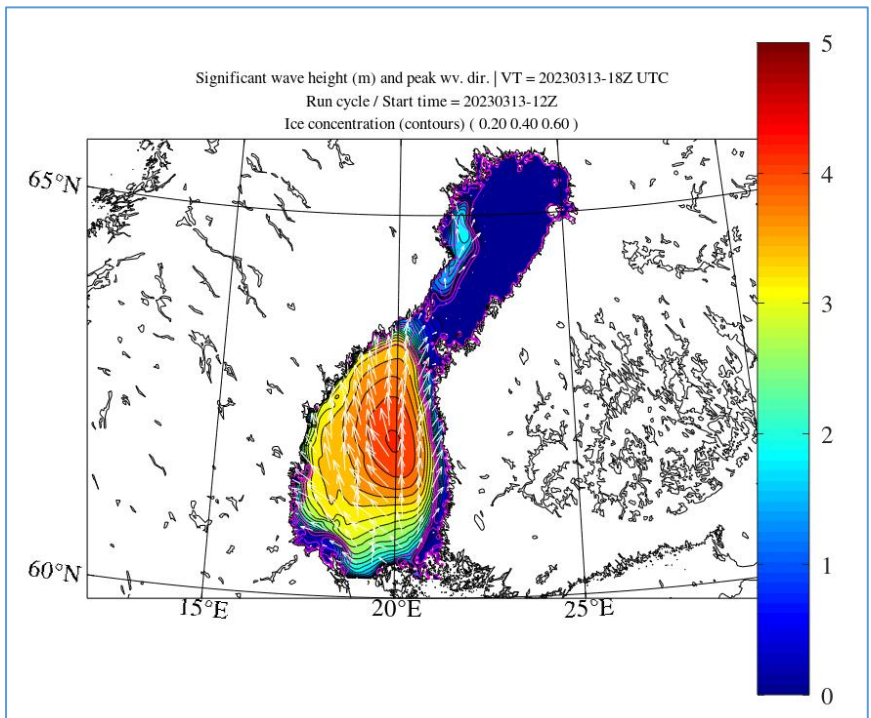

Figure 11. Example of a wave forecast from the 2 km Gulf of Bothnia SWAN demo. Ice concentration contours are shown as magenta lines. Color scale is significant wave height $H_s$ in meters. Peak wave direction is indicated with white arrows. The valid time (VT) of the plot is 1800 UTC 13 March 2023, being the 6-hour forecast from run cycle 1200 UTC 13 March 2023.

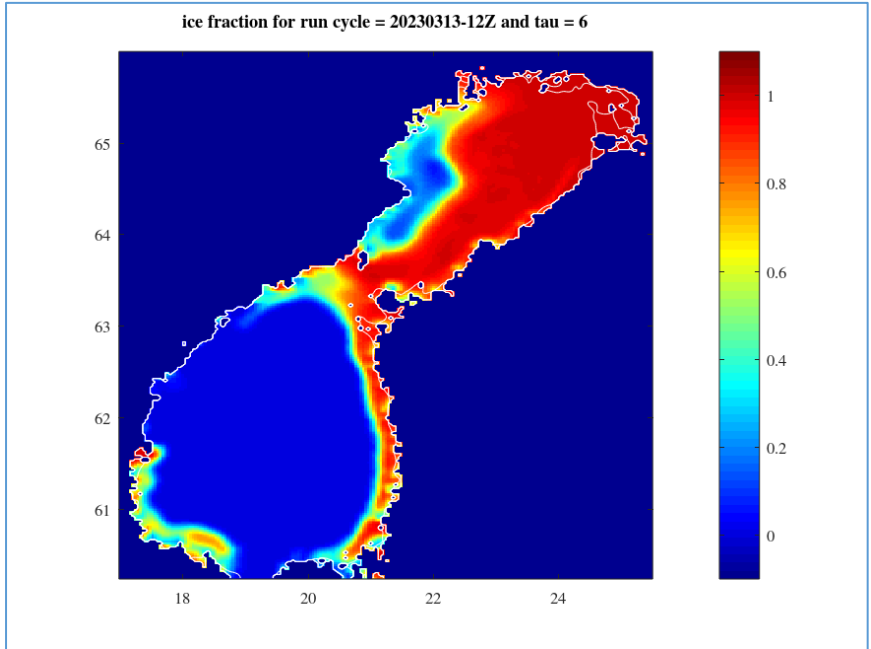

Figure 12. Like Figure 11, but showing the ice fraction, $a_{ice}$. The forcing is taken from the ice model, but the $a_{ice}$ field shown is on the SWAN model's computational grid, i.e., after re-gridding.

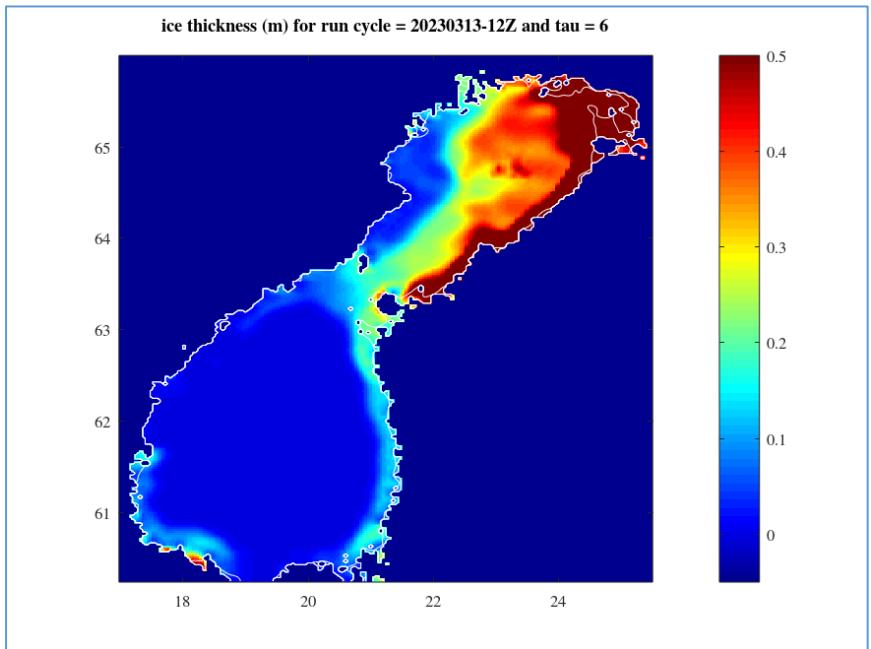

Figure 13. Like Figure 12, but showing ice thickness $h_{ice}$.

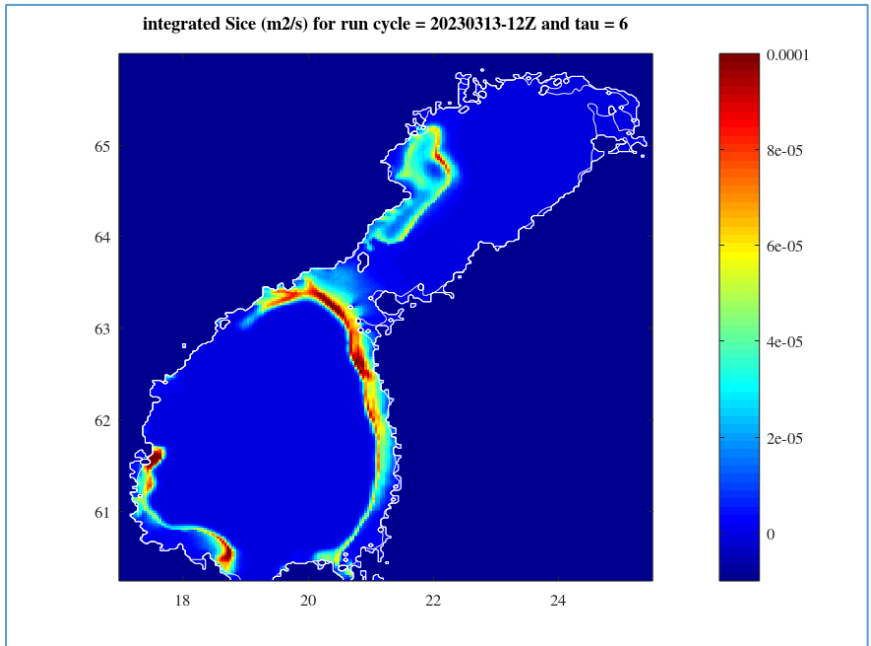

Figure 14. Like Figure 12, but showing the spectrum-integrated value of the source term for dissipation of wave energy by sea ice.

## 4.7. Demo case 3: Sea of Okhotsk (WW3)

### 4.7.1. Settings and run times

As mentioned in Section 3.4.1, directory names are for internal documentation and may be disregarded by readers outside NRL 7320.

- Local directory: /net/chosin/export/backup_rogers/Okhotsk
- "work directory" of record: work_Oct3C ($S_{ice}$ routine=IC4M9)
- Duration 10 run cycles: 20230115T0000Z to 20230119T1200Z, 12 hours each
- Surface currents, water levels, ice fraction, and ice thickness from local HYCOM/CICE archives: /u/HYCOM/GLBy0.08/nc
- Surface winds from local NAVGEM archives, 0.5°
- Spacing
    - Minimum: 5.88 km
    - Median: 5.99 km
    - Maximum: 6.02 km
- Grid specifications (COAMPS notation is used here)
    - m    = 327 ! longitudes, nx
    - n    = 411 ! latitudes, ny
    - nproj = 2     ! 2 indicates Lambert-Conformal grid
    - delx  = 6000.0
    - dely  = 6000.0
    - iref  = 163    !Grid i-index for rlon
    - jref  = 1      !Grid j-index for rlat
    - rlon  = 150.0

- o   rlat  =  41.5
- o   phnt1 =  48.0
- o   phnt2 =  56.0
- o   alnnt = 150.0
* 89984 sea points
* Boundary forcing time spacing: 3 hours.
* Timings:
  * o   run time 41 minutes
  * o   total time from one cycle to another: 245 sec for a 12-hour cycle, of which 183 sec is the main compute
  * o   72-hour forecast would be approximately 24.5 minutes
  * o   120-hour (5 day) forecast: would be approximately 40.8 minutes.
* sample grid input file:  ./run/2023011500/ww3_grid.inp.ww3g1
* sample grid output file:  ./run/2023011500/log.wav_setup.moddef.ww3g1
* sample multi stdout file: ./run/2023011912/log.wav_fcst
* sample multi log files: ./run/2023011912/log.ww3g1 ./run/2023011912/log.mww3
* sample post output file: ./2023011912/log.wav_post.point
* ST4 wind input setting: BETAMAX = 1.33, matching the value used in ESPC
* Bottom friction setting: gamma: -0.0380
* non-vanilla variants of Demo 3:
  * o   a run with $S_{ice}$ routine=IC4M6, local name: 'work_Oct4A'

### 4.7.2. *Verification*

Verification plots are shown in Figure 15 to Figure 22. In Figure 17 to Figure 22, we indicate a 3×4 grid of locations where spectral output was extracted for evaluation. These locations are given in Table I.

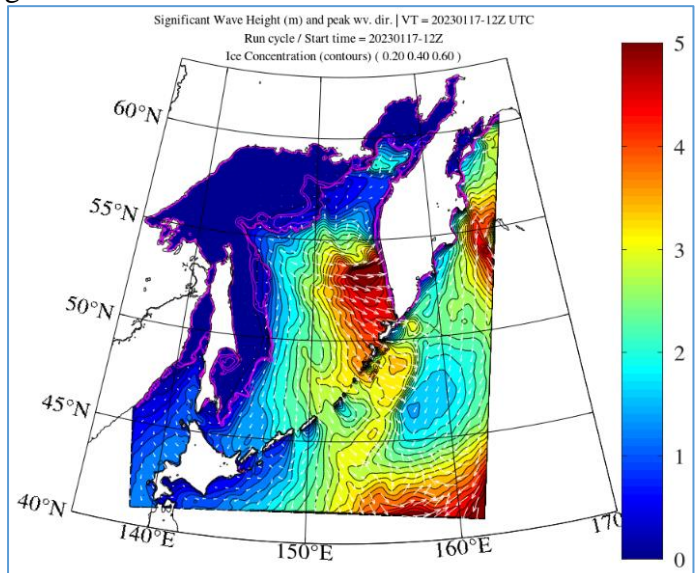

Figure 15. Example from the Sea of Okhotsk WW3 demo. Ice concentration contours are shown as magenta lines. Color scale is significant wave height $H_s$ in meters. Peak wave direction is indicated with white arrows. The valid time (VT) of the plot is 1200 UTC 17 Jan. 2023, being the analysis from run cycle 1200 UTC 17 Jan. 2023.

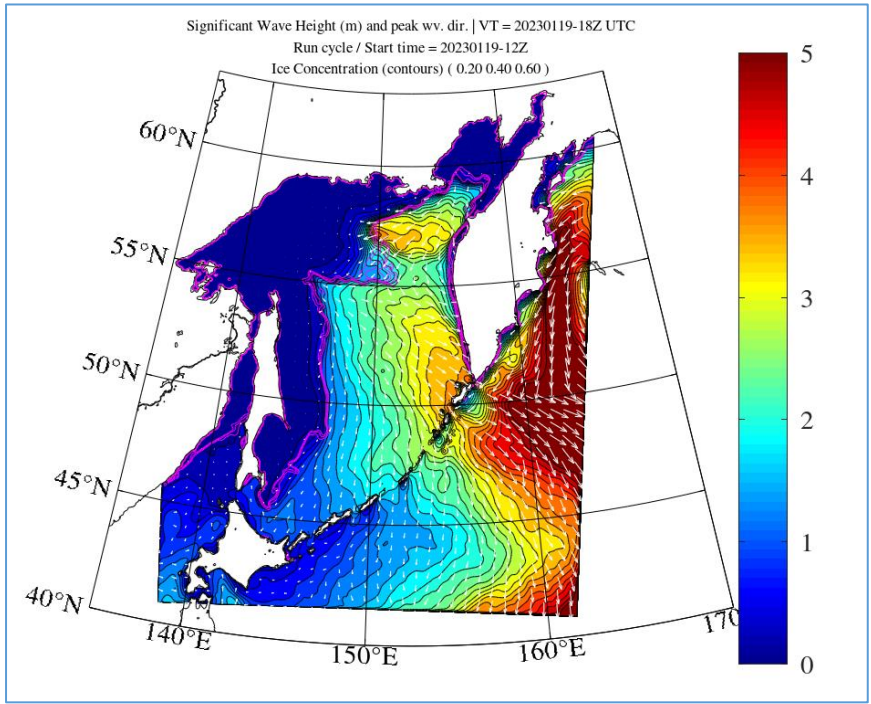

Figure 16. Like Figure 15, but the valid time (VT) of the plot is 1800 UTC 19 Jan. 2023, being the 6-hour forecast from run cycle 1200 UTC 19 Jan. 2023.

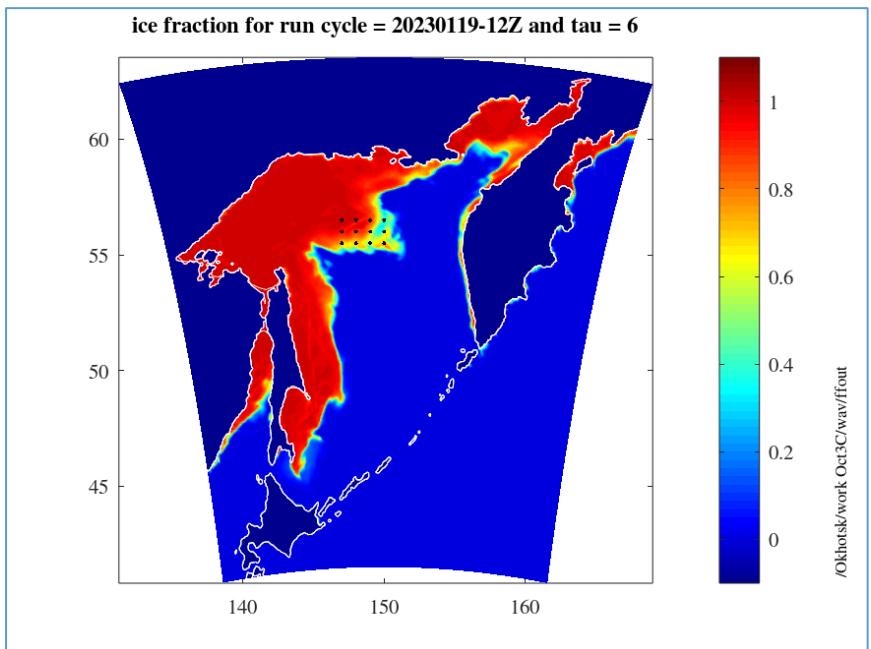

Figure 17. Like Figure 16, but showing the ice fraction, $a_{ice}$. The forcing is taken from the ice model, but the $a_{ice}$ field shown is on the WW3 model's computational grid, i.e., after re-gridding. The black dots in a 3×4 grid indicate locations where spectral output was extracted for evaluation (Table I).

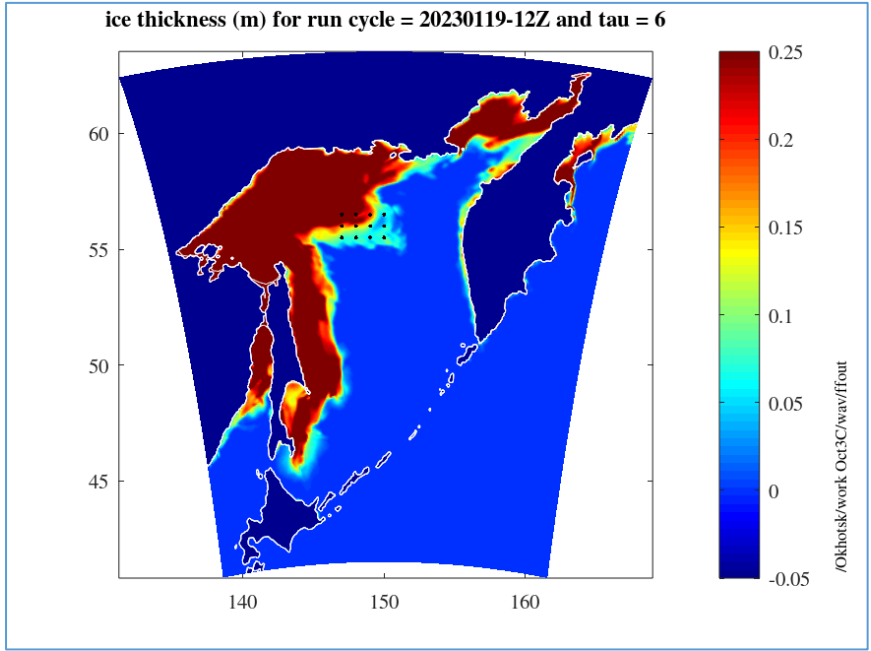

Figure 18. Like Figure 17, but showing ice thickness $h_{ice}$.

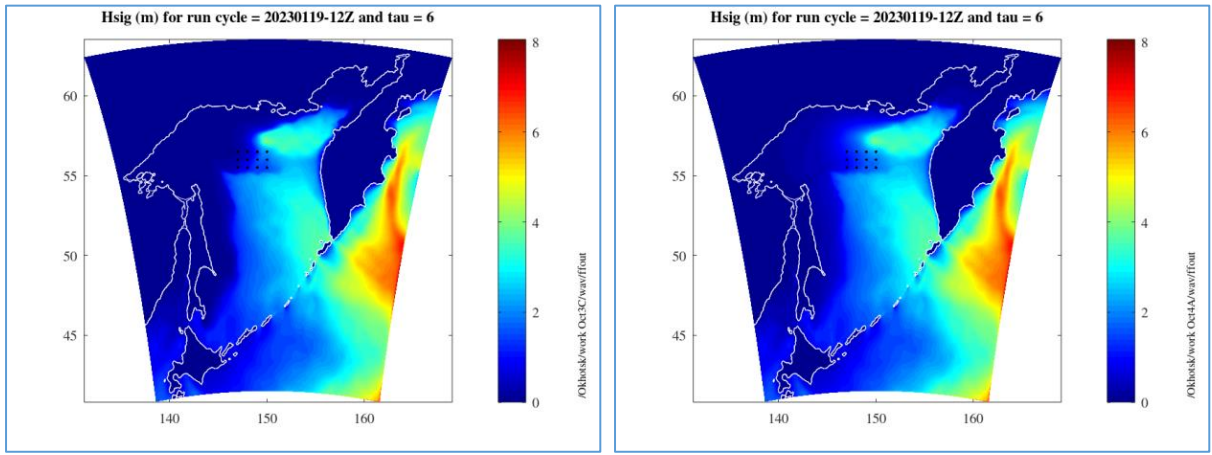

Figure 19. Like Figure 17, but comparing significant waveheight $H_s$ from two model runs. Left panel: $S_{ice}$=IC4M9. Right panel: $S_{ice}$=IC4M6.

In the spectral comparison (Figure 23), the most notable difference between the two models is that within the relatively thick ice at point PNT09, in the 0.08 to 0.10 Hz range, more dissipation is evident with IC4M9 than with IC4M6, consistent with Figure 2.

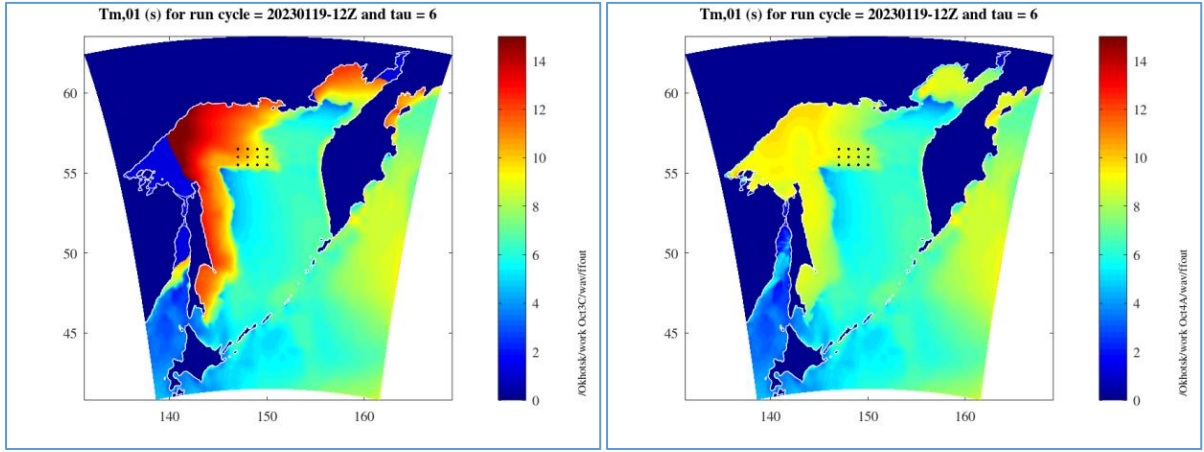

Figure 20. Like Figure 19, but comparing mean period $T_{m01}$. Left panel: $S_{ice}$=IC4M9. Right panel: $S_{ice}$=IC4M6.

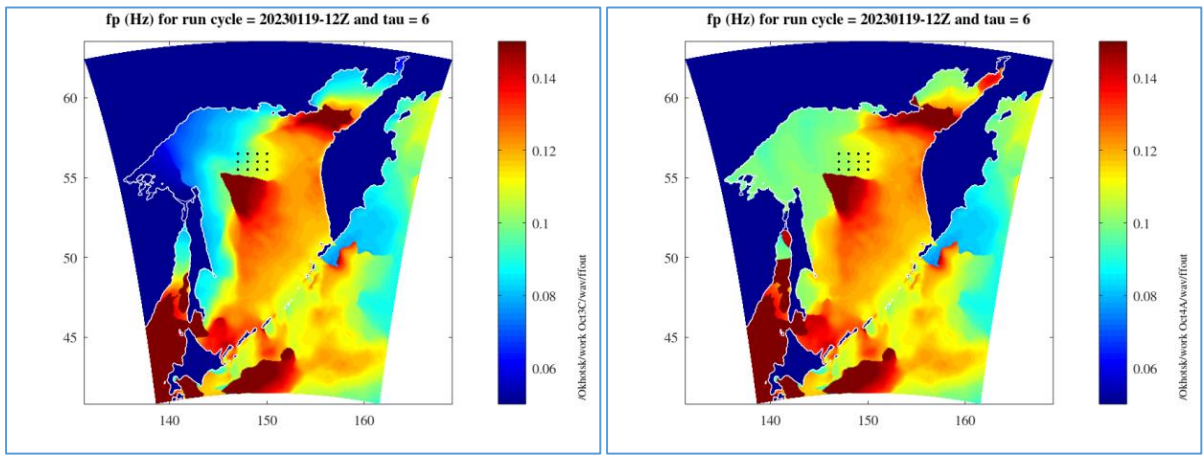

Figure 21. Like Figure 19, but comparing peak period $f_p$. Left panel: $S_{ice}$=IC4M9. Right panel: $S_{ice}$=IC4M6.

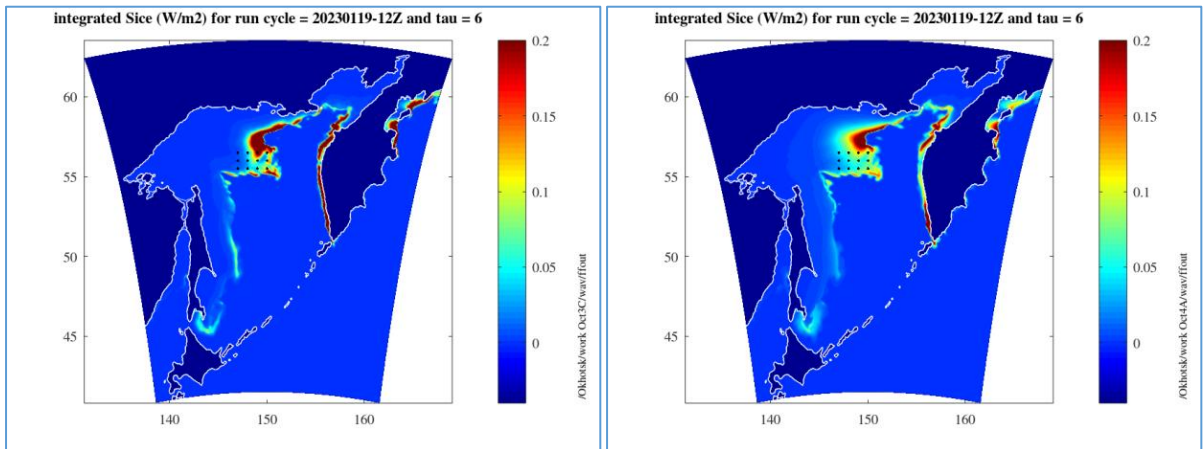

Figure 22. Like Figure 19, but comparing the spectrum-integrated value of the source term for dissipation of wave energy by sea ice. Left panel: $S_{ice}$=IC4M9. Right panel: $S_{ice}$=IC4M6.

Table I. Locations at which spectral output was extracted for the Sea of Okhotsk case. The southwest, southeast, northwest, and northeast corners were used for Figure 23.

```
lon  lat    label    relative pos.
147  55.5  'PNT01'  SW (southwest)
148  55.5  'PNT02'  SC
149  55.5  'PNT03'  SC
150  55.5  'PNT04'  SE (southeast)
147  56.0  'PNT05'  CW
148  56.0  'PNT06'  CC
149  56.0  'PNT07'  CC
150  56.0  'PNT08'  CE
147  56.5  'PNT09'  NW (northwest)
148  56.5  'PNT10'  NC
149  56.5  'PNT11'  NC
150  56.5  'PNT12'  NE (northeast)
```

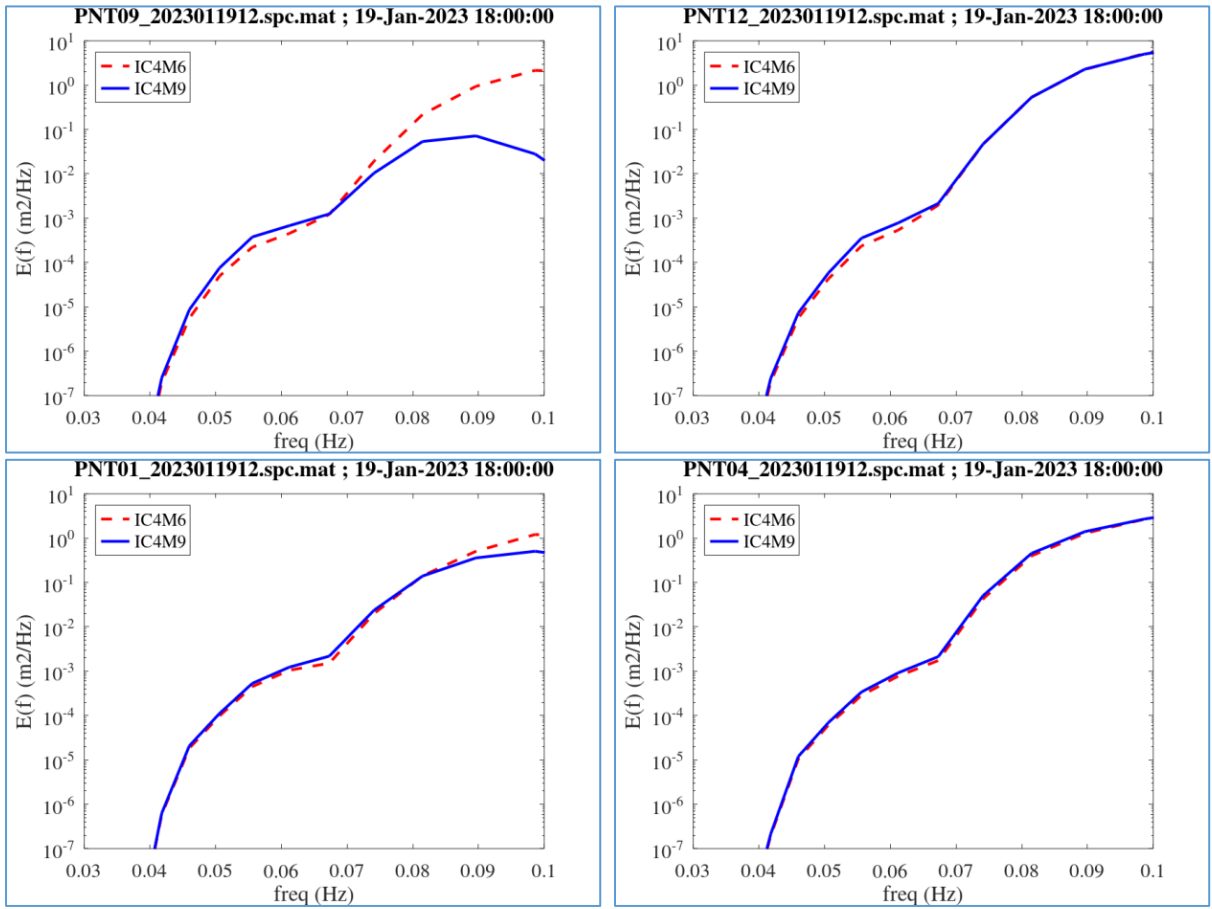

Figure 23. Comparison of spectra (ocean wave energy level as a function of frequency) at four locations where spectral output was extracted for evaluation (Table I). Valid time is 1800 UTC 19 Jan. 2023. Top left: northwest corner of 3×4 grid. Top right: northeast corner. Lower left: southwest corner. Lower right: southeast corner. Blue solid line is the non-directional wave spectrum from the IC4M9 model. Red dashed line is the non-directional wave spectrum from the IC4M6 model.

**Demo case 4: Barents Sea, WW3 (Introduction)**

The "Barents Sea" grid is shown in Figure 24. It includes the Barents Sea, Kara Sea, Svalbard, Franz Josef Land, and the eastern Norwegian Sea. Demo case 4B includes the Gulf of Bothnia but demo case 4A does not. It is a 6 km Lambert Conformal grid, created using COAMPS utilities using our settings in the file 'wav.rc'. Boundary forcing for COAMPS WW3 comes from a global WW3 hindcast. Restart files were saved from the latter and used to create the boundary forcing for the COAMPS-WW3 runs, using the "spritzer" program. Two time periods, February 10 to 20 2024 (demo 4A) and April 4 to 30 2024 (demo 4B) were created for the COAMPS-WW3 Barents Sea. The durations after removing spin-up are 7 and 23 days, respectively: these are the durations available for comparison to observations.

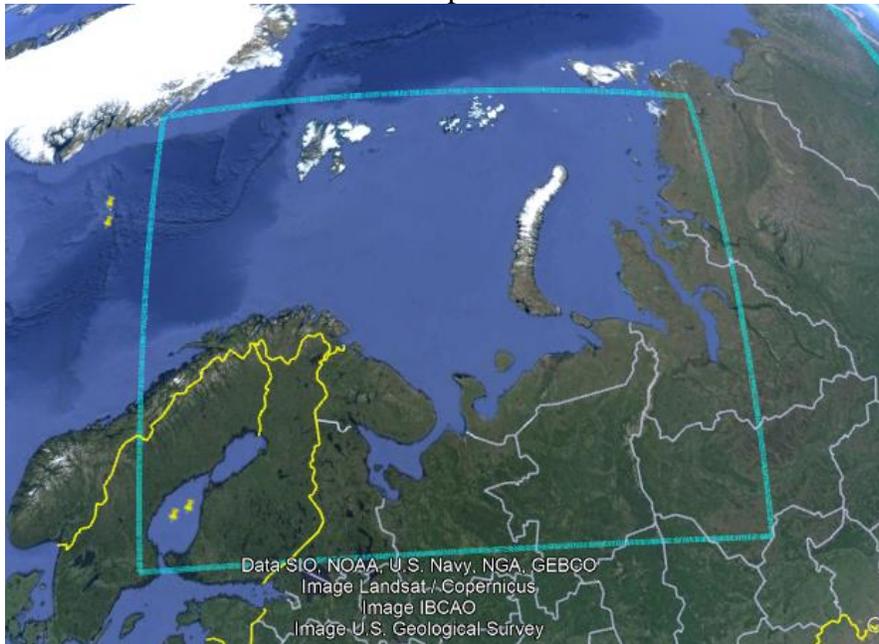

Figure 24. The Barents Sea grid, for COAMPS/WW3 demo case 4.

## 4.8. Demo case 4A: Barents Sea, WW3, February 2024

### 4.8.1. Settings and run times

As mentioned in Section 3.4.1, directory names are for internal documentation and may be disregarded by readers outside NRL 7320.
- Local directory: /net/chosin/export/backup_rogers/Barents
- "work directory" of record: work_May17 (normal output); work_May28 (re-run w/spectral output)
- Duration 20 run cycles: 20240210T0000Z to 20240219T1200Z
- Surface currents, water levels, ice fraction, and ice thickness from local HYCOM/CICE archives: /u/HYCOM/GLBy0.08/nc
- Surface winds from local NAVGEM archives, 0.5°
- Spacing (actual resolution)
    - Minimum: 5.87 km
    - Median: 5.99 km
    - Maximum: 6.01 km

- Grid specifications (COAMPS notation is used here)
    - m    = 471 ! longitudes, nx
    - n    = 381 ! latitudes, ny
    - nproj = 2    ! 2 indicates Lambert-Conformal
    - delx  = 6000.0
    - dely  = 6000.0
    - iref  = 235     !Grid i-index for rlon
    - jref  = 1       ! Grid j-index for rlat
    - rlon  = 43.0
    - rlat  =  62.0
    - phnt1 =  68.0 ! 1/3 of the way up the center line
    - phnt2 =  75.0 ! 2/3 of the way up
    - alnnt = 43.0
- 95277 sea points
- Boundary forcing time spacing: 3 hours.
- Timings:
    - run time: 90 minutes
    - total time from one cycle to another: 262 sec (12-hour  cycle), of which 228 sec is the main compute
    - 72-hour forecast: approximately 26.2 minutes
    - 120-hour (5 day) forecast: approximately 43.7 minutes
- sample grid input file:   ./run/2024021000/ww3_grid.inp.ww3g1
- sample grid output file:  ./run/2024021000/log.wav_setup.moddef.ww3g1
- sample multi stdout file: ./run/2024021000/log.wav_fcst
- sample multi log files: ./run/2024021000/log.ww3g1 and ./run/2024021000/log.mww3
- sample post output file: ./run/2024021912/log.wav_post.field.ww3g1
- ST4 wind input setting:   BETAMAX = 1.45 (note that this is higher than the value used in the Okhotsk case)
- Bottom friction with setting:
- gamma: -0.0190 (note that this is different from Okhotsk)
- non-vanilla variants of Demo 4A (here, we use the convention that the run with vanilla settings is "COAMPS1"):
    - COAMPS2, /work_May29/: switch to NAVGEM 0.18°, processed separately
    - COAMPS3, /work_May29B/: like prior, but omit current and wlev
    - COAMPS4, /work_May30/: like prior, but uses the older $S_{ice}$ routine, IC4M6
    - COAMPS5, /work_May30B/: like prior, but use NAVGEM 0.18 $a_{ice}$
    - COAMPS7, /work_May31/: like prior, but $S_{ice} = 0$.
    - COAMPS8, /work_May31B/: like prior, but with currents
    - COAMPS9, /work_May31C/: like prior, but using $S_{ice}$ routine=IC4M6 instead of $S_{ice} = 0$.
    - COAMPS11, /work_Jun3/: like prior, but using $S_{ice}$ routine=IC4M9, with $h_{ice}$ from GOFS (thus $a_{ice}$ and $h_{ice}$ come from different products).

NB: In COAMPS, processing of spectral output is not possible and must be performed by the user, separate from the COAMPS job, after the job is completed. Above, the "work directory of record" for "normal output" (meaning, output of parameters such as waveheight, ice fraction, and integrated dissipation) is different from the "work directory of record" for spectral output, because post-processing of the normal output errored out when spectral output was included. This failure occurred because COAMPS post-processing for WW3 does not understand 3-d fields such as $E(f,x,y)$. We have since addressed this in the COAMPS software by instructing it to ignore 3d fields. Spectral output must still be processed by the user separately.

### *4.8.2. Verification*

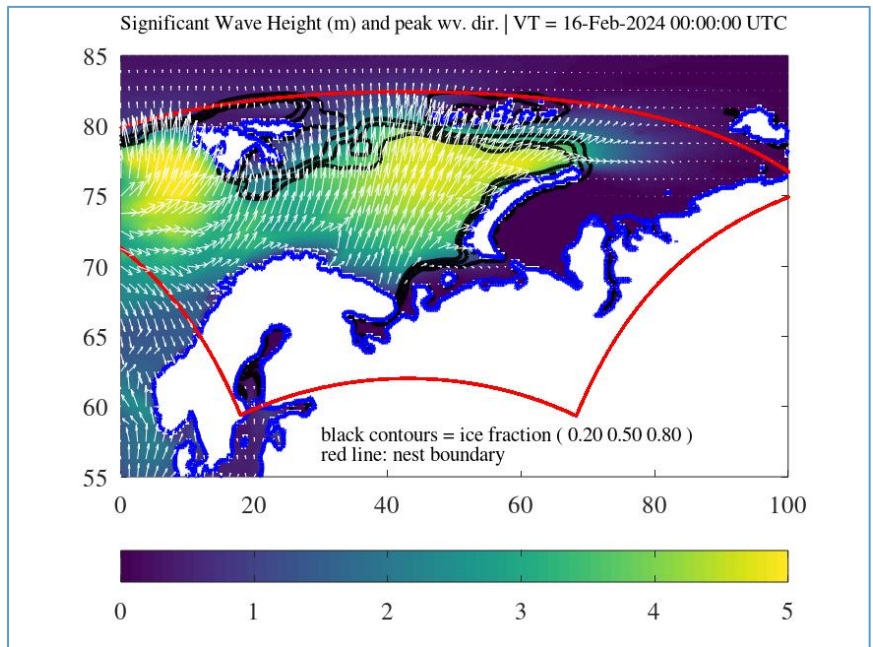

Figure 25. A plot created from a global WW3 hindcast, showing the regional COAMPS-WW3 grid in red. Ice concentration contours are shown as black lines. Color scale is significant wave height $H_s$ in meters. Peak wave direction is indicated with white arrows. The valid time (VT) of the plot is 0000 UTC 16 February 2024.

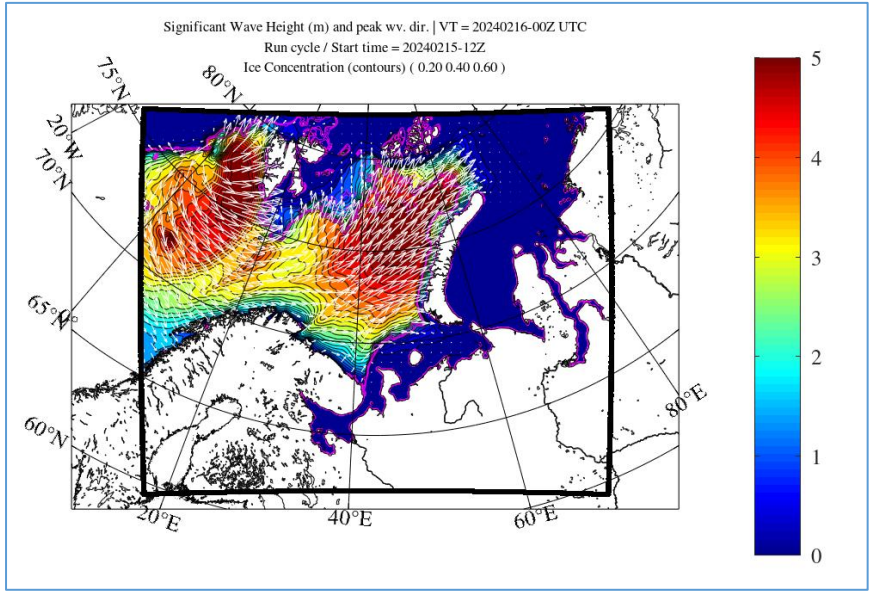

Figure 26. Example result from demo case 4A, which is the February 2024 Barents Sea case. The Gulf of Bothnia is not included in this demo. Ice concentration contours are shown as magenta lines. Color scale is significant wave height $H_s$ in meters. Peak wave direction is indicated with white arrows. The valid time (VT) of the plot is 0000 UTC 16 February 2024, being the 12-hour forecast from run cycle 1200 UTC 15 February 2024.

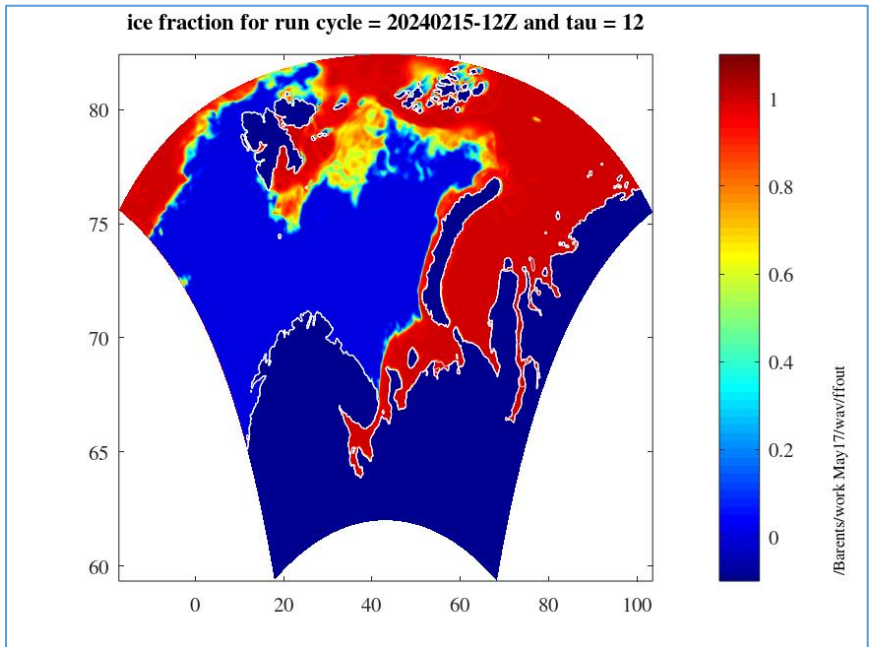

Figure 27. Like Figure 26, but showing the ice fraction, $a_{ice}$. The forcing is taken from the ice model, but the $a_{ice}$ field shown is on the WW3 model's computational grid, i.e., after re-gridding.

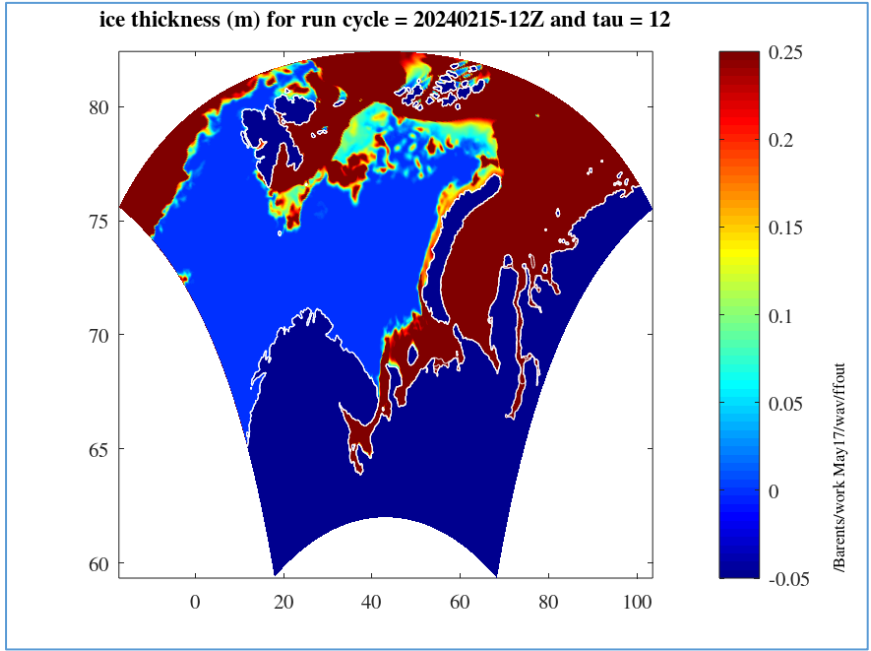

Figure 28. Like Figure 27, but showing ice thickness $h_{ice}$.

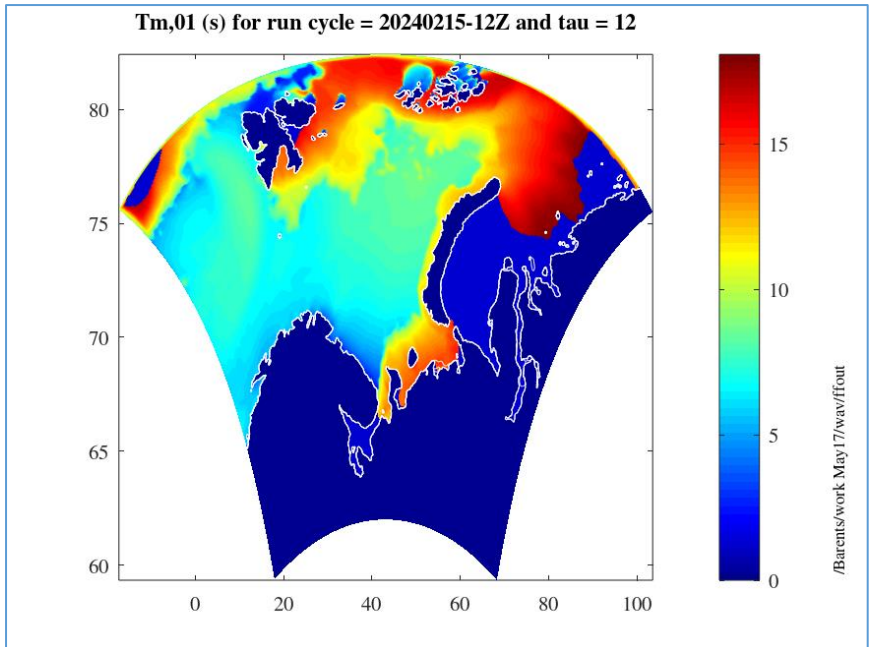

Figure 29. Like Figure 27, but showing the mean period $T_{m01}$.

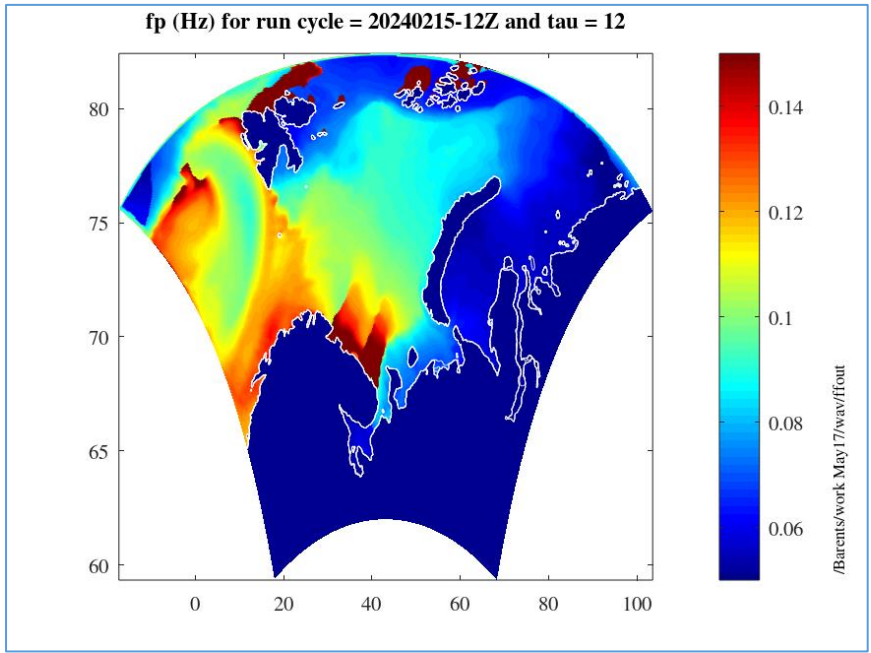

Figure 30. Like Figure 27, but showing the peak frequency, $f_p$.

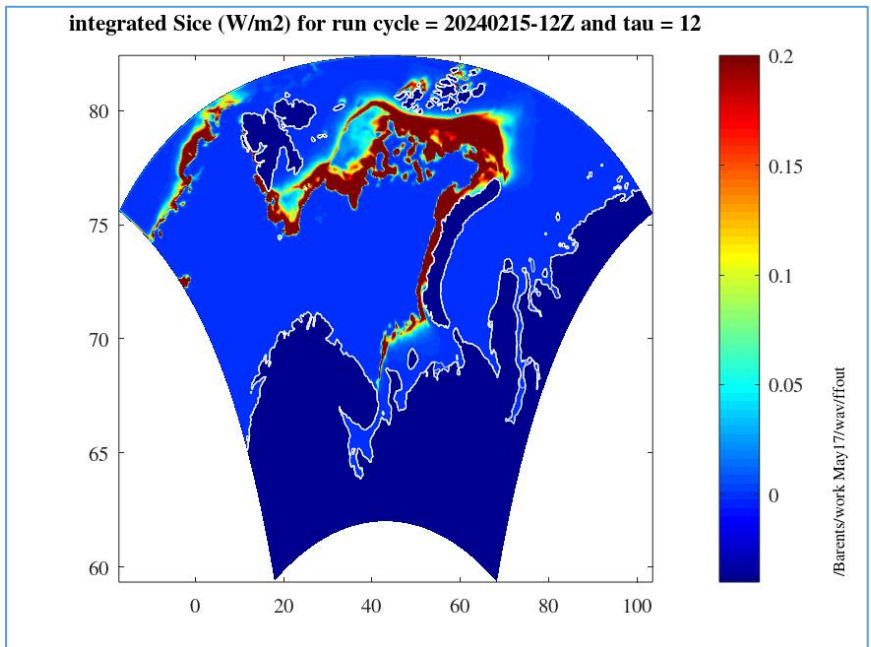

Figure 31. Like Figure 27, but showing the spectrum-integrated value of the source term for dissipation of wave energy by sea ice.

## 4.9. Demo case 4B: Barents Sea, WW3, April 2024

### 4.9.1. Settings and run times

As mentioned in Section 3.4.1, directory names are for internal documentation and may be disregarded by readers outside NRL 7320.

- Local directory: /net/chosin/export/data/rogers/Barents2/
- "work directory" of record: work_Jun21B
- Duration 54 run cycles: 20240404T0000Z to 20240430T1200Z
- Surface currents, water levels, ice fraction, and ice thickness from local HYCOM/CICE archives: /u/HYCOM/GLBy0.08/nc
- Surface winds from local NAVGEM archives, 0.5°
- grid is identical to demo case 4A, except that Gulf of Bothnia is included by setting those regions as "active sea points"
- 98182 sea points (different from case 4A due to adding Gulf of Bothnia)
- Boundary forcing time spacing: 3 hours.
- Timings:
    - total run time 4.23 hours or 4 hours, 14 minutes
    - total time from one cycle to another= 269 sec (12-hour cycle), of which 205 sec is the main compute
    - 72-hour forecast: 26.9 minutes
    - 120-hour (5 day) forecast: 44.8 minutes
- sample grid input file:   ./run/2024040400/ww3_grid.inp.ww3g1
- sample grid output file:   ./run/2024040400/log.wav_setup.moddef.ww3g1
- sample multi stdout file:   ./run/2024043012/log.wav_fcst
- sample multi log files:   ./run/2024043012/log.ww3g1 and ./run/2024043012/log.mww3
- sample post output file: ./run/2024043012/log.wav_post.field.ww3g1
- ST4 wind input setting:   BETAMAX = 1.45 (same as 4A, but different from Okhotsk)
- Bottom friction with setting: Gamma=-0.0190 (same as 4A, but different from Okhotsk)
- non-vanilla variants of Demo 4B (here, we use the convention that the run with vanilla settings is "COAMPS1")
    - COAMPS2, /work_Jun21C/:
        - $S_{ice}$ routine=IC4M9
        - winds and aice from NAVGEM 0.18 processed separately.
        - $h_{ice}$ copied from vanilla COAMPS set-up.
    - COAMPS3, /work_Jun24/:
        - $S_{ice}$ routine=IC4M6 ($h_{ice}$ not used).
        - Otherwise, it is similar to COAMPS2.
    - COAMPS4, /work_Jun28/: $S_{ice}$=0

*4.9.2. Verification*

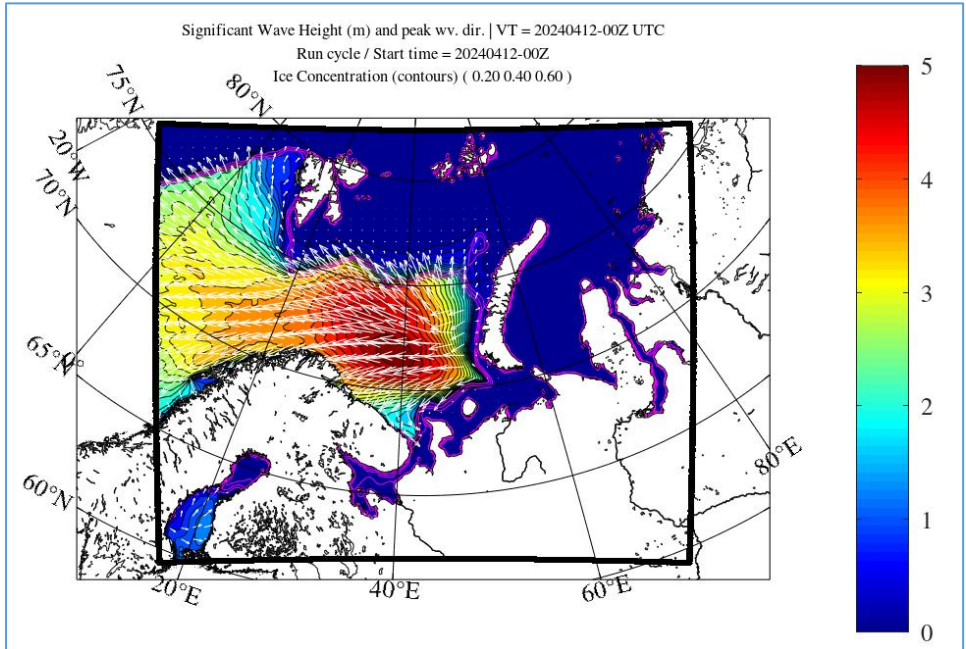

Figure 32. Example from demo case 4B, which is the April 2024 Barents Sea case. Gulf of Bothnia is included in this demo. Ice concentration contours are shown as magenta lines. Color scale is significant wave height $H_s$ in meters. Peak wave direction is indicated with white arrows. The valid time (VT) of the plot is 0000 UTC 12 April 2024, being the analysis from the run cycle of the same date/time. The "COAMPS2" case is shown, where winds and ice concentration fields are taken from 0.18° NAVGEM.

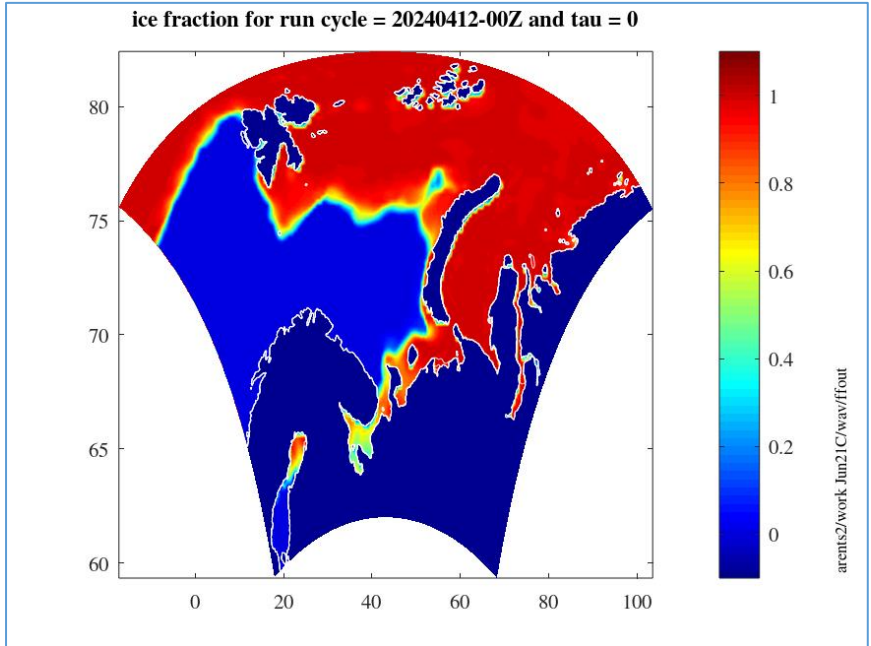

Figure 33. Like Figure 32, but showing the ice fraction, $a_{ice}$. This is an input (forcing) field, but the $a_{ice}$ field shown is on the WW3 model's computational grid, i.e., after re-gridding.

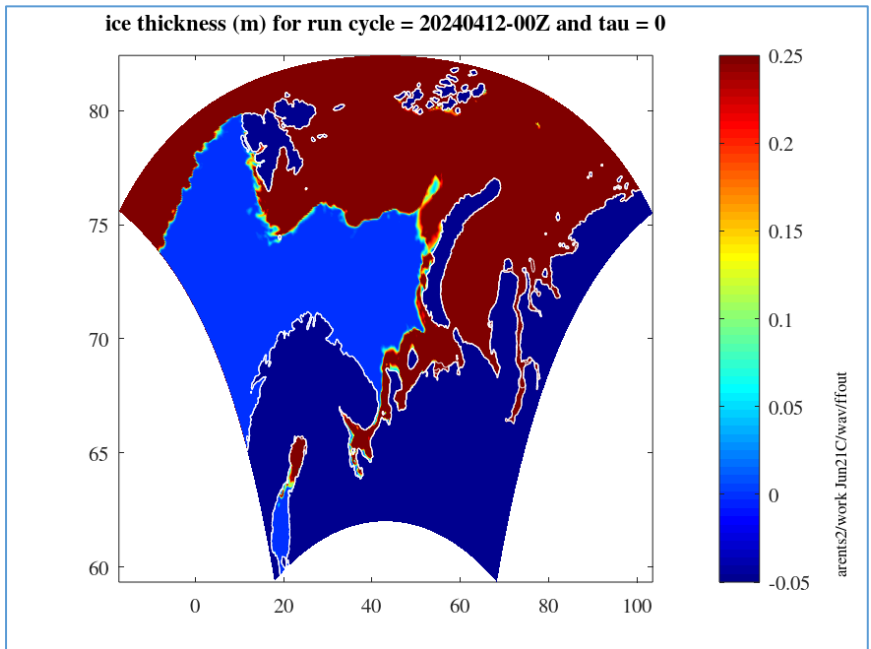

Figure 34. Like Figure 32, but showing ice thickness $h_{ice}$.

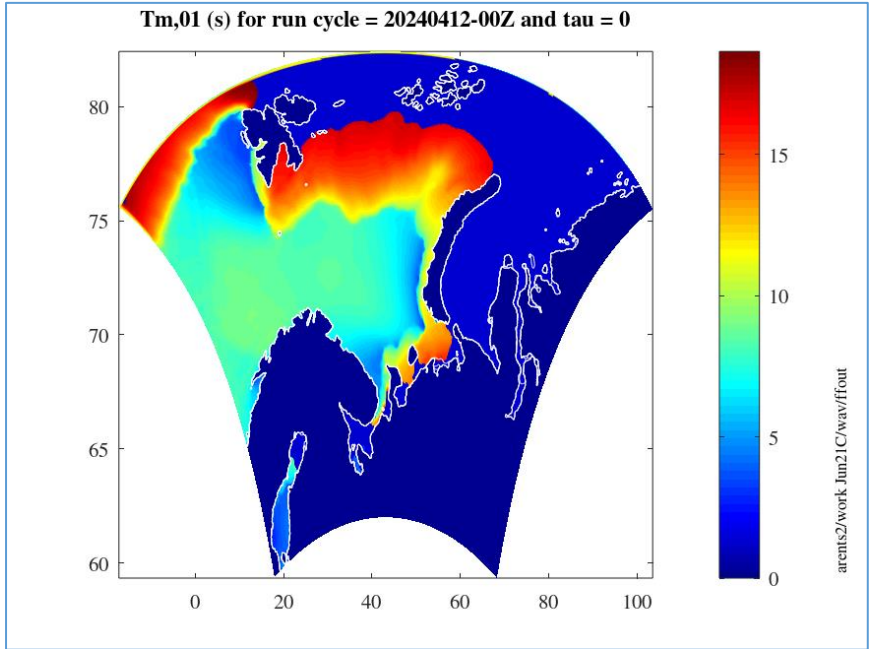

Figure 35. Like Figure 32, but showing the mean period $T_{m01}$.

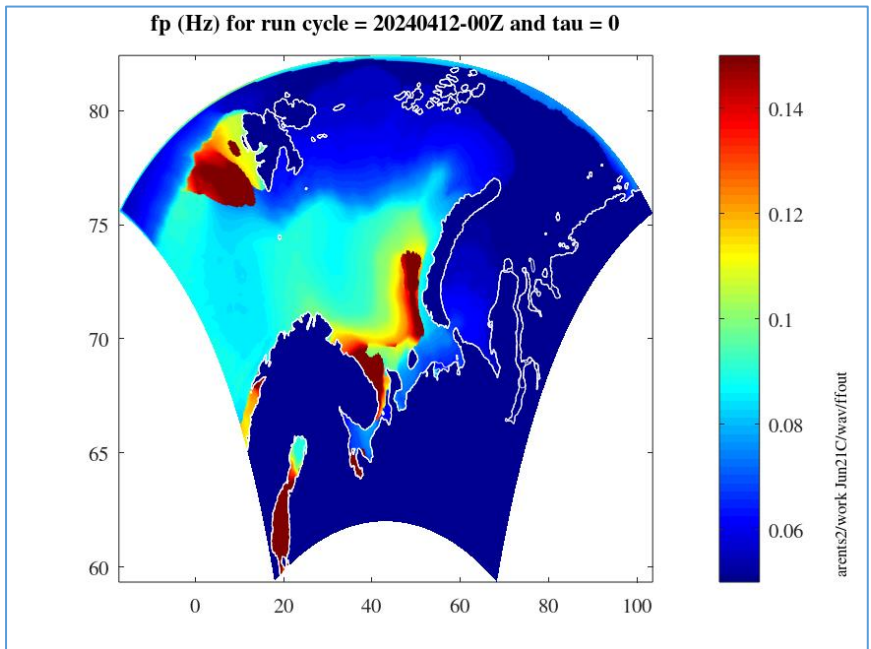

Figure 36. Like Figure 32, but showing the peak frequency, $f_p$.

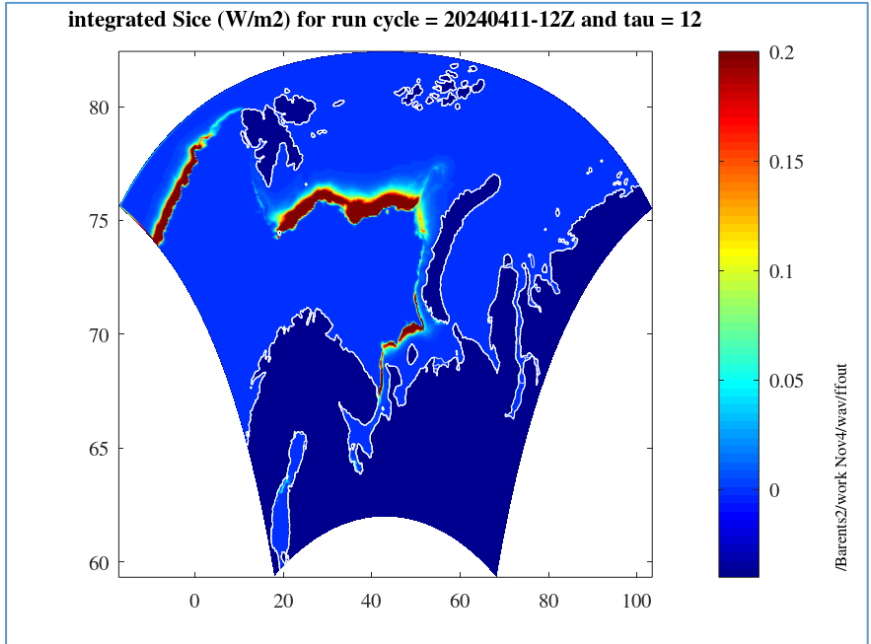

Figure 37. The spectrum-integrated value of the source term for dissipation of wave energy by sea ice. The valid time (VT) of the plot is 0000 UTC 12 April 2024, being the 12-hour forecast for the run cycle 1200 UTC 11 April 2024. NB: This has the same valid time (VT) as prior plots, but here we show tau=12 instead of tau=0, since at time of output for tau=0, WW3 does not have data stored for integrated source terms. For fields derived from wave spectra (e.g., waveheight, mean period) and forcing fields (e.g., ice thickness), output at tau=0 is available and identical to output at tau=12 for prior cycle.

### 4.10. Summary of run times for all demo cases

Run times on the local linux workstation (see Section 4.1) were given in individual sections and are tabulated in Table II here for direct comparison. Some of the cases were also run on the Navy DSRC (DoD Supercomputing Resource Center) system 'narwhal', which is a Cray EX with 128 cores per node.

The pre-processing (setup) time for the DSRC is relatively large. We did not observe this on the local system, and it may be at least partly explained by the fact that we use higher resolution ocean and ice (1/25°) on the DSRC, while we use lower resolution ocean and ice (1/12°) archives on the local network.

Table II. Run-time summary on local linux workstation. The # of cores reported here is for the main compute. The 5-day forecast is a sequence of 12-hour run cycles, rather than a single 5-day run cycle.

|  | # cores | single 12-hour run cycle (seconds) | 5-day forecast (minutes) |
|---|---|---|---|
| Bering Strait, 2.3 km SWAN | 32 | 304 | 51.5 |
| Gulf of Bothnia, 1 km SWAN | 32 | 746 | 124.3 |
| Gulf of Bothnia, 2 km SWAN |  | 216 | 37.1 |
| Sea of Okhotsk, 6 km WW3 | 48 | 245 | 40.8 |
| Barents Sea WW3, 95k sea points | 48 | 262 | 43.7 |
| Barents Sea WW3, 98k sea points | 48 | 269 | 44.8 |

Table III. Timings on Navy DSRC system 'narwhal'. In the case of WW3, the total time is broken up into setup, forecast (main compute), and post-processing.

| Domain | # cores | timings in seconds for 48-hr forecast | | | | extrapolated total time for 5-day forecast, in minutes |
|---|---|---|---|---|---|---|
|  |  | setup | fcst | post | total |  |
| Bothnia 2km SWAN | 32 | - | - | - | 581.6 | 24.2 |
| Bering Strait SWAN | 32 | - | - | - | 922.4 | 38.4 |
| Barents WW3 | 128 | 224.0 | 301.4 | 18.0 | 543.4 | 22.6 |
| Okhotsk WW3 | 128 | 153.0 | 282.4 | 21.8 | 457.2 | 19.0 |

## 5. Comparisons to observations

For the Barents cases, we performed our first comparisons against wave observations. Output from the February 2024 case (4A) is compared to data from the SWIM instrument on the CFOSAT satellite. For the April 2024 case (4B), we use both SWIM/CFOSAT and drifting buoy data provided by the Norwegian Meteorological Institute.

### 5.1. Comparison to satellite wave observations

The SWIM (Surface Waves Investigation and Monitoring) instrument on the CFOSAT (China-France Oceanography SATellite) satellite provides wave spectra. This allows us to evaluate model performance based on total energy and energy within frequency bands.

The SWIM/CFOSAT data are provided by "AVISO+" which is a service of the French national space agency, CNES. SWIM is a near-nadir-pointing Ku-Band real-aperture scanning radar, described as "wave spectrometer" or "wave scatterometer". CFOSAT has a sun-synchronous orbit (inclination 97°), with "almost global coverage" within 13 days, according to CNES. The geographic resolution is relatively coarse: a wave spectrum corresponds to a 70 km x 90 km cell. Though latency is not relevant to our present usage, the product is delivered in NRT (Near Real Time) within three hours of collection. It provides directional spectra, but these are not used here. Instead, we use the non-directional wave spectra. SWIM uses the radar cross-section of long waves to compute a spectrum of the modulation of the radar signal, which is in turn used to compute a slope spectrum, $F(k)$. Non-directional spectra are provided in this $F(k)$ form (here, $k$ is wavenumber), and we convert to frequency spectrum $E(f)$ by assuming deep water dispersion relation. The range of wavenumber $k$ provided with the NRT Level 2 (L2) $F(k)$ spectra corresponds to frequencies $f$ from 0.056 - 0.263 Hz in deep water.

SWIM uses six rotating beams with incidence 0°, 2°, 4°, 6°, 8°, 10° in a conical scan. Wave spectral information such as $F(k)$ is derivable from beams with incidence 6°, 8°, 10°. In this report, we use the "combined" form. This uses all three beams, thereby reducing statistical noise. [NB: The 10° product has advantages over the other two beams, e.g., reduced "speckle noise", implying that the 10° product may be superior to the combined product. Hauser et al. (2020) found best results with 10°, and we have adopted the 10° product in another study. We have also compared the two products, 10° vs. "combined", and found that the impact of the choice on wave model validation is very slight.] Further description of SWIM and CFOSAT can be found in Hauser et al. (IEEE 2020), Le Merle et al. (JGR 2021), and Aouf et al. (IEEE 2019; GRL 2021).

We obtained files from the AVISO+ ftp site '`ftp-access.aviso.altimetry.fr`', directory '`/cfosat_products/swim_l2_op06/2024`'[3].

For total energy, we use the traditional metric of significant waveheight:

$$H_{m0} = 4\sqrt{\int_{f_{min}}^{f_{max}} E(f) df},$$

where $f_{min}$ and $f_{max}$ are determined by the frequency limits of the instrument, which are more restrictive than that of the model. $E(f)$ is the spectral density of the sea surface elevation. The integral $m_0 = \int E(f) df$ is the total variance of the sea surface elevation which is proportional to total energy. As a rule, we never extrapolate observations to match the frequency range of the model; instead, we truncate the model spectra to match the limits of the observations.

For energy within frequency bands, the energy is quantified using a similar calculation, $H_{m0B}$, where "$B$" stands for "band" or "between":

$$H_{m0B} = 4\sqrt{\int_{f_1}^{f_2} E(f) df}.$$

When deciding how many bands to use, there is a compromise between detail (more bands) vs. manageability (fewer bands). We use four bands, and they are defined in Table IV.

Table IV. Frequency bands used for $H_{m0}$ and $H_{m0B}$ in SWIM/CFOSAT comparisons. Band #1 of $H_{m0B}$ is denoted as $H_{m0B,1}$ and so on. For $H_{m0}$, $f_1 \equiv f_{min}$ and $f_2 \equiv f_{max}$.

| Parameter | $f_1$ (Hz) | $f_2$ (Hz) |
|---|---|---|
| $H_{m0}$ | 0.056 | 0.263 |
| $H_{m0B,1}$ | 0.056 | 0.08 |
| $H_{m0B,2}$ | 0.08 | 0.11 |
| $H_{m0B,3}$ | 0.11 | 0.15 |
| $H_{m0B,4}$ | 0.15 | 0.263 |

Since the frequency limits of SWIM, 0.056 to 0.263 Hz, are more restricted than that of the model, 0.038 to 0.73 Hz, and that of wind-generated gravity waves in general, the $H_{m0}$ here may be considered as a definition of a "narrow" $H_{m0}$. In some places herein, we refer to this total $H_{m0}$ as the "narrow" $H_{m0}$.

---

[3] At time of writing, '`swim_l2p_op06`' is renamed as '`swim_l2`'

The primary validation was to compare SWIM against COAMPS output, but wave spectra from the global model were also compared against SWIM for the region and time of the COAMPS simulations, to verify that COAMPS provides higher skill than the global model. Other comparisons were made, and together they quantify impact of
1. using high-resolution (0.18°) vs. low-resolution (0.5°) wind vector forcing, of
2. using NAVGEM 0.18° vs. GOFS 0.08° ice fraction forcing,
3. including vs. excluding effect of surface currents,
4. included vs. omitting dissipation by sea ice $S_{ice}$, and
5. including vs. omitting dependence on ice thickness in the $S_{ice}$ formula (IC4M9 vs. IC4M6, respectively).
6. Using the 6 km resolution of the COAMPS run, vs. the 18 km resolution of the global model.

Conclusions about each of these numbered comparisons are given later in this section.

Scatter plots were created for each COAMPS run, for the global model, and for each of the five wave parameters, so there were 80 scatter plots. For each scatter plot, skill metrics—including correlation, scatter index, bias, normalized bias, and RMS error—were computed. It is not possible to present all those results here. Figure 38 and Figure 39 are scatter plots from one of the simulations for case 4A. Figure 40 and Figure 41 show scatter plots from one of the simulations for case 4B. Table V and Table VI summarize the correlation statistics for the case 4A runs and case 4B runs, respectively.

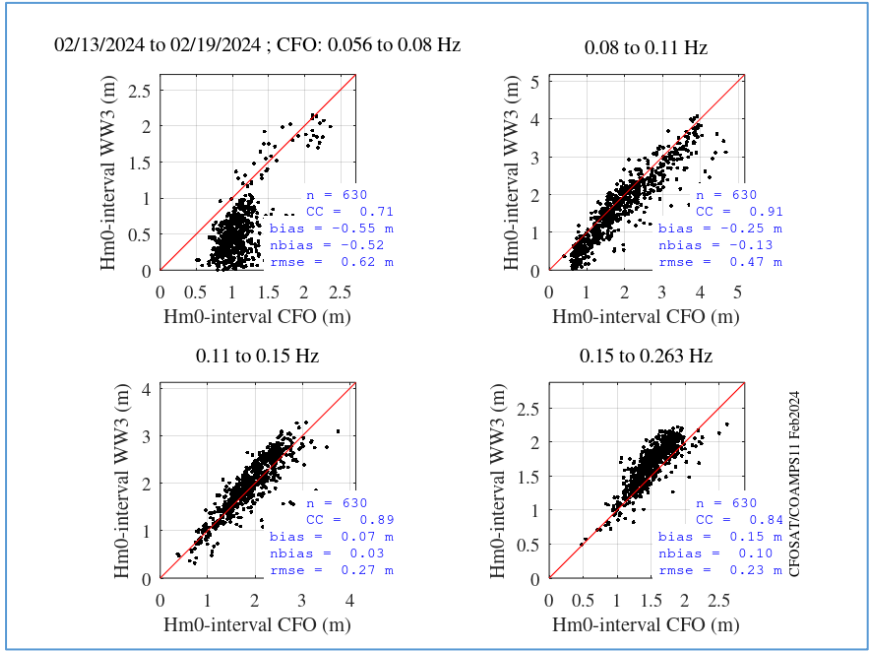

Figure 38. Comparison of COAMPS/WW3 to SWIM/CFOSAT observations for four frequency bands, for the "COAMPS11" run of case 4A, i.e., the February 2024 Barents Sea case. "Hm0-interval" is four times the square root of the sea surface elevation variance in that band, making it a "partial wave height", denoted as $H_{m0B}$ in the text. Here, the vertical axis is WW3 for the case with high resolution wind and ice fraction forcing, surface currents in forcing, and $S_{ice}$ that has dependence on ice thickness (IC4M9). Much of the mismatch in the lowest frequency band is probably due to instrument error (e.g., the so-called speckle noise). We are investigating instrument bias correction methods within a separate project; one common feature is that SWIM tends to have too much energy in this first band, for cases of $H_{m0B,1} < 2$ m.

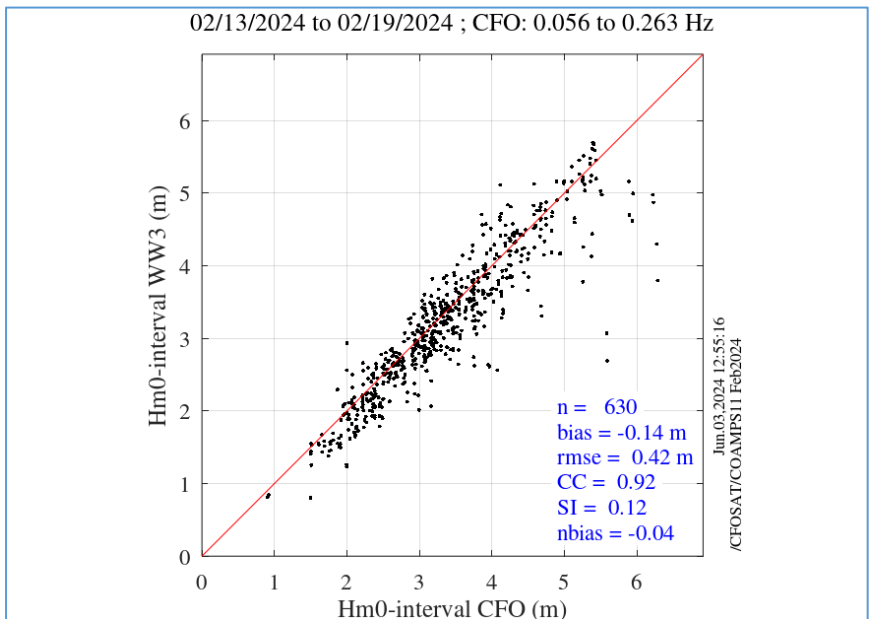

Figure 39. Like prior figure but showing wave height computed from energy in all four frequency bands.

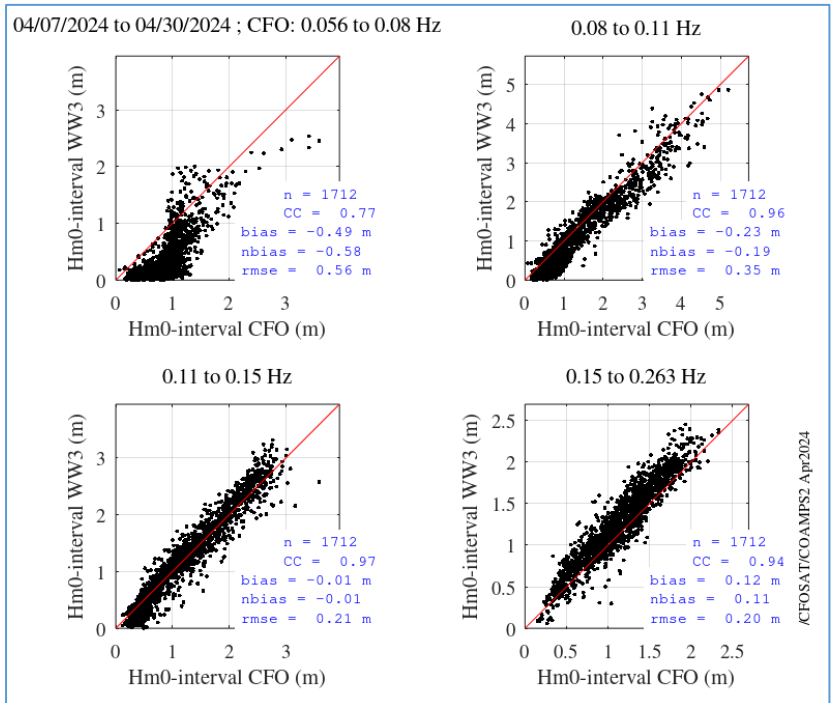

Figure 40. Comparison of COAMPS/WW3 to CFOSAT observations for four frequency bands, for the "COAMPS2" run of the April 2024 case. Here, the vertical axis corresponds to the model with high resolution wind and ice fraction forcing, surface currents in forcing, and $S_{ice}$ that has dependence on ice thickness (IC4M9).

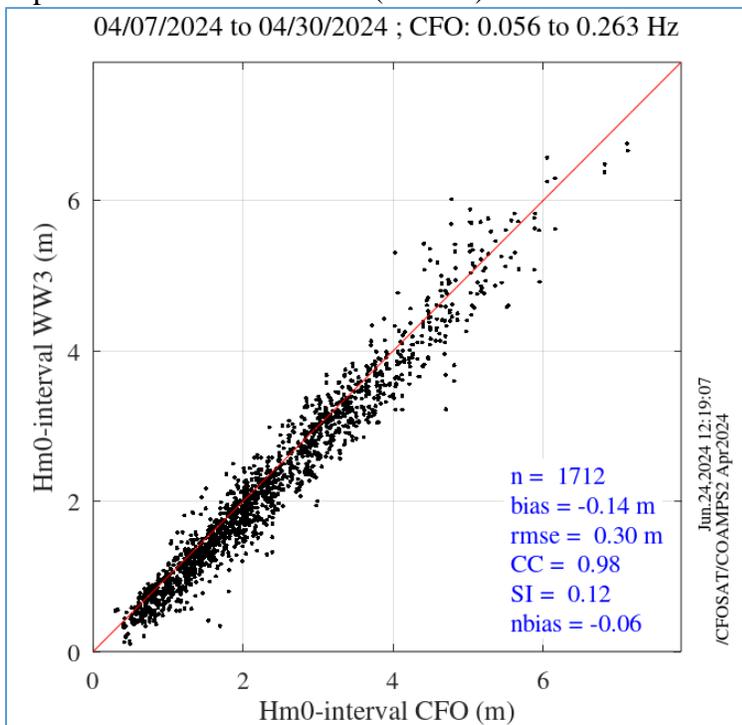

Figure 41. Like prior figure but showing wave height computed from energy in all four frequency bands.

Table V. Table of Pearson correlation, "CC", for case 4A, which is the February 2024 Barents Sea case. The notation such as "COAMPS2" was introduced in Section 4.8.1. The column for winds indicates whether winds are from a 0.18° or 0.5° NAVGEM product. The column for ice indicates whether ice fraction information is taken from the 0.18° NAVGEM product or from the 0.08° GOFS product. The column for ocean ("ocn") indicates whether ocean forcing from GOFS is included. Where ice thickness is used, it is always taken from GOFS. The column for IC4 indicates the formula for $S_{ice}$ that was used, either IC4M6, or IC4M9, or $S_{ice} = 0$. "MISC" refers to miscellaneous settings which are unlikely to make much difference, but which were included in the "COAMPS6" test for purposes of mimicking the global model run as closely as possible. These miscellaneous settings include: first direction, gamma (bottom friction), and anti-GSE settings. In these experiments, we set them to either standard IRI settings or standard COAMPS settings. In all cases, CC is computed using 630 colocations. "Band #" refers to the band of $H_{m0,B}$. "Narrow" is the total $H_{m0}$ computed from all four bands together. The last column is a summation of the prior five columns (for perfect correlation, this would equal 5.00).

```
                                              <----------- CC ------------>
                                              <--- Band # --->
                                               1    2    3    4    Narrow total
run #     winds         ice           ocn? IC4 MISC*
Global    NAVGEM 0.18  NAVGEM 0.18   no    M6  IRI     0.68 0.90 0.86 0.84 0.91   4.19
COAMPS1   NAVGEM 0.5   GOFS          yes   M9  COAMPS  0.64 0.85 0.82 0.73 0.85   3.89
COAMPS2   NAVGEM 0.18  GOFS          yes   M9  COAMPS  0.70 0.91 0.87 0.75 0.90   4.13
COAMPS3   NAVGEM 0.18  GOFS          no    M9  COAMPS  0.66 0.89 0.85 0.74 0.89   4.03
COAMPS4   NAVGEM 0.18  GOFS          no    M6  COAMPS  0.67 0.90 0.86 0.83 0.91   4.17
COAMPS5   NAVGEM 0.18  NAVGEM 0.18   no    M6  COAMPS  0.68 0.89 0.86 0.84 0.91   4.18
COAMPS6   NAVGEM 0.18  NAVGEM 0.18   no    M6  IRI     0.68 0.89 0.86 0.83 0.90   4.16
COAMPS7   NAVGEM 0.18  NAVGEM 0.18   no    0   COAMPS  0.66 0.88 0.86 0.84 0.90   4.14
COAMPS8   NAVGEM 0.18  NAVGEM 0.18   yes   0   COAMPS  0.69 0.89 0.88 0.85 0.91   4.22
COAMPS9   NAVGEM 0.18  NAVGEM 0.18   yes   M6  COAMPS  0.71 0.91 0.89 0.84 0.91   4.26
COAMPS11  NAVGEM 0.18  NAVGEM 0.18   yes   M9  COAMPS  0.71 0.91 0.89 0.84 0.92   4.27
```

Table VI. Table of Pearson correlation, "CC", for case 4B, which is the April 2024 Barents Sea case. The notation such as "COAMPS2" was introduced in Section 4.9.1. The numbers of colocations are not identical for the global IRI case and the COAMPS cases, as indicated by "# colocs", and so the comparison to CC from the global must be regarded as approximate. The table is otherwise similar to Table V.

```
                                                         <----------- CC ------------>
                                                         <--- Band # --->
                                                          1    2    3    4    Narrow total
run #     winds        ice          ocn? IC4 MISC*  # colocs
Global    NAVGEM 0.18  NAVGEM 0.18  no   M6  IRI    1873    0.77 0.96 0.96 0.94 0.97   4.60
COAMPS1   NAVGEM 0.5   GOFS         yes  M9  COAMPS 1712    0.74 0.94 0.94 0.91 0.95   4.48
COAMPS2   NAVGEM 0.18  NAVGEM 0.18  yes  M9  COAMPS 1712    0.77 0.96 0.97 0.94 0.98   4.62
COAMPS3   NAVGEM 0.18  NAVGEM 0.18  yes  M6  COAMPS 1712    0.78 0.97 0.97 0.94 0.98   4.64
COAMPS4   NAVGEM 0.18  NAVGEM 0.18  yes  0   COAMPS 1712    0.78 0.96 0.96 0.94 0.98   4.62
```

These comparisons indicate the following, following the same enumeration used earlier in this section:

1. The higher resolution wind forcing has the largest positive impact on model skill, e.g., Pearson correlation coefficient (CC) improves from 0.85 to 0.91 at band #2 and 0.82 to 0.87 at band #3. [4A and 4B: COAMPS1 vs. COAMPS2]
2. The higher resolution ice fraction forcing improves skill very slightly. [4A: COAMPS4 vs. COAMPS5]
3. Including currents improves skill, e.g., CC improves from 0.89 to 0.91 at band #2 and 0.85 to 0.87 at band #3). [4A: COAMPS2 vs. COAMPS3]
4. Omitting ice dissipation has an overall small negative impact on correlation. [4A: COAMPS8 vs. COAMPS9; 4B: COAMPS3 vs. COAMPS4]
5. Including ice thickness in forcing has an overall small impact on correlation (positive in February 2024, negative in April 2024). [4A: COAMPS9 vs. COAMPS11; 4B: COAMPS2 vs. COAMPS3]
6. WW3 geographic resolution has a counter-intuitive impact on correlation (higher resolution➔slightly lower correlation). [4A: Global vs. COAMPS6]

Further, accuracy of the COAMPS/WW3 run is modestly improved relative to the global WW3 model (e.g., CC improves from 0.86 to 0.89 at band #3 [4B: Global vs. COAMPS11]), possibly due to the inclusion of surface currents. Most of these results are expected and positive, but results (4), (5) and (6) were disappointing, and there are two takeaways. First the SWIM dataset does not include parts of the ocean that are strongly affected by $S_{ice}$, presumably due to ice flags used in the QC, making it non-optimal for this type of evaluation. Second, the $S_{ice}$ formula that depends on ice thickness was calibrated for a case where the ice thickness is taken from experimental SMOS (radiometer) estimates (Rogers et al., 2021a), while in our case, ice thickness comes from CICE. We expect that the two products have bias of different sign (negative and positive, respectively), suggesting that a recalibration is necessary for the case of using CICE forcing.

### 5.2. Comparison to buoy wave observations

Though the comparisons to SWIM observations in Section 5.1 proved useful for studying the factors affecting accuracy of the Barents Sea demo (case 4) in open water, the weak sensitivity to the experiments of setting the source term $S_{ice} = 0$ indicates that only a small number of the satellite observations were taken in areas affected by $S_{ice}$ (i.e., within the sea ice or in regions containing wave energy that has previously passed through the sea ice). In the case of the April 2024 Barents demo, we have acquired buoy observations which will help to address this limitation. We were notified of the availability of this dataset by Ana Carrasco of the Norwegian Meteorological Institute (NMI, www.met.no), and data were provided to us by Malte Muller and Jean Rabault, also of the NMI. The data were collected during the Svalbard Marginal Ice Zone 2024 Campaign (https://openmetbuoy-arctic.com/svalmiz2024.html). The buoys were of an inexpensive, open-source design (OpenMetBuoy, Rabault et al. 2022), deployed directly on the ice. They are capable of continuing measurements after falling into water, though they are not built to withstand strong crushing forces such as might happen when caught in interstitial water between large floes. Each buoy recorded non-directional wave spectra and temperatures of the air, snow, and ice from four thermistors. A listing of the contents of the data file provided to NRL is shown in Figure 42. The buoys have identification numbers KVS-01 through KVS-35, and data were available from 33 of the 35 buoys. Herein, we primarily refer to the buoy by their position (column number) in the data file (Figure 42), 1 through 33, rather than their buoy ID,

KVS-01 through KVS-35. There are a total of 4273 buoy-derived wave spectra, $E(f)$, with $f_{min} = 0.0439$ Hz and $f_{max} = 0.308$ Hz. This is a broader frequency range (at either end) than the SWIM spectra, but still more restricted than the wave model. The data start 0030 UTC 07 April 2024 and end 2230 UTC 30 April 2024.

The buoys deployed in the ice north and west of Svalbard are predominately within the Marginal Ice Zone (MIZ), though many drifted into open water during the experiment (Figure 43).

```
Name                                  Size              Bytes    Class      Attributes

Hs0                                   552x33            145728   double
T02                                   552x33            145728   double
T24                                   552x33            145728   double
accel_energy_spectrum                 55x552x33         8015040  double
elevation_energy_spectrum             55x552x33         8015040  double
frequencies_waves_imu                 55x1                 440   double
lat                                   2191x33           578424   double
lon                                   2191x33           578424   double
pHs0                                  552x33            145728   double
pT02                                  552x33            145728   double
pT24                                  552x33            145728   double
pcutoff                               552x33            145728   double
processed_elevation_energy_spectrum   55x552x33         8015040  double
temp_10cm_calibrated                  2199x33           580536   double
temp_10cm_raw                         2199x33           580536   double
temp_1m_calibrated                    2199x33           580536   double
temp_1m_raw                           2199x33           580536   double
temp_snowice_calibrated               2199x33           580536   double
temp_snowice_raw                      2199x33           580536   double
temp_snowsurface_calibrated           2199x33           580536   double
temp_snowsurface_raw                  2199x33           580536   double
time                                  2191x33           578424   int64
time_temp                             2199x33           580536   int64
time_waves_imu                        552x33            145728   int64
```

Figure 42. Contents of buoy datafile '2024_KVS_deployment.nc', as imported to Matlab/Octave. The dimension corresponding to the number of buoys is 33. The dimension corresponding to the number of frequencies is 55. Note that the wave spectra have 552 records and there is corresponding time information (`time_waves_imu`), but no corresponding position information. The position information (`lat, lon`) has 2191 records with a corresponding time array (`time`), which made it possible to infer the position corresponding to each wave measurement. Also, note that though the size of the information is the same for each buoy (either 552 or 2199 records), the time information is not identical for each buoy: time arrays include a buoy dimension '× 33'. A variable 'trajectory', which is not visible here (it is readable using 'ncdump' but not Matlab), contains a list of the 33 buoy IDs, e.g. 'KVS-01', ... , 'KVS-35', with 'KVS-05' and 'KVS-33' being absent.

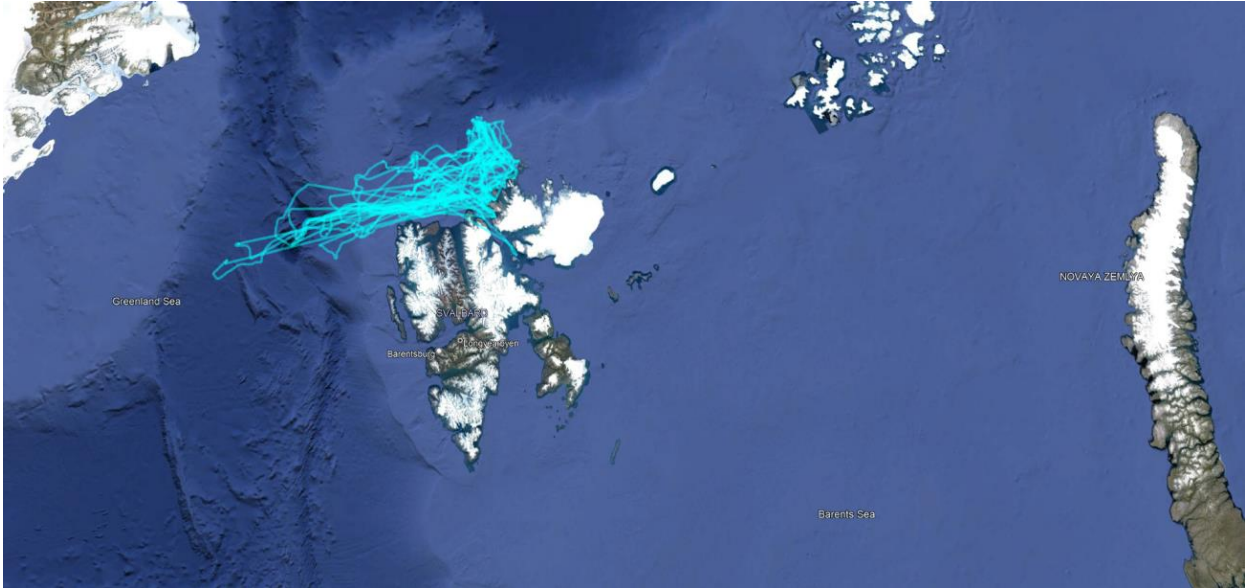

Figure 43. Buoy tracks during the Svalbard Marginal Ice Zone 2024 experiment, April 7 to 30 2024.

### 5.2.1. Comparisons of energy bands: time series

Plots of time series of $H_{m0}$ and $H_{m0B}$ were made for each of the 33 buoys, comparing the observations with three models for case 4B, COAMPS2 ($S_{ice}$=IC4M9), COAMPS3 ($S_{ice}$=IC4M6), and COAMPS4 ($S_{ice}$=0). Four of the 33 plots are given as examples in Figure 44-Figure 47. There were significantly different outcomes for different buoys, and these four buoys were selected with an intent to represent the variety of outcomes. Those four results are discussed here. In this discussion, it is important to keep in mind that dissipation by sea ice is not local to the buoy: energy is dissipated as it is advected toward the buoy, making the energy lower at the buoy. This implies that the relevant sea ice is not at the buoy itself, but is within the advection region "up-wave" from the buoy.

- Figure 44 shows the time series for buoy #25. In total energy, the three models are roughly similar for much of the time and match the data well. During the first half of the large wave event April 19-20, all models do well, but the IC4M9 model overpredicts dissipation during the second half of the event. This overprediction is most severe in the 3$^{rd}$ and 4$^{th}$ bands. In the 4$^{th}$ band, we see that the IC4M6 is also overpredicting dissipation though not as severely as IC4M9. In this band, the $S_{ice}$=0 model is the most accurate during the event, implying that the ice information provided to the wave model is overpredicted. More specifically, the wave observations indicate that there is little (possibly zero) ice at or near the buoy, but the wave model indicates that there is significant ice dissipating wave energy before it reaches the buoy location.
- Figure 45 shows the relatively short (6-day) time series for buoy #7. Here the $S_{ice} = 0$ model is again most skillful for much of the time series, except for the first two days of the deployment, when the buoy is presumably in ice, and the IC4M9 model has the most skill then. In the 4$^{th}$ band, there is a clear but not great separation between the models for April 9-13, with $S_{ice} = 0$ performing best, suggesting that the model forcing is light ice cover, while the buoy is seeing little or none.

- Figure 46 shows the time series for buoy #4. In this case, the IC4M9 is outperforming the other models in all four bands. For most of the time series, April 8-15, there is significant separation between the models, and only IC4M9 predicts the strong dissipation implied by the buoy. However, around April 17, the models are similar and all underpredict dissipation, implying that ice forcing is underpredicted during this time, e.g., the model believes the buoy to be in open water, when it is really in (or at least, down-wave from) ice.
- Figure 47 shows the time series for buoy #29, which is relatively long, over 15 days. Here the dissipation implied by the buoy is very strong, and so IC4M9 strongly outperforms the other models, though even this model underpredicts dissipation during the two wave events April 20 and April 22.

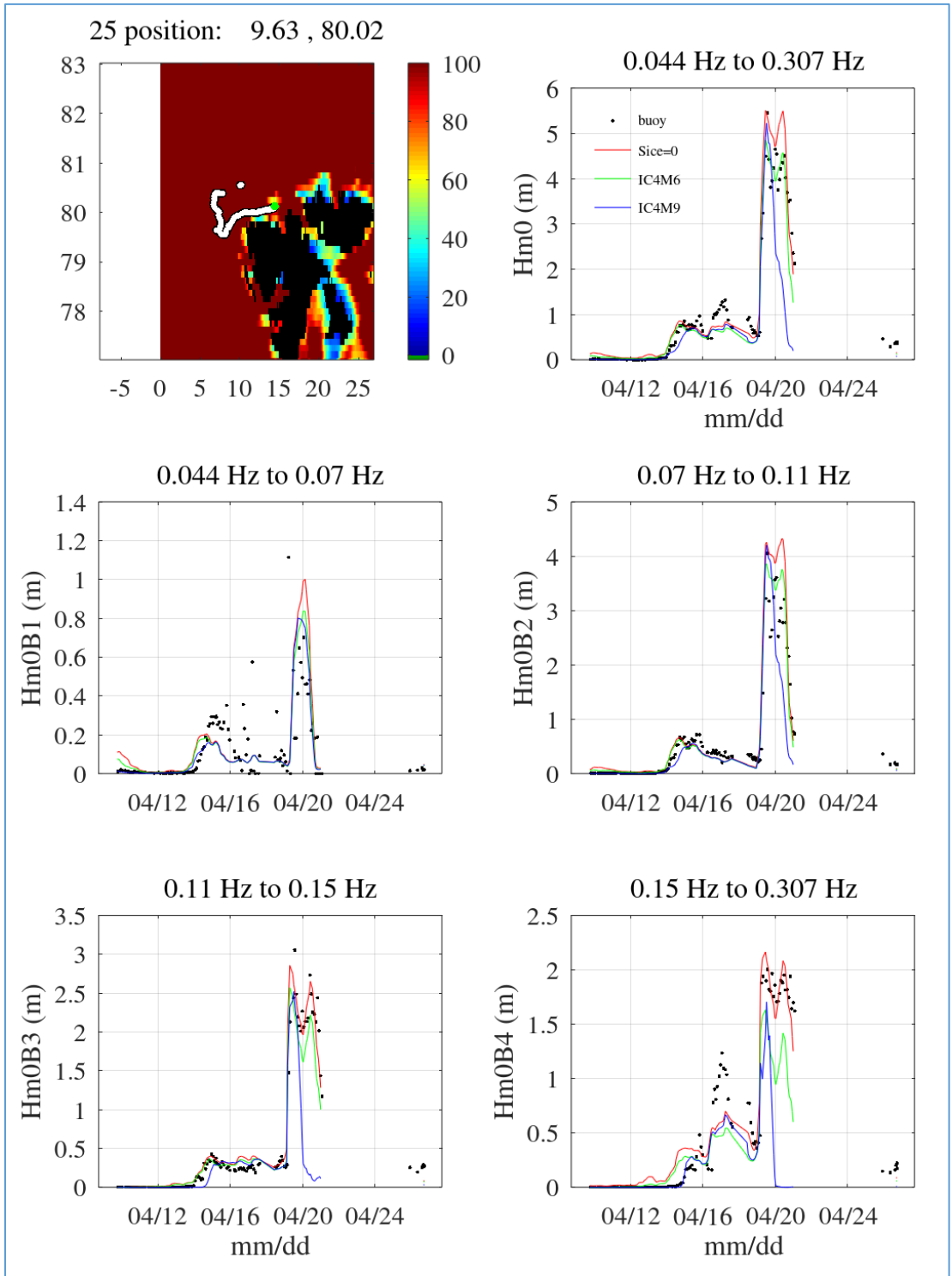

Figure 44. Buoy comparison for buoy # 25 (buoy ID, by column # in data array of buoy .nc file). Top left: buoy position. Green dot marks position at start of time series. Top right: time series of total waveheight measured by buoy, compared to three models. Center and lower panels: $H_{m0B}$ time series.

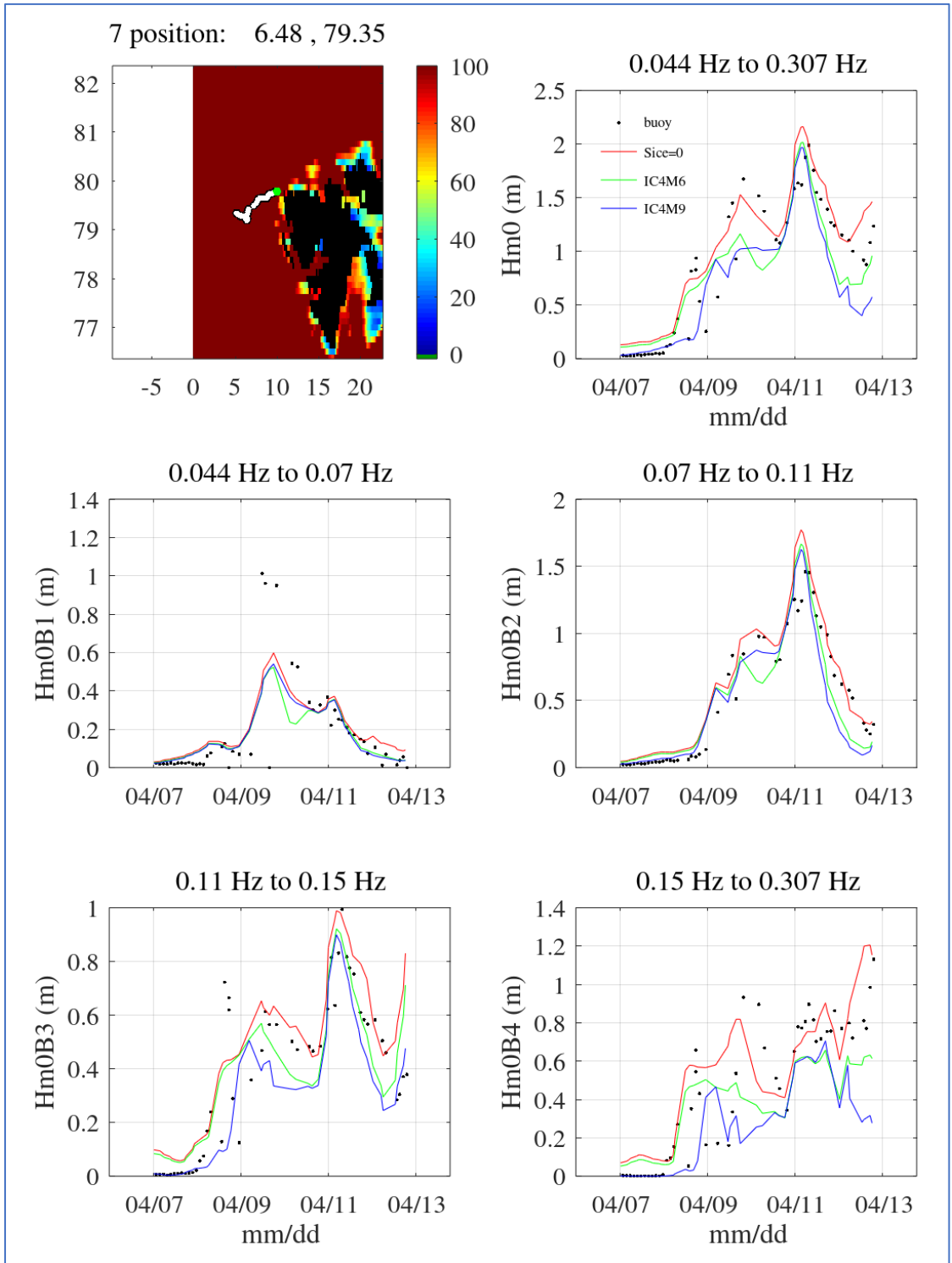

Figure 45. Like prior figure but for buoy #7.

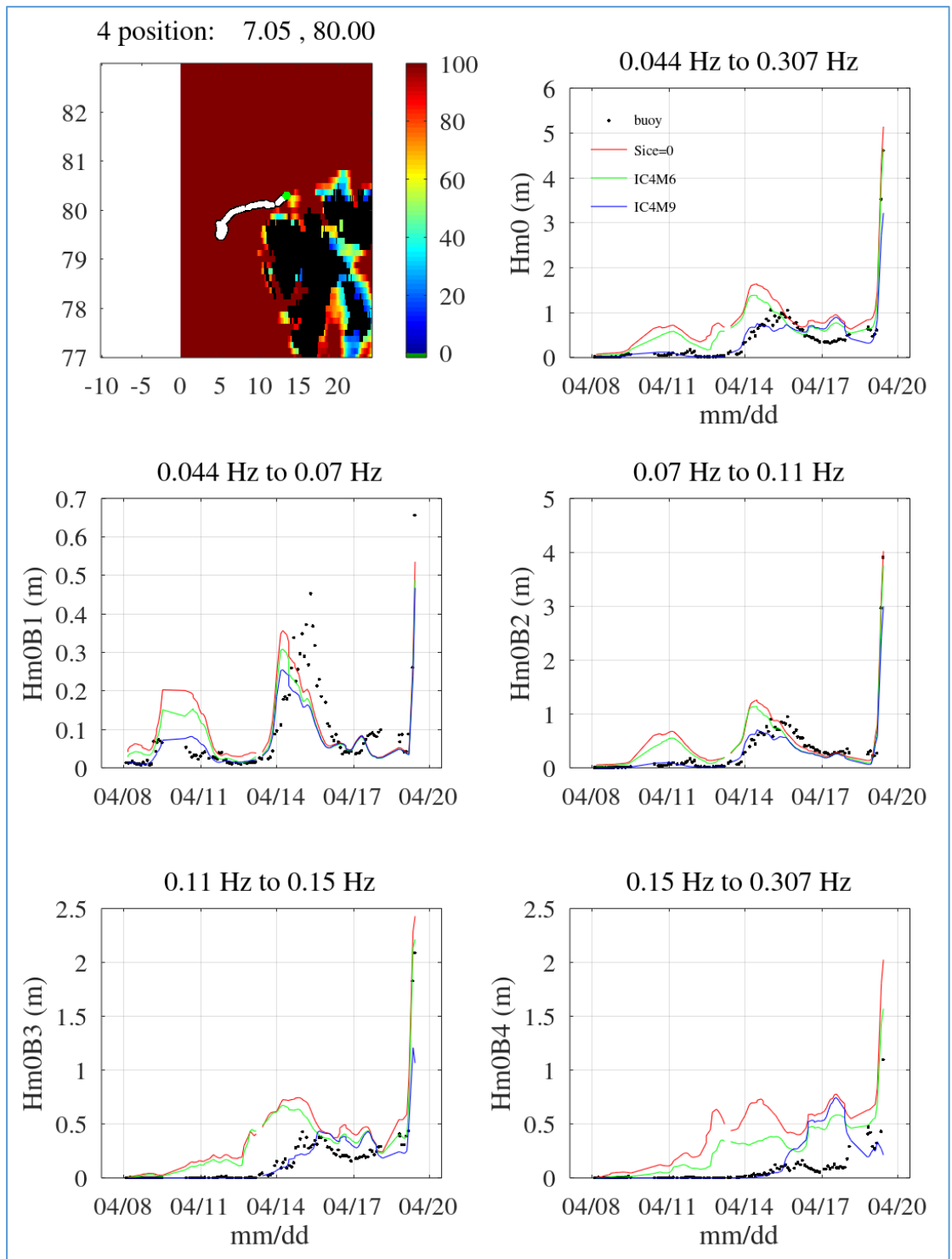

Figure 46. Like prior figure but for buoy #4.

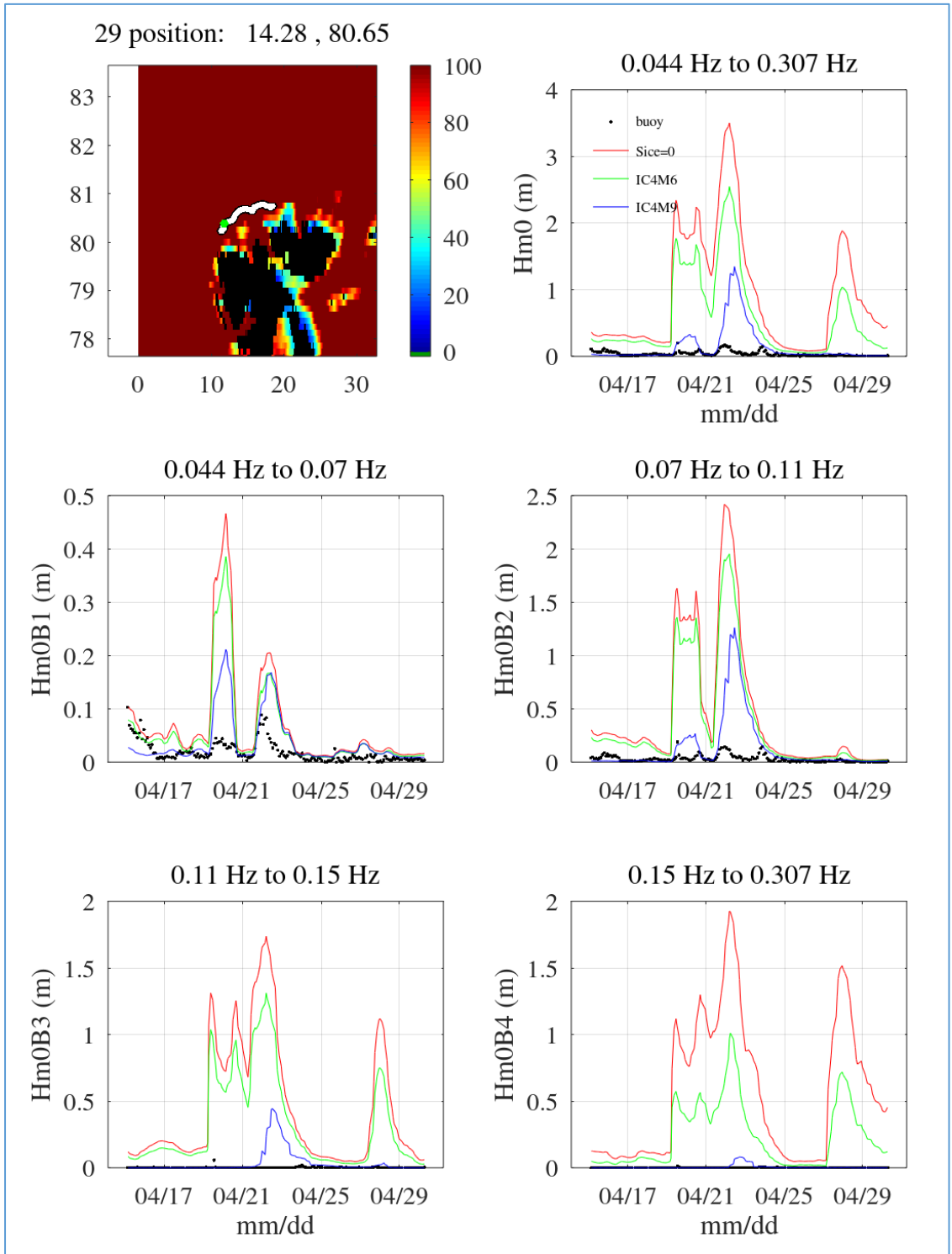

Figure 47. Like prior figure but for buoy #29.

*5.2.2. Comparison of energy bands: bias tables*

Given the variation in outcomes from one buoy to another, it is useful to look at them individually, and while we cannot present 33 plots here, we can tabulate the error metrics for individual buoys. Bias is selected for comparison here. We present bias for the three models, and also include a fourth "model", a notional model which always predicts zero energy, $E = 0$. Thus, its bias is simply the negative of the mean value observed by the buoy. By including this "model" in our tables, we can see both extremes, $S_{ice} = 0$ and $E = 0$, and consider where the two physical parameterizations IC4M6 and IC4M9 fall between the two extremes. Table VII shows bias for total waveheight. Table VIII shows bias for $H_{m0B1}$ and $H_{m0B2}$. Table IX shows bias for $H_{m0B3}$ and $H_{m0B4}$. The four buoys used in examples in Figure 44 - Figure 47 are highlighted. In all tables, the rows are sorted by the value of the bias of the $E = 0$ model. This is equivalent to sorting by the buoy-measured value, with more energetic cases at the top.

With a few exceptions, the models bound the observations, with $S_{ice} = 0$ model having positive bias and the $E = 0$ model having negative bias. Those exceptions are cases where even the $S_{ice} = 0$ model has negative bias, which indicates that the model is generally underpredicting energy for that band/buoy combination, regardless of how ice is treated, perhaps due to underpredicted wind forcing, underprediction by the wind input source function, or overprediction of open water whitecapping dissipation.

The tabulated bias values permit a general narrative about the outcome for any band/buoy combination. For example:
- Small |bias| of $S_{ice} = 0$ model & medium |bias| of IC4 models & large |bias| of $E = 0$ model. This indicates buoy-observed energy is significant, and the ice forcing to the model is over-predicted.
- Small |bias| of $S_{ice} = 0$ model & small |bias| of IC4 models & large |bias| of $E = 0$ model. This indicates that buoy-observed energy is significant, and the impact of ice at that buoy is small, and that "light" ice cover is represented well in the forcing.
- Medium |bias| of $S_{ice} = 0$ model & smallest |bias| with IC4M6 model & medium |bias| of $E = 0$ model. This indicates an important role of the $S_{ice}$ term, with IC4M6 outperforming IC4M9, implying either a) ice forcing is too thick, or b) ice forcing is correct, but there is a calibration problem with IC4M9 (dissipation too strong).
- Medium |bias| of $S_{ice} = 0$ model & smallest |bias| with IC4M9 model & medium |bias| of $E = 0$ model. This indicates an important role of the $S_{ice}$ term, with IC4M9 outperforming IC4M6, with the latter underpredicting dissipation.
- Medium or large |bias| of $S_{ice} = 0$ model & smallest |bias| with IC4M9 model & small |bias| of $E = 0$ model. This indicates that buoy-observed energy is low, and that the $S_{ice}$ term has an important role, with IC4M9 doing a good job of predicting the strong dissipation.

There is a drawback to the bias tables. From the time series plots, we can see that the narrative can be different in different parts of the time series. The bias tables only allow us to infer the dominant narrative.

Table VII. Bias for $H_{m0}$, in meters. 'icol' refers to the buoy ID, by column in the data array of the buoy .nc file. Shaded rows correspond to example plots. Rows are sorted by bias value of the '$E = 0$' model, which is, by definition, the negative of the mean value of the observations.

| bias | Hm0 | | | |
|---|---|---|---|---|
| icol | Sice=0 | IC4M6 | IC4M9 | E=0 |
| 16 | 0.27 | -0.09 | -0.98 | -1.30 |
| 17 | 0.42 | 0.07 | -0.71 | -1.02 |
| 25 | 0.09 | -0.10 | -0.37 | -1.00 |
| 12 | 0.52 | 0.14 | -0.64 | -0.94 |
| 20 | -0.07 | -0.32 | -0.79 | -0.91 |
| 7 | 0.13 | -0.06 | -0.19 | -0.84 |
| 15 | 0.45 | 0.19 | -0.49 | -0.77 |
| 31 | 0.04 | -0.07 | -0.19 | -0.72 |
| 21 | 0.43 | 0.20 | -0.27 | -0.51 |
| 2 | 0.13 | 0.06 | -0.15 | -0.42 |
| 8 | 0.77 | 0.50 | -0.20 | -0.40 |
| 4 | 0.42 | 0.26 | 0.01 | -0.40 |
| 13 | 0.17 | 0.05 | -0.18 | -0.33 |
| 30 | 0.22 | 0.14 | -0.08 | -0.14 |
| 3 | 0.36 | 0.20 | 0.03 | -0.10 |
| 22 | 0.27 | 0.19 | -0.06 | -0.09 |
| 23 | 0.49 | 0.28 | 0.04 | -0.07 |
| 18 | 0.56 | 0.33 | 0.07 | -0.06 |
| 1 | 0.28 | 0.20 | -0.02 | -0.05 |
| 29 | 0.88 | 0.55 | 0.11 | -0.04 |
| 33 | 0.84 | 0.52 | 0.10 | -0.03 |
| 27 | 0.56 | 0.32 | 0.00 | -0.03 |
| 32 | 0.18 | 0.11 | -0.02 | -0.02 |
| 14 | 1.35 | 0.85 | 0.02 | -0.02 |
| 10 | 0.93 | 0.57 | 0.01 | -0.02 |
| 19 | 0.77 | 0.42 | 0.02 | -0.02 |
| 9 | 0.92 | 0.55 | 0.04 | -0.02 |
| 24 | 0.86 | 0.49 | 0.03 | -0.02 |
| 11 | 0.96 | 0.57 | 0.04 | -0.02 |
| 28 | 0.59 | 0.29 | 0.01 | -0.02 |
| 5 | 0.54 | 0.25 | 0.02 | -0.01 |
| 26 | 0.57 | 0.26 | 0.01 | -0.01 |
| 6 | 0.06 | 0.01 | 0.00 | -0.01 |

Table VIII. Bias for $H_{m0B1}$ (left-hand table) and $H_{m0B2}$ (right-hand table), in meters.

| bias | Hm0B1 | | | | bias | Hm0B2 | | | |
|---|---|---|---|---|---|---|---|---|---|
| icol | Sice=0 | IC4M6 | IC4M9 | E=0 | icol | Sice=0 | IC4M6 | IC4M9 | E=0 |
| 20 | -0.03 | -0.07 | -0.11 | -0.19 | 16 | 0.03 | -0.13 | -0.70 | -0.97 |
| 7 | 0.02 | -0.01 | -0.01 | -0.17 | 17 | 0.12 | -0.05 | -0.58 | -0.84 |
| 16 | 0.01 | -0.02 | -0.07 | -0.17 | 12 | 0.13 | -0.04 | -0.47 | -0.74 |
| 12 | 0.03 | -0.01 | -0.05 | -0.16 | 15 | 0.13 | 0.01 | -0.42 | -0.68 |
| 17 | 0.07 | 0.02 | -0.03 | -0.15 | 25 | 0.12 | 0.03 | -0.15 | -0.61 |
| 15 | 0.03 | -0.01 | -0.05 | -0.15 | 31 | 0.03 | -0.02 | -0.12 | -0.49 |
| 13 | 0.05 | 0.01 | -0.04 | -0.14 | 7 | 0.09 | -0.01 | -0.04 | -0.49 |
| 31 | -0.01 | -0.03 | -0.04 | -0.13 | 21 | 0.21 | 0.09 | -0.24 | -0.45 |
| 25 | 0.02 | 0.00 | -0.01 | -0.13 | 20 | 0.08 | -0.03 | -0.32 | -0.41 |
| 21 | 0.04 | 0.01 | -0.03 | -0.12 | 8 | 0.34 | 0.24 | -0.15 | -0.32 |
| 4 | 0.01 | 0.00 | -0.02 | -0.09 | 4 | 0.16 | 0.10 | -0.06 | -0.32 |
| 8 | 0.02 | 0.00 | -0.02 | -0.09 | 2 | 0.10 | 0.06 | -0.05 | -0.28 |
| 22 | 0.01 | 0.00 | -0.03 | -0.06 | 13 | 0.15 | 0.06 | -0.15 | -0.26 |
| 2 | 0.03 | 0.02 | 0.01 | -0.05 | 30 | 0.08 | 0.05 | -0.06 | -0.11 |
| 1 | 0.03 | 0.02 | 0.00 | -0.03 | 22 | 0.22 | 0.17 | -0.04 | -0.06 |
| 30 | 0.03 | 0.01 | 0.00 | -0.03 | 3 | 0.10 | 0.07 | 0.02 | -0.06 |
| 14 | 0.11 | 0.07 | 0.01 | -0.02 | 18 | 0.22 | 0.17 | 0.06 | -0.06 |
| 27 | 0.06 | 0.04 | 0.00 | -0.02 | 23 | 0.14 | 0.11 | 0.03 | -0.06 |
| 32 | 0.03 | 0.01 | -0.01 | -0.02 | 1 | 0.18 | 0.14 | -0.02 | -0.04 |
| 29 | 0.05 | 0.04 | 0.02 | -0.02 | 29 | 0.41 | 0.33 | 0.10 | -0.03 |
| 10 | 0.13 | 0.08 | 0.01 | -0.02 | 33 | 0.39 | 0.31 | 0.09 | -0.03 |
| 18 | 0.03 | 0.02 | 0.01 | -0.02 | 27 | 0.32 | 0.24 | 0.00 | -0.01 |
| 33 | 0.05 | 0.04 | 0.02 | -0.02 | 32 | 0.14 | 0.10 | -0.01 | -0.01 |
| 3 | 0.01 | 0.01 | 0.00 | -0.02 | 19 | 0.35 | 0.26 | 0.02 | -0.01 |
| 23 | 0.02 | 0.02 | 0.01 | -0.02 | 9 | 0.45 | 0.34 | 0.03 | -0.01 |
| 9 | 0.06 | 0.04 | 0.01 | -0.02 | 28 | 0.22 | 0.17 | 0.01 | -0.01 |
| 19 | 0.05 | 0.03 | 0.00 | -0.01 | 24 | 0.41 | 0.31 | 0.02 | -0.01 |
| 24 | 0.06 | 0.04 | 0.01 | -0.01 | 11 | 0.47 | 0.36 | 0.04 | -0.01 |
| 11 | 0.06 | 0.04 | 0.01 | -0.01 | 14 | 0.87 | 0.63 | 0.02 | -0.01 |
| 28 | 0.03 | 0.02 | 0.00 | -0.01 | 5 | 0.16 | 0.12 | 0.02 | -0.01 |
| 5 | 0.02 | 0.02 | 0.00 | -0.01 | 26 | 0.19 | 0.14 | 0.01 | -0.01 |
| 26 | 0.03 | 0.02 | 0.00 | -0.01 | 10 | 0.65 | 0.46 | 0.01 | -0.01 |
| 6 | 0.01 | 0.00 | 0.00 | -0.01 | 6 | 0.03 | 0.01 | 0.00 | 0.00 |

Table IX. Bias for $H_{m0B3}$ (left-hand table) and $H_{m0B4}$ (right-hand table), in meters.

| bias | Hm0B3 | | | | bias | Hm0B4 | | | |
|---|---|---|---|---|---|---|---|---|---|
| icol | Sice=0 | IC4M6 | IC4M9 | E=0 | icol | Sice=0 | IC4M6 | IC4M9 | E=0 |
| 16 | 0.12 | -0.01 | -0.61 | -0.67 | 25 | 0.00 | -0.16 | -0.31 | -0.50 |
| 25 | 0.04 | -0.05 | -0.25 | -0.50 | 7 | 0.08 | -0.08 | -0.19 | -0.45 |
| 20 | -0.09 | -0.15 | -0.46 | -0.46 | 20 | -0.02 | -0.24 | -0.44 | -0.44 |
| 17 | 0.23 | 0.10 | -0.41 | -0.44 | 16 | 0.39 | 0.05 | -0.36 | -0.37 |
| 12 | 0.32 | 0.17 | -0.36 | -0.42 | 31 | 0.04 | -0.06 | -0.11 | -0.32 |
| 7 | 0.08 | 0.01 | -0.08 | -0.38 | 12 | 0.59 | 0.26 | -0.18 | -0.19 |
| 31 | 0.05 | 0.01 | -0.07 | -0.29 | 17 | 0.47 | 0.15 | -0.18 | -0.18 |
| 15 | 0.29 | 0.20 | -0.24 | -0.28 | 2 | 0.07 | 0.00 | -0.14 | -0.17 |
| 2 | 0.05 | 0.01 | -0.13 | -0.23 | 15 | 0.46 | 0.23 | -0.12 | -0.12 |
| 21 | 0.24 | 0.17 | -0.12 | -0.17 | 8 | 0.55 | 0.30 | -0.08 | -0.08 |
| 4 | 0.23 | 0.17 | 0.00 | -0.17 | 4 | 0.36 | 0.21 | 0.08 | -0.07 |
| 8 | 0.44 | 0.34 | -0.11 | -0.13 | 21 | 0.39 | 0.18 | -0.03 | -0.04 |
| 13 | 0.04 | 0.01 | -0.10 | -0.10 | 13 | 0.10 | 0.04 | -0.03 | -0.03 |
| 30 | 0.10 | 0.09 | -0.05 | -0.06 | 3 | 0.34 | 0.16 | 0.02 | -0.02 |
| 3 | 0.14 | 0.10 | 0.01 | -0.05 | 30 | 0.18 | 0.09 | -0.01 | -0.01 |
| 23 | 0.21 | 0.16 | 0.02 | -0.02 | 23 | 0.42 | 0.20 | 0.01 | 0.00 |
| 18 | 0.28 | 0.20 | 0.03 | -0.01 | 18 | 0.41 | 0.17 | 0.01 | 0.00 |
| 29 | 0.45 | 0.32 | 0.03 | 0.00 | 29 | 0.57 | 0.24 | 0.00 | 0.00 |
| 33 | 0.42 | 0.30 | 0.03 | 0.00 | 19 | 0.52 | 0.17 | 0.00 | 0.00 |
| 28 | 0.26 | 0.16 | 0.00 | 0.00 | 5 | 0.45 | 0.14 | 0.00 | 0.00 |
| 5 | 0.22 | 0.14 | 0.00 | 0.00 | 28 | 0.45 | 0.14 | 0.00 | 0.00 |
| 26 | 0.23 | 0.15 | 0.00 | 0.00 | 33 | 0.56 | 0.23 | 0.00 | 0.00 |
| 1 | 0.18 | 0.14 | 0.00 | 0.00 | 11 | 0.61 | 0.23 | 0.00 | 0.00 |
| 22 | 0.15 | 0.11 | 0.00 | 0.00 | 22 | 0.13 | 0.06 | 0.00 | 0.00 |
| 11 | 0.48 | 0.33 | 0.00 | 0.00 | 24 | 0.56 | 0.20 | 0.00 | 0.00 |
| 24 | 0.42 | 0.28 | 0.00 | 0.00 | 26 | 0.46 | 0.14 | 0.00 | 0.00 |
| 19 | 0.36 | 0.24 | 0.00 | 0.00 | 27 | 0.34 | 0.10 | 0.00 | 0.00 |
| 27 | 0.28 | 0.18 | 0.00 | 0.00 | 1 | 0.14 | 0.08 | 0.00 | 0.00 |
| 6 | 0.02 | 0.00 | 0.00 | 0.00 | 6 | 0.04 | 0.00 | 0.00 | 0.00 |
| 14 | 0.73 | 0.48 | 0.00 | 0.00 | 9 | 0.58 | 0.22 | 0.00 | 0.00 |
| 9 | 0.46 | 0.31 | 0.00 | 0.00 | 14 | 0.68 | 0.26 | 0.00 | 0.00 |
| 10 | 0.48 | 0.30 | 0.00 | 0.00 | 32 | 0.07 | 0.02 | 0.00 | 0.00 |
| 32 | 0.08 | 0.05 | 0.00 | 0.00 | 10 | 0.42 | 0.15 | 0.00 | 0.00 |

### 5.2.3. Comparison of energy bands: combined data

We created two additional methods of appraising model performance, with two general goals. The method should...
1) ...be more concise than the methods of Sections 5.2.1 and 5.2.2, and
2) ...convey the same general impression that the reader would get from laboriously studying the 33 plots, of which Figure 44 - Figure 47 are examples.

The first criterion makes it difficult or impossible to evaluate performance of each buoy separately. Therefore, with both approaches we combine the data. Individual co-locations are retained, but they are not organized by buoy. A benefit to this approach, besides being more concise, is that every co-location has the same weight. This contrasts with the tabulated bias of Section 5.2.2, where the impact of an individual co-location on the reported bias values is effectively weighted by $1/n$, where $n$ is the number of co-locations for that buoy.

*Method of pass/fail*

The first method assigned a pass or fail condition to each co-location, and then computes the % of pass and fail. The criterion uses the following relations for each co-location, using the symbol "➔" to represent "then" in "if/then" logic:

$$e_H = H_{model} - H_{obs}$$
$$H_{obs} \geq 0.1 \text{ m} \rightarrow e_{max\_allowed} = H_{obs}/2$$
$$H_{obs} \leq 0.1 \text{ m} \rightarrow e_{max\_allowed} = 0.05 \text{ m}$$
$$|e_H| < e_{max\_allowed} \rightarrow \text{pass}$$
$$e_H \geq e_{max\_allowed} \rightarrow \text{fail high (model overpredicting)}$$
$$-e_H \geq e_{max\_allowed} \rightarrow \text{fail low (model underpredicting)}$$

In all cases, $H$ can be $H_{m0,B1}$, $H_{m0,B2}$, $H_{m0,B3}$, or $H_{m0,B4}$. The maximum allowed error $e_{max\_allowed}$ is made conditional, because low values of $H$ occur often, and we do not want to consider a model to have failed if, e.g., $H_{model} = 10^{-5}$ m and $H_{obs} = 10^{-6}$ m.

The drawback to this method is that it does not make any distinction between "pass with exact match" vs. "barely pass" or "fail with enormous error" vs. "barely fail".

Table X shows the pass/fail results using all the co-locations. Unsurprisingly, the $S_{ice} = 0$ has a "fail high" condition much of the time for three of the four bands. The exception is the low frequency band, and this makes sense, as low frequency wave energy is dissipated the least by sea ice. The IC4M9 and $E = 0$ models are the only ones that fail low a significant fraction of the time, e.g., 14% for $H_{m0,B1}$ with IC4M9, and the IC4M9 model strongly outperforms the $S_{ice} = 0$ and IC4M6 models in bands 2 to 4. The tendency of IC4M6 to fail high increases with frequency, indicating that $S_{ice}$ may be too weak with IC4M6 in higher frequencies.

There exists a potential criticism of this comparison: insofar as many of the observations are in a highly dissipative environment, the comparison will naturally favor the most dissipative model. We explore the validity of the criticism by two means. First, we include the $E = 0$ model as an alternative model. We see that the % passing for IC4M9 is only slightly better than that of $E = 0$ for bands 2 to 4. At band 1, the % passing of IC4M9 is better than that of $E = 0$, 88% vs. 79%. Second, we include a table which only uses co-locations where $H_{m0B}>0.1$ m: Table XI. For this subset of the data, IC4M9 fails low for a large fraction of cases, e.g. 68% for band 3. For band 2 to 4, IC4M9 performs worse than $S_{ice} = 0$ and IC4M6 but better than $E = 0$.

Table X. Pass/Fail results for all co-locations. The four bands here are the bands of $H_{m0B}$.

```
% fail (low) / % pass / % fail (high)
using all observations
          Sice=0       IC4M6       IC4M9       E=0
band # 1  3/75/22      4/83/14     6/88/ 6     21/79/ 0
band # 2  1/39/60      2/50/48     14/75/11    29/71/ 0
band # 3  0/31/68      0/43/57     13/83/ 4    19/81/ 0
band # 4  0/23/77      1/46/53     9/89/ 3     13/87/ 0
```

Table XI. Pass/Fail results for the subset of cases with $H_{m0B}>0.1$ m.

```
% fail (low) / % pass / % fail (high)
using only observations with Hm0B>=10 cm
          Sice=0       IC4M6       IC4M9       E=0
band # 1  22/54/24     29/56/15    39/56/ 5    100/ 0/ 0
band # 2  3/64/33      8/68/24     57/35/ 8    100/ 0/ 0
band # 3  2/55/44      2/70/28     68/28/ 4    100/ 0/ 0
band # 4  1/46/53      13/50/37    70/25/ 5    100/ 0/ 0
```

*Method of declared winners*

In this method, for each co-location, the "best" is the one with the lowest $|e_H|$. We declare a winner if $|e_H|$ of the best model is less than half the $|e_H|$ of second-best model. Of course, this results in many of cases where no single model is clearly better than the others, and this is recorded as a "no winner" outcome. The positive feature of this approach is that it explicitly addresses the *relative strength* of the models: a model is credited with a win only if that win is difficult to achieve. Conversely, it does not credit any model with a win in cases where a match to the observations is easy to achieve. A drawback to this approach is that it does not distinguish between a 2-way tie and a 4-way tie: both are considered as "no winner".

As with the pass/fail method, we attempt to address the criticism that the outcome may be determined a priori simply by the question of whether the co-location is in a highly dissipated regime or not. This is accomplished by two means. First, we again include the $E = 0$ model as an alternative model, and so the IC4M9 model cannot win simply by dissipating the energy level to zero. Second, we subdivide into "low energy" and "high energy" cases, with $H = 0.10$ m used as the threshold. Since the outcomes are simple tallies, we do not need to include an "all cases" table: that would simply be the sum of the two tables presented. It is not obvious to us whether it is best to use $H_{m0}$ or $H_{m0B}$ as a threshold, so we present the first method of dividing the populations in Table XII and the second method of dividing the populations in Table XIII.

Remarkably, for the low-energy cases, IC4M9 records the most wins, even though the $E = 0$ model is included. This is a favorable outcome, indicating that the model can often correctly distinguish the difference between strong dissipation and complete dissipation.

Summarizing the results in these four tables, Table X to Table XIII:
- All cases or low-energy cases, at band 1, IC4M9 outperforms the other three models.
- All cases or low-energy cases, at bands 2 to 4, IC4M9 outperforms the $S_{ice} = 0$ and IC4M6 models, but the $E = 0$ also performs well.
- Not-low-energy cases at band 1, performance of IC4M9 is equivalent to that of the $S_{ice} = 0$ and IC4M6 models (Table XI) or better than that of those two models (Table XIII). The $E = 0$ model performs poorly.
- Not-low-energy cases at bands 2 to 4, performance of IC4M9 is worse than that of the $S_{ice} = 0$ and IC4M6 models. The $E = 0$ model performs poorly.

Table XII. Counts of wins by each model, as explained in the text. The co-locations are separated into two populations, according to the total waveheight, $H_{m0}$, as measured by the buoy.

```
Hm0_buoy<10 cm case:     Sice=0    IC4M6    IC4M9     E=0   no winner
            band # 1        47      213      904     656       1371
            band # 2        18       34      724     647       1768
            band # 3        16       61      226     393       2495
            band # 4         0      121      119     133       2818
Hm0_buoy>=10 cm case:    Sice=0    IC4M6    IC4M9     E=0   no winner
            band # 1        96      121      168      58        655
            band # 2       191      146      132      40        589
            band # 3       105      191       76      93        633
            band # 4        99       81       64      81        773
```

Table XIII. Like Table XII, except that the co-locations are separated according to the band waveheight, $H_{m0B}$, as measured by the buoy.

```
Hm0B_buoy<10 cm case:    Sice=0    IC4M6    IC4M9     E=0   no winner
            band # 1        95      288     1001     711       1701
            band # 2        22       36      751     648       1807
            band # 3        20       76      260     485       2750
            band # 4         2      124      158     210       3353
Hm0B_buoy>=10 cm case:   Sice=0    IC4M6    IC4M9     E=0   no winner
            band # 1        48       46       71       3        325
            band # 2       187      144      105      39        550
            band # 3       101      176       42       1        378
            band # 4        97       78       25       4        238
```

**5.3. Comparison to ice thickness observations**

In Sections 5.2.1 and 5.2.2, we mentioned how the overprediction of dissipation by the IC4M9 model may be partly attributed by overprediction of ice thickness by CICE. Therefore, it is useful to attempt an independent evaluation of the model ice thickness.

*5.3.1. SMOS*

Satellite-based ice thickness estimates are derived from the MIRAS radiometer onboard the European Space Agency's SMOS satellite. Ice thickness from SMOS is a relatively new, first-generation product, e.g., starting from Kaleschke et al. (2012) and Huntemann et al. (2014), and is sometimes referred to as an "experimental" product, meaning that it cannot be expected to have the same level of accuracy as, say, ice concentration from passive microwave, wind speed from scatterometer, or waveheight from altimeter.

Processed files are provided by AWI (Alfred Wegener Institute) (https://www.meereisportal.de[4], Kaleschke et al. (2012, 2016) and Tian-Kunze et al. (2014)). The primary authors are affiliated with U. Hamburg, so this is the "UH" product. A related SMOS ice thickness product from U.

---

[4] Downloaded files have names such as 'SMOS_Icethickness_v3.3_north_20240406.nc' and are approximately 22 MB in size. Website for several thickness products: https://data.seaiceportal.de/relaunch/thickness?lang=en . Website for this specific thickness product: https://data.seaiceportal.de/data/smos_awi/v3.3/n/2024 .

Bremen (UB) is discussed in Huntemann et al. (2014) and Pațilea et al. (2019). Both products are on an approximately 12 km irregular grid, with one analysis per day. For high values of ice thickness ($h_{ice}$), the instrument saturates, but that saturation may be different for the UH product used here, vs. the UB product used by Rogers et al. (2021a,b).

Kaleschke et al. (2016) evaluate the AWI SMOS product for March 2014 in a region east of Svalbard using ship-based measurements. They conclude "The UB SMOS product provides ice thickness only up to a maximum of 0.5 m while the UH product resolves thicknesses up to about 1.5 m."

Huntemann et al. (2014) state that the product is intended for the "freeze-up period"[5]. Combining this with the restriction of $h_{ice} \leq 50$ cm for the U. Bremen product, from a remote sensing point of view, we can guess that the most benign situation would be sheet ice (nilas, grey ice, cemented pancake ice, and other types of level, first year ice).

The daily AWI SMOS data files contain the following fields:
- Sea ice thickness, in meters.
- Sea ice thickness 'total uncertainty', in meters.
- Saturation ratio, which is the ratio of retrieved ice thickness and maximal retrievable ice thickness, as a percentage.
- Brightness temperature intensity, "(TBh+TBv)/2" in K.
- Brightness temperature uncertainty.
- Number of TBh, TBv pairs available.
- Snow surface temperature, in K.
- Bulk ice salinity, in psu.
- Percent of "RFI-contaminated" measurements.

### 5.3.1. GOFS

GOFS 3.1 uses the CICE (Community Ice CodE, Hunke and Lipscomb 2008) model, with assimilation of ice concentration observations, as described in Metzger et al. (2017).

The accuracy of ice thickness from CICE is not well established (see, e.g., Xu and Li 2023). There are many routine satellite-derived ice concentration and ice edge products, but relatively few products for ice thickness, and those that do exist are relatively new (CryoSat-2 altimeter, SMOS passive microwave radiometer).

GOFS 3.1 is validated by Metzger et al. (2017) using NASA IceBridge data, which is for the Arctic region north of Alaska, Canada, and Greenland but does not include the Barents Sea. These are cases with ice significantly thicker than our MIZ case, e.g., $h_{ice}$=1.5 to 7 m from IceBridge 'Central Arctic' case (Metzger et al. (2017), Fig. 23). The comparisons do not indicate any systematic bias, and in fact, compared to data from the easternmost flight, '20150319', GOFS 3.1 exhibited a bias of -45 cm.

---

[5] Specifically, U. Bremen suggest October to April for the northern hemisphere. Thus the SMOS dataset should be suited for our April timeframe.

### 5.3.2. *In situ measurements*

In situ ice observations were provided by Dr. Malte Muller of the Norwegian Meteorological Institute. These were taken during the deployment of the 'OpenMetBuoy' wave and temperature buoys described in Section 5.2, deployed for the Svalbard Marginal Ice Zone 2024 Campaign. When each buoy was deployed on a floe, a hole was drilled in the floe and the snow and ice thickness was recorded. The cruise report, Muller et al. (2024), states: "Sea-ice thickness, snow depth, and density observations were collected at a location close-by (ca. 5 meter) to the OpenMETBuoy station." Figure 11 in that report includes a spatial plot of these 34 $h_{ice}$ measurements, and they are tabulated in Table 1 of that report. The buoys are numbered KVS-01 through KVS-35, but KVS-33 is not listed, so 34 buoys are in the list. The Table includes date/time information but not position, so we needed to determine the position information indirectly by looking up the buoy ID and nearest date/time in the primary data file, 2024_KVS_deployment.nc. The buoy KVS-05 was included in the table of in situ ice thickness measurements, but since it is missing from 2024_KVS_deployment.nc, we were not able to determine its position, and thus it is omitted from the $h_{ice}$ comparisons; thus, we have 33 buoy ice thicknesses.

### 5.3.3. *Summary of limitations*

We have access to three independent estimates of ice thickness. All three have limitations, which we summarize here.

- <u>CICE model in GOFS</u>. Our primary concern about the model is that we do not know anything about the accuracy of its ice thickness within the MIZ. We are not aware of any prior evaluations which would shed light on this.
- <u>In situ from SvalbardMIZ2024 experiment</u>. The primary concern with the in situ measurements is that they are highly local (one drill hole each) and may not be representative of a larger area, e.g., a 12 km ×12 km grid cell (SMOS resolution), 9 km × 9 km grid cell (1/12° CICE resolution), 6 km × 6 km grid cell (WW3 resolution), or 4.5 km × 4.5 km grid cell (1/25° CICE resolution). The drill holes were made near the center of each floe which may be the thickest part of the floe (M. Muller, J. Rabault, personal communication). It is also possible that there was a bias associated with selection of floes for buoy deployment.
- <u>SMOS</u>. As we discussed in Section 5.3.1, SMOS has several concerns. First, it was calibrated for cases of new, level ice. Second, the product was calibrated for 100% ice concentration. We do not know how accurate it is for broken ice floes in the MIZ, e.g., with 50% ice concentration. Third, the radiometer signal saturates when $h_{ice}$ is much more than 50 cm. Though U. Hamburg claims that it can be used for thicknesses up to 150 cm, we still suspect that the product is less reliable beyond 50 cm. For example, if the true ice thickness is 150 cm, what would SMOS report? We do not know.

### 5.3.4. *Comparisons coincident with in-situ measurements*

The $h_{ice}$ colocations are compared in Figure 48 to Figure 50. The SMOS product was missing on some days, so there are 31 co-locations with SMOS, rather than 33. At all but three of the co-locations, we find that $h_{ice,model} > h_{ice,in\_situ} > h_{ice,SMOS}$, and for all co-locations, $h_{ice,model} > h_{ice,in\_situ}$. In Figure 48 and Figure 49, we see that $h_{ice,in\_situ}$ is negatively correlated with both $h_{ice,model}$ and $h_{ice,SMOS}$. There is positive correlation between $h_{ice,model}$ and $h_{ice,SMOS}$, however (Figure 50). These correlation outcomes may be associated with our question that the in situ measurement may not be representative of the larger area, while in the case of $h_{ice,model}$ and $h_{ice,SMOS}$, both are representative of a larger area (9 to 12 km squared). It is especially useful to keep in mind our belief that if the $h_{ice,in\_situ}$ is biased, it is probably biased high, and yet the $h_{ice,model}$ is *even thicker, at all locations*. This tells us that the model ice thickness is almost certainly biased high.

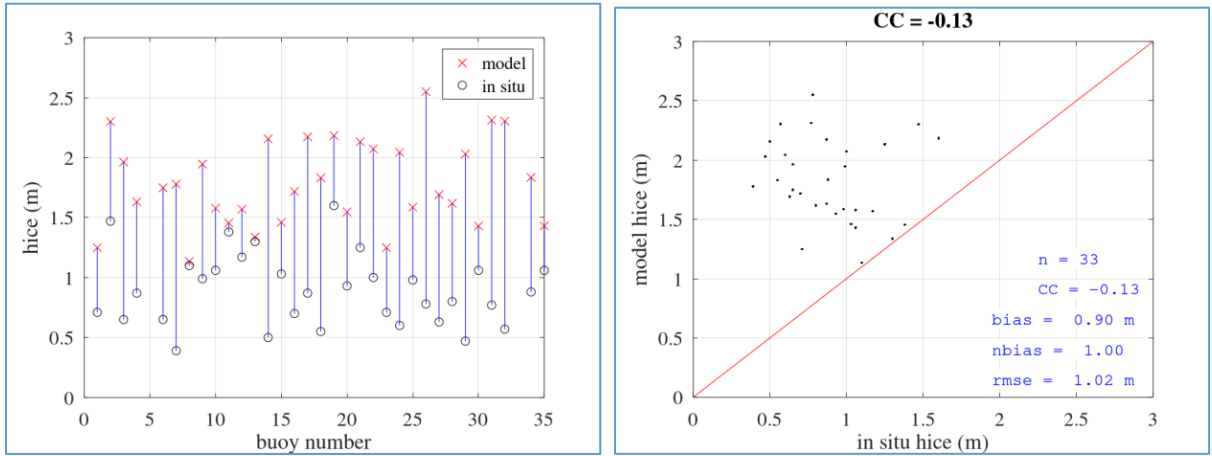

Figure 48. Comparison of ice thickness reported by GOFS/CICE vs. that reported from in situ measurement during buoy deployment. Left panel: Vertical axis is ice thickness and horizontal axis is the integer buoy number. Right panel: a scatter plot; vertical axis = model ice thickness and horizontal axis = in situ ice thickness.

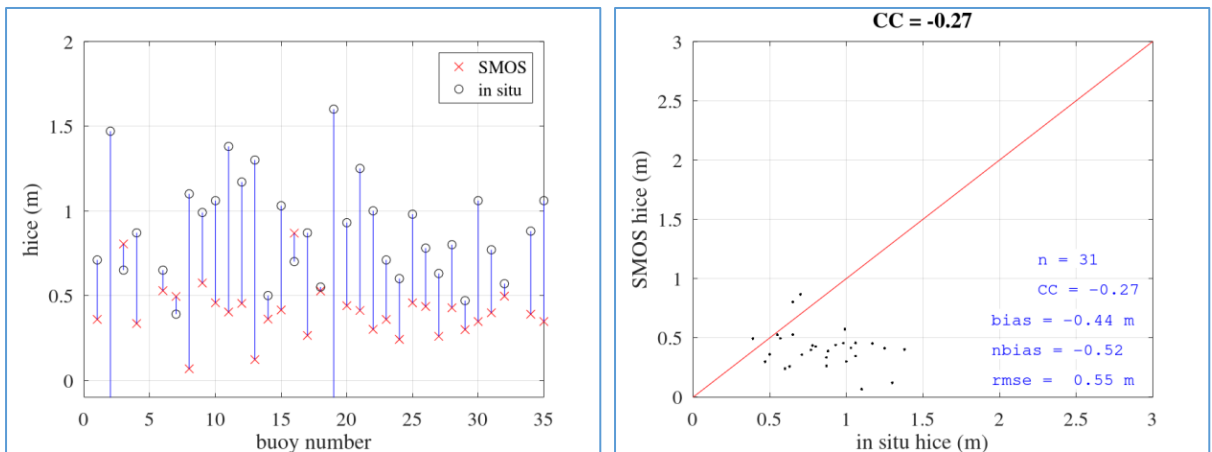

Figure 49. Similar to Figure 48, but comparing SMOS ice thickness vs. in situ ice thickness.

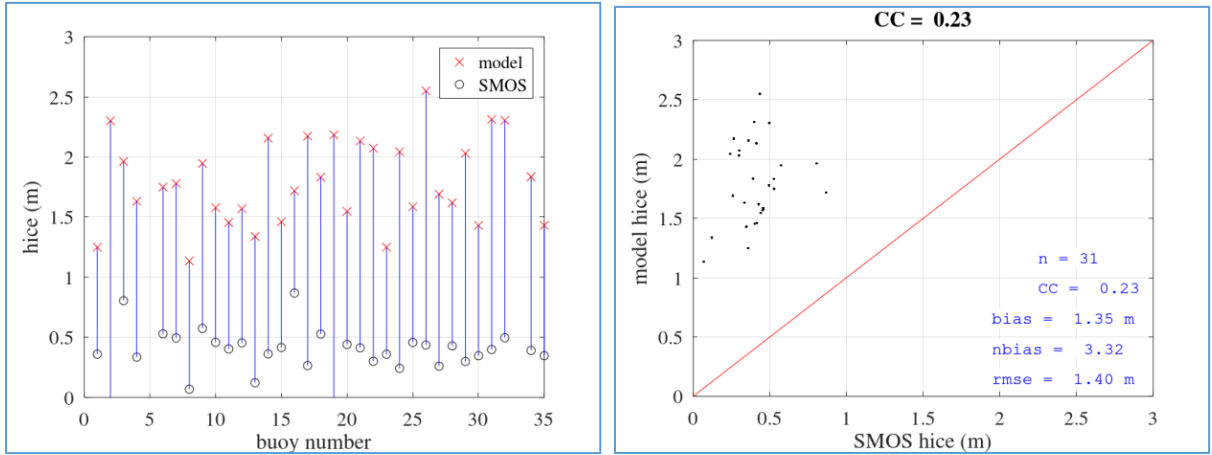

Figure 50. Similar to Figure 48, but comparing model ice thickness vs. SMOS ice thickness.

The figures above do not convey information about the spatial variability of $h_{ice}$ or the location of the in situ measurement. We created a plot for each of the 33 *in situ* deployments with the location marked on a plot of $h_{ice,model}$. Four examples are provided here (Figure 51). Notably, in some cases the buoy is very close to the coastline, e.g., within one grid cell's width.

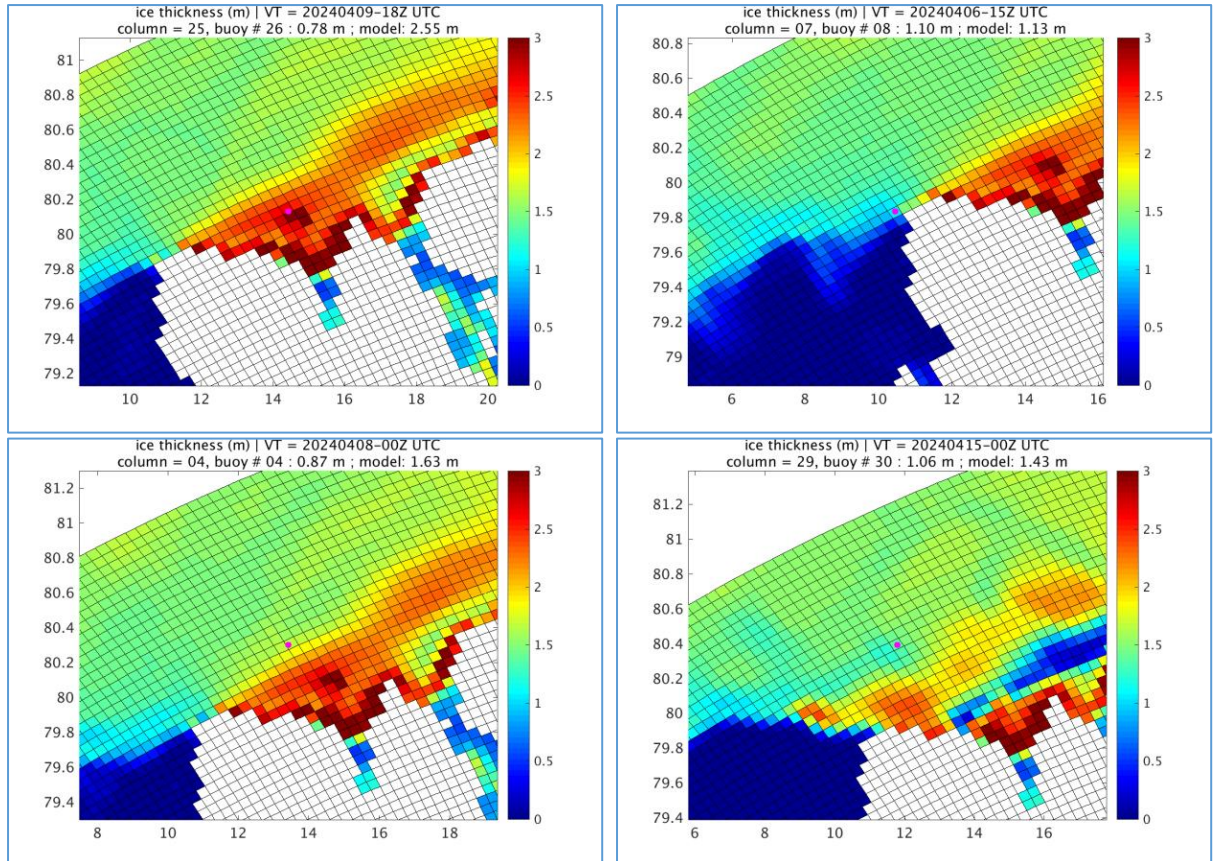

Figure 51. Model ice thickness (m) at four times (examples from 33 such plots created). The ice thickness from CICE has been regridded to the WW3 6 km grid. The times are indicated above each plot, and each time is within two hours of the time of an in situ measurement of ice (i.e., a buoy deployment). The buoy number, in situ ice thickness measurement $h_{ice,in\_situ}$ and model ice thickness $h_{ice,model}$ at the in situ location is also indicated above each plot. The location of the in situ measurement is indicated with a magenta dot.

### 5.3.5. SMOS vs. model: large-scale comparisons

In the case of SMOS and the model, it is not necessary to limit our comparison to the locations where in situ measurements were made. Many co-locations are possible. Though it is possible to perform long-term global comparison, we limit our scope to the location of the WW3 6 km grid. We remapped SMOS to this grid, as shown in the top panels of Figure 52. The SMOS gridded field represent a one-day period. Since the SMOS field has a median relevant time of 1200 UTC, we compare with the model field on the same day at 1200 UTC, also remapped to the WW3 grid (lower panel).

With the two estimates on a common grid, i.e., co-located, it is then possible to make scatter plots. Two examples are shown in Figure 53. As noted already, the SMOS estimates are unreliable for thick ice. Recall from Section 5.3.1 that the SMOS data files include two parameters relevant to this:

- Sea ice thickness 'total uncertainty', in meters
- Saturation ratio, which is the ratio of retrieved ice thickness and maximal retrievable ice thickness, as a percentage

We experimented with three methods of quality control: 1) limit on $h_{ice}$, 2) limit on the 'total uncertainty, and 3) limit on 'saturation ratio'. Methods (1) and (2) would be redundant if 'maximal retrievable ice thickness' is a fixed value. Experiments with the three methods gave mostly similar results, implying that the three methods are semi-redundant, so in Figure 53, we only show the result using the saturation ratio for QC. The outcome without any QC applied is also shown. We find that by applying the QC, the Pearson correlation increases from 0.67 to 0.78, which seems to justify the application. Interestingly, even without QC, the correlation is much higher than the SMOS vs. model comparison shown in Figure 50 (CC=0.23). We performed experiments to determine whether this discrepancy may be associated with the small sample size (n=33) of Figure 50 by taking many random sub-samples of size 33 from the larger dataset of Figure 53. They robustly indicate a correlation between 0.68 and 0.88, indicating that the sample size did not the cause the low correlation in Figure 50 (CC=0.23). The apparent discrepancy may be associated with the fact that the 33 sample locations used in Figure 50 were in and near the MIZ, whereas the comparison of Figure 53 includes many points further inside the ice pack.

In any case, the fair correlation (CC=0.78) between $h_{ice,model}$ and $h_{ice,SMOS}$ is a positive outcome and may be exploited to our advantage. In the remote sensing community, it is common to make minor "slope and offset" corrections to a satellite product, based on historical comparisons of the satellite product against a more trustworthy ground truth dataset and performing linear regression. Something similar could be done here, to answer the question "SMOS is not available for the time/location that I need to force the wave model, but given a value of $h_{ice,model}$, what is the most likely $h_{ice,SMOS}$ that would have been reported if SMOS *was* available for this time/location?" This approach could be taken using scatter plots like Figure 53. A best fit line (or curve) could be used to make a "correction" to the model ice thickness. Admittedly we have not proven that $h_{ice,SMOS}$ is more accurate than $h_{ice,model}$, but the three-way comparison with $h_{ice,SMOS}$ in Section 5.3.4 does suggest that $h_{ice,model}$ is biased high. Moreover, we have two reasons to believe that our $S_{ice}$ formula IC4M9 will perform better with $h_{ice,SMOS}$, regardless of the accuracy of $h_{ice,SMOS}$ relative to that of $h_{ice,model}$: 1) the formula was calibrated using $h_{ice,SMOS}$, and 2) wave spectrum evaluations in Section 5.2 suggest many cases with overprediction of dissipation by IC4M9 which would be improved with a thinner ice forcing. Such an ad hoc "correction" could be introduced either by pre-processing of $h_{ice,model}$ or as an adjustment to the parameter within the IC4 routine.

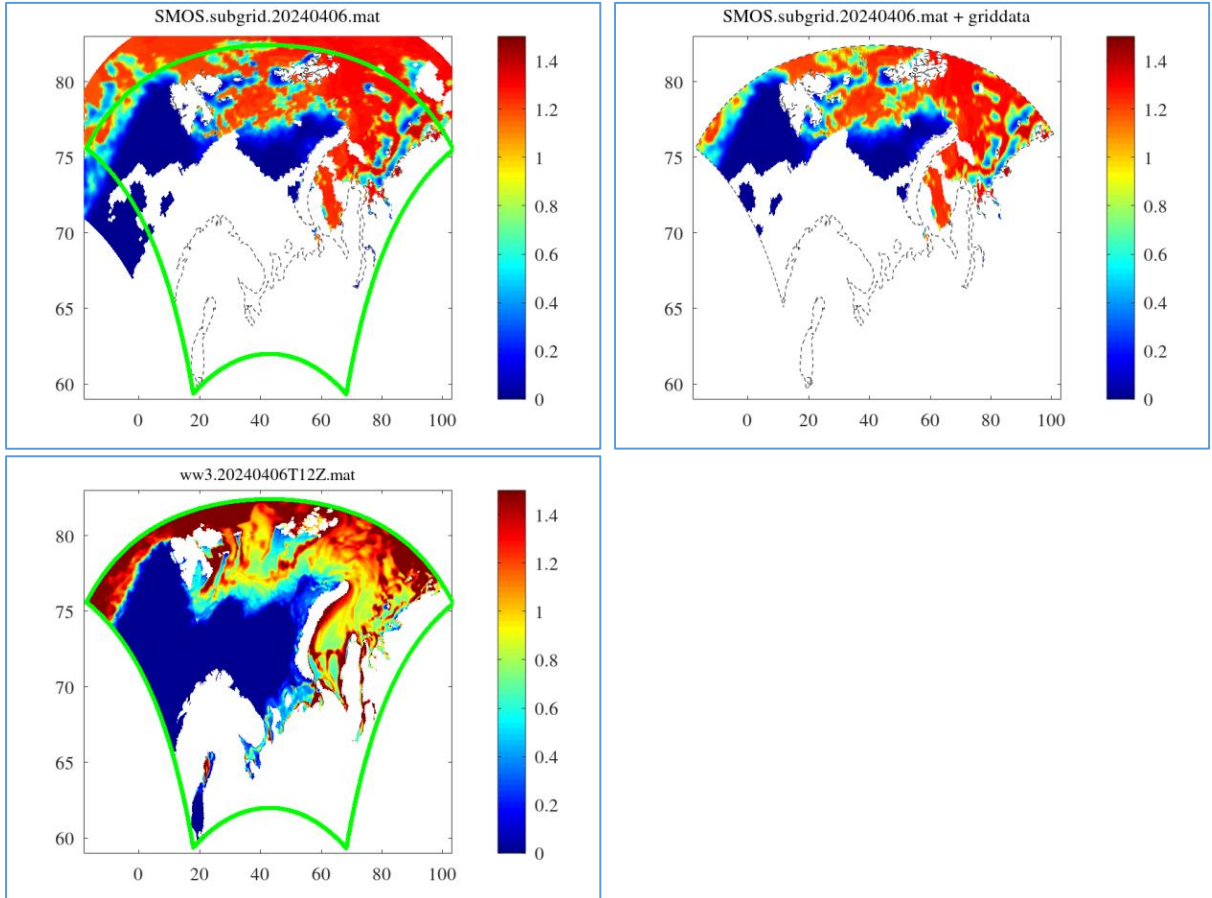

Figure 52. Ice thickness in meters. Upper left: SMOS product on its native 12 km grid for 6 April 2024. White areas indicate that no valid information is available for that day, either because of open water, absence of observations, or data quality reasons. Thick light green line indicates the bounds of the WW3 Barents Sea grid. Upper right: SMOS product remapped to the Barents Sea 6 km WW3 grid. Lower left: GOFS/CICE output, remapped to the Barents Sea 6 km WW3 grid, for 1200 UTC 6 April 2024.

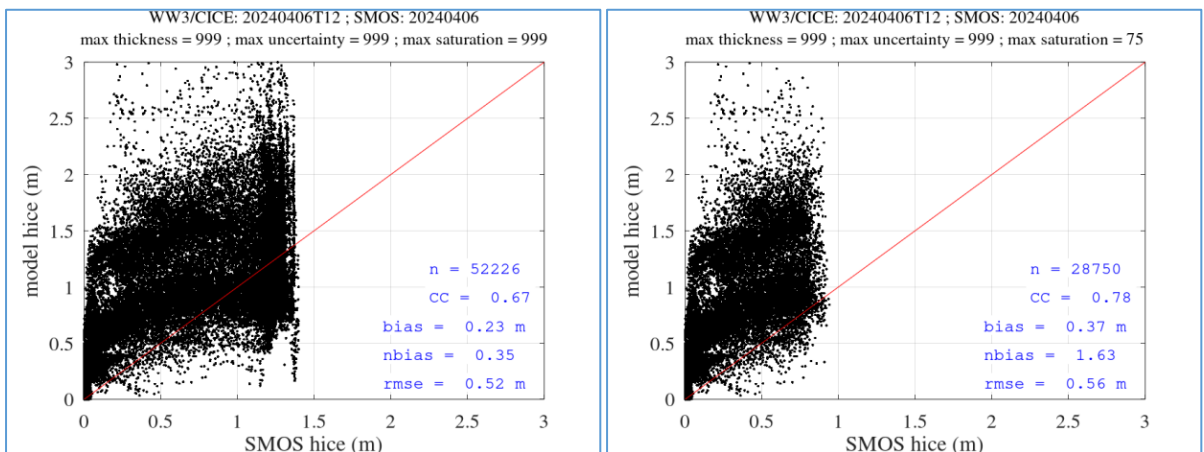

Figure 53. Comparison of ice thickness in meters of the SMOS remapped to WW3 grid (upper right panel of Figure 52) vs. CICE remapped to WW3 grid (lower panel of Figure 52), for 6

April 2024. Left panel: comparison without any QC applied. Right panel: comparison after excluding cases where the saturation ratio is greater than 75%.

## 5.4. Comparisons to observations: Summary

The primary outcomes of the comparisons to observations follow.
- Though it is not definitively proven, the preceding comparisons strongly suggest that the application of the IC4M9 routine for dissipation by sea ice here could be improved by using an ice thickness product with values that are thinner and therefore more consistent which those given by SMOS. A large fraction of the comparisons to SWIM spectra indicates an overprediction of dissipation that is consistent with this.
- Model accuracy is clearly improved by using higher resolution NAVGEM fields.
- Model accuracy is not clearly dependent on the resolution of WW3 (comparing global WW3 against a COAMPS WW3 that is otherwise similar).
- Wave model accuracy is improved by including surface currents in forcing. The difference is clear, though not large.

**Acknowledgements**

The production of the SMOS sea ice thickness data was funded by the ESA project *SMOS & CryoSat-2 Sea Ice Data Product Processing and Dissemination Service*, and data from April 6 to 15 2024 were obtained from https://www.meereisportal.de (grant: REKLIM-2013-04).

We thank Drs. Malte Muller and Jean Rabault of the Norwegian Meteorological Institute (Meteorologisk institutt) for sharing wave and ice data from SvalbardMIZ2024, and Dr. Ana Carrasco, also of NMI for initiating this collaboration, and all three for feedback on our comparisons to the NMI data.

This work was funded by the 6.4 project 'High-Latitude Operational Situational Awareness'. Richard Allard is the Principal Investigator of the project. The work is sponsored by the Oceanographer and Navigator of the Navy, N2N6E, and the program manager is Dr. William Schulz, Office of Naval Research. This is NRL contribution number NRL/MR/7320--2025/1 and is approved for public release.